# DOES GREEN INNOVATION CROWDS OUT OTHER INNOVATION OF FIRMS? - BASED ON EXTENDED CDM MODEL AND UNCONDITIONAL QUANTILE REGRESSIONS

*Yi Yiang*[1*] *and Richard S.J. Tol*[1,2,3,4,5,6,7]

[1] *Department of Economics, University of Sussex, Falmer, BN1 9SL, UK*
[2] *Institute for Environmental Studies, Vrije Universiteit, Amsterdam, The Netherlands*
[3] *Department of Spatial Economics, Vrije Universiteit, Amsterdam, The Netherlands*
[4] *Tinbergen Institute, Amsterdam, The Netherlands*
[5] *CESifo, Munich, Germany*
[6] *Payne Institute for Public Policy, Colorado School of Mines, Golden, CO, USA*
[7] *College of Business, Abu Dhabi University, UAE*
\* <u>maggiejiang16@gmail.com</u>

## Abstract

In the era of sustainability, firms grapple with the decision of how much to invest in green innovation and how it influences their economic trajectory. This study employs the Crepon, Duguet, and Mairesse (CDM) framework to examine the conversion of R&D funds into patents and their impact on productivity, effectively addressing endogeneity by utilizing predicted dependent variables at each stage to exclude unobservable factors. Extending the classical CDM model, this study contrasts green and non-green innovations' economic effects. The results show non-green patents predominantly drive productivity gains, while green patents have a limited impact in non-heavy polluting firms. However, in high-pollution and manufacturing sectors, both innovation types equally enhance productivity. Using unconditional quantile regression, I found green innovation's productivity impact follows an inverse U-shape, unlike the U-shaped pattern of non-green innovation. Significantly, in the 50th to 80th productivity percentiles of manufacturing and high-pollution firms, green innovation not only contributes to environmental sustainability but also outperforms non-green innovation economically.

# 1. INTRODUCTION

Technological innovation is pivotal in elevating environmental performance, enhancing resource efficiency, minimizing production pollution, and fostering sustainable products. While firms are integral to the propagation and integration of green innovation (GI), and assume environmental responsibility (Marin, 2014), the economic ramifications of GI on firm productivity significantly influence a company's inclination towards environmental research and development (R&D) investments. This paper seeks to ascertain if environmental R&D can offer companies tangible benefits that harmonize environmental conservation with economic competitiveness.

To unravel this, the study adopts the venerable CDM model, offering a meticulous examination of R&D endeavors through a structured three-step equation framework, scrutinizing the dynamics between R&D investments, R&D outcomes, and overall productivity. My inquiry stands unique as it pioneers the exploration of GI's crowding-out effects through an extended CDM paradigm that discovers economic opportunity costs of green and non-green innovations, leveraging firm-level data from China. Significantly, I delve into the intricate, non-linear interplay between innovation and productivity via unconditional quantile regression (UQR) estimations.

The CDM model, originally developed by Crépon, Duguet, and Mairessec in 1998, plays a pivotal role in elucidating the intricate nexus between firm-level productivity, innovation, and R&D engagement. Its robustness lies in its adeptness at mitigating challenges like endogeneity issues, self-selection and omitted variable bias, simultaneous equation discrepancies, and reverse causality through a gamut of predictions based on independent covaraites in three equations of the framework.

The first R&D equation of the model leverages a Heckman selection model to rectify potential self-selection biases arising from the R&D investments information disclosure, allowing for the prediction of R&D intensity from a firm's R&D investment and a set of extrogenous variables. The second patent equation involves the utilization of count data models, specifically designed to handle the discrete nature of patent outputs, to estimate patent intensity from the sanitized prediction of lagged R&D intensity, thus overcoming issues of simultaneity and omitted variable bias. The final productivity equation deftly combines the predicted patent intensity with firm productivity, providing a refined estimation of the influence of innovation outputs on productivity levels, while meticulously controlling for the endogeneity that often plagues concurrent assessments of innovation and productivity.

This three-tiered econometric approach is uniquely suited for the current study due to its capability to provide a nuanced understanding of the innovation-productivity paradigm, particularly within the context of Chinese listed firms, where the traditional methodologies fall short in capturing the complex dynamics at play. By integrating the extended CDM framework that disentangle the influences of green and non-green innovations, this study stands on solid methodological ground, paving the way for robust and reliable inferences about the economic opportunity costs of green innovation.

This study also extend the classic CDM model to explore the complex dynamics affecting firm-level R&D input and outcomes, emphasizing the profound influence of environmental regulations, market conditions, and ownership structures. It pioneers the use of provincial pollution charges and industrial pollution treatment investment as nuanced proxies to dissect the impact of environmental policies on R&D intensity. The findings suggest a negative correlation, indicative of a crowding-out effect where environmental protection costs could deter broader innovation resources.

In addressing the critical evaluation of R&D outcomes, this study scrutinizes the limitations of the patent equation in conventional CDM model, which estimates firm-level patent intensity as a function of R&D intensity and observable controls. A notable deficiency of this approach is its oversight of unobserved, time-invariant factors influential in the patenting process. To rectify this, the current analysis presents two novel methodological enhancements aimed at refining the prediction accuracy of patent intensity.

First, the study innovatively calibrates the predicted patent counts by incorporating a firm's average patenting activity observed from 2010 to 2018, thus encapsulating the impact of firm's consistent, unobserved patenting preference. This refinement of patent equation prediction preserves the exogeneity of the productivity equation by ensuring the time-invariant factors, captured within a firm's average patent history, remain uncorrelated with the temporal variability of the error term in the productivity estimation.

Second, to rectify the zero-value issue in patent intensity calculations, the method introduces a nominal constant (0.001) to all predicted patent counts, ensuring valid logarithmic transformations for zero values. Normally, patent intensity is calculated by dividing patent counts by employee numbers. However, for artificially adjusted zero values, this division is omitted to avoid introducing non-existent variance, thereby maintaining the integrity of the patent intensity measurement across the spectrum of innovative activity.

These methodological adjustments, pivotal for the accurate computation of patent intensity, are central to the extended CDM framework and subsequent estimations, including the Unconditional Quantile Regression (UQR), Regression-based Inference Function (RIF) treatment effects and Conditional Quantile Regression (CQR) models.

In advancing the understanding of the relationship between GI and productivity, this study diverges from Marin's (2014) linear perspective, offering a nuanced examination through UQR. It unveils a U-shaped relationship between non-green innovation intensity and productivity, contrasting with an inverse U-shaped relationship for GI across firms' productivity levels.

This analysis introduces a novel perspective on the impact of GI on firm productivity, challenging Marin's earlier findings that suggested a negative impact of GI on polluting firms' productivity. The results of this study are more nuanced, revealing that non-polluting firms face a crowding-out effect from GI due to its high opportunity costs relative to non-green innovations. In contrast, polluting firms demonstrate similar marginal effects from both GI and other innovations on productivity.

The study also incorporates a RIF treatment effects approach, introducing a binary indicator for the presence of green patents as a treatment variable. Model estimations, with and without Inverse Probability Weighting (IPW), suggest that high-polluting firms only witness GI's productivity benefits at or above the median of the productivity distribution, implying a threshold of operational efficiency is necessary for realizing the gains from GI. In stark contrast, firms not facing pollution concerns require major non-green innovations to maintain productivity growth. Notably, for these firms, GI inversely correlates with productivity at the highest echelons (top 20%), signaling a reversal from benefit to burden.

This study's findings not only challenge the established narrative but also signal strategic implications for firms with varied pollution profiles, suggesting that a one-size-fits-all approach to innovation may not be universally beneficial.

## 2. LITERATURE REVIEW

The intersection of GI with a firm's core competencies is nuanced. While traditional product or process innovations are directly tied to a firm's competitive edge, GI is often viewed as an ancillary commitment that could impinge on profitability, given that GI is not typically a primary factor in consumer choice, particularly in developing nations where consumers may be less inclined to absorb the additional costs tied to environmental stewardship.

Conversely, unchecked pollution generates negative externalities that distort market prices. This underscores the importance of consistent and comprehensive environmental policies to correct market failures by attributing economic value to environmental goods and services (Fankhauser et al., 2013). Without mandated environmental policies, firms that do not invest in environmental protection may benefit from free-riding on the efforts of others. This places companies that do invest in environmental protection and GI at a competitive disadvantage. The Porter hypothesis suggests that well-structured environmental policies could incentivize GI, thereby fostering both firm efficiency and environmental preservation, potentially leading to a win-win scenario (Porter and Linde, 1995).

The literature on GI has historically centered on its drivers, economic impacts, environmental benefits, and policy incentives (Barbieri et al., 2017). Recently, some studies have begun to examine the possible trade-offs between GI and other forms of innovation, invoking the theory of opportunity costs (Popp and Newell, 2009; Marin, 2014). At the corporate level, the private opportunity costs of GI are the forgone investments in alternative innovation pursuits (Popp and Newell, 2009). The resource-based view posits that firms are constrained by limited funding, time, and human resources, which implies that diverting resources to GI could detract from the development of product and process innovations, with small and entrepreneurial firms in developing economies being especially susceptible to such constraints (Edeh and Acedo, 2021).

Scholars have approached the notion of crowding-out effects from varied perspectives, focusing on both the quantitative (input/output) and economic returns of GI (Popp and Newell, 2009; Marin, 2014; Marin and Lotti, 2017). Popp and Newell (2009) propose that a rise in environmental patents at the expense of other types of innovation may signal a crowding-out effect. They scrutinize this phenomenon from three angles: inter-sectoral dynamics, intra-sectoral allocation, and the comparative social value of different R&D investments. Marin (2014), on the other hand, measures crowding-out in terms of economic returns, suggesting that investments in GI that yields fewer patents or smaller productivity gains than other innovations indicates an inefficiency and a potential crowding-out effect.

The empirical evidence on GI's crowding-out effects on other forms of innovation is mixed, with studies showing varied outcomes based on the subjects and contexts investigated. Popp and Newell (2009) find crowding-out effects of energy innovations on other innovations within industries but no evidence of cross-industry crowding-out effects. Marin (2014) reports a slight negative impact of GI on overall manufacturing productivity and points out that green patents are less beneficial than other patents. Moreover, the study highlights a lower efficiency in converting R&D investments into GI as compared to non-green innovations. Contrasting these findings, echoing the positive impacts of environmental policies, Zhu et al. (2019) find that China's emission trading pilots enhance low-carbon innovation by 5-10% in firms, without crowding out other technological innovation.

When it comes to environmental regulation, firms may experience crowding-out effects if the costs of complying with such regulations—through investments in pollution control, for

instance—divert resources away from R&D. Yuan and Zhang (2017) observed that environmental regulations could have a short-term negative impact on industrial R&D investment intensity but may promote R&D in the longer term.

Previous applications of the CDM model have primarily examined the interplay between innovation and productivity in contexts such as Italian manufacturing sectors (Marin, 2014; Marin and Lotti, 2017) and SMEs in Sub-Saharan Africa (Edeh and Acedo, 2021), with further industry-focused studies in China (Yuan and Zhang, 2017; Yuan and Xiang, 2018). However, these studies have not delved into firm-level analysis within China, leaving a significant gap in understanding the specific impacts of environmental innovation at this granular level. This study addresses this gap by adapting the CDM model to scrutinize the nuanced effects of GI on firm productivity in China, thereby contributing a novel perspective to the existing body of research.

This paper is structured to first estimate the relationship between innovation activities and productivity using the CDM model, then explore the crowding-out effects of GI using an extended CDM model, and finally, apply an unconditional quantile regression (UQR) to assess the non-linear impacts of patents on firm productivity. Robustness tests that include Regression-based Inference Function (RIF) and conditional quantile regression (CQR) will further scrutinize the relationship between GI and productivity.

## 3. METHODOLOGY AND DATA

The CDM model, established by Crépon et al. (1998), offers a structural framework to dissect firms' innovation processes, examining how R&D investments yield patents and, ultimately, enhance productivity (Hall and Mairesse, 2009). This three-stage model employs instrumental variable techniques by using predicted value of R&D input and outcomes, to circumvent endogeneity, simultaneity and reverse causality, thus clarifying the causal links between R&D, innovation outputs, and productivity. Marin et al. (2014) expanded this model, distinguishing between environmental and other forms of innovation to infer crowding-out effects. Applying the classic and extended CDM models, this study analyzes the full sample of China's listed firms and six sub-samples, differentiating high-pollution from non-pollution firms and contrasting high-tech with low-tech, and manufacturing with non-manufacturing sectors.

3.1  R&D equation

In addressing R&D investment decision biases, this study employs a modified Heckman model. In the CDM framework, the Heckman model originally addresses self-selection bias in R&D investment decisions. This study modifies the model to focus on the self-selection bias in the disclosure of R&D expenditure among Chinese listed firms, all of which engage in R&D activities[1]. Approximately 32.9% of firm-year observations between 2010 and 2018 lack R&D expenditure data disclosure, likely influenced by firms' reputation concerns or investor confidence. The modified approach uses a probit model for the Inverse Mill's Ratio (IMR) in its first step, estimating the probability of R&D expenditure disclosure based on various controls and exclusion criteria. The IMR then corrects for bias in the second step's R&D intensity assessment. This approach, previously adopted in studies like Edeh and Acedo (2021) and Wang et al. (2021), effectively handles the non-disclosure biases in R&D expenditure. Details of the equations for the IMR calculation and bias correction are provided.

---

[1] Being large entities, all Chinese listed firms included in this study engage in R&D activities. The absence of R&D expenditure data in certain cases is attributed to non-disclosure rather than a lack of R&D activity, as evidenced by their fully-disclosed and positive patent data.

$$RD = \beta X + \eta D + \mu \quad (1)$$

$$D^* = \alpha_0' Z + \alpha_1' X + \varepsilon \quad (2)$$

$$RD = \beta' X + \eta D + \rho IMR + \epsilon \quad (3)$$

Equation (1) represents the R&D investment ($RD$) model that includes the control variables $X$ and R&D disclosure dummy variable $D$, without utilizing the Heckman model. If unobservable self-selection factors that affecting $RD$ and $D$ simultaneously exist (reflected by the correlated error term $\mu$ in (1) and $\varepsilon$ in (2)), the estimation of $\eta$ may be biased. The Heckman model is employed to control for this bias by constructing IMR through equation (2) and substituting it into equation (3). If the coefficient $\rho$ on the IMR is statistically significant, it indicates the presence of a self-selection problem, validating the correction for bias using the Heckman model.

To ensure equation (3)'s validity and address potential multicollinearity between control variable $X$ and the IMR, equation (2) includes exclusion restriction variables ($Z$). These variables, uncorrelated with RD, solely impact D, helping address the potential multicollinearity problem between $X$ and the IMR in equation (3), as the estimation of the IMR also relies on $X$ in equation (2). Finally, to confirm no severe multicollinearity in equation (3) due to IMR, I use the Variance Inflation Factors (VIF) indicator, seeking VIF values less than 10 in the test results.

In R&D equation of CDM model, the selection of multiple exclusion restrictions, crucial for model sensitivity analysis (Lennox et al., 2012), includes environmental performance disclosure (EPD) as a dummy variable, logged total assets, and IPO age. EPD, a novel exclusion restriction in this study, shares similarities in R&D information disclosure tendencies but doesn't directly affect R&D intensity. Following Marin (2014), IPO age and total assets are also used, under the premise that while they may correlate with R&D disclosure, they don't directly impact

R&D intensity. R&D intensity ($RDINT$) is then predicted as logged R&D expenditure per thousand employees, using the Heckman model's estimation.

$$RDINT_{i,t} = \alpha_0 lnPPC_{i,t-1} + \alpha_1 LEV_{i,t} + \alpha_2 lnEMP_{i,t} + \alpha_3 lnCAPINT_{i,t} + \alpha_4 lnPCINT_{i,t} + \alpha_5 CR4_{i,t} + \alpha_6 SOE_i + \eta D_{i,t} + \rho IMR_{i,t} + \omega_t + \delta_t + \varepsilon_{i,t} \quad (4)$$

The environmental policy indicators use the logarithm of lagged provincial pollutant charge ($lnPPC_{i,t-1}$), a fee levied on businesses for pollutant emissions, including sewage, waste gas, and hazardous waste[2] (Guo et al, 2019). A higher $PPC$ indicates greater environmental costs for firms, thereby serving as a valid indicator for measuring the crowding-out effects of environmental policy on a firm's R&D investments.

To enhance robustness, the analysis incorporates a two-year lagged PPC ($lnPPC_{i,t-2}$) as an alternative proxy, inspired by Yuan and Zhang (2017) who found that a one-year lag in environmental regulations reduced R&D spending, while a two-year lag enhanced it. Additionally, the one-year lagged provincial investments in industrial pollution treatment ($lnIPT_{i,t-1}$) is used, which capture the total expenses incurred by industrial firms for pollution treatment at the provincial level. These expenses are funded by pollution charges, government subsidies, and enterprise self-financing. This measure not only reflects the financial burden of environmental protection on firms but also indicates the intensity of local environmental policy implementation.

---

[2] Under China's 2003 "Measures for the Administration of Pollutant Discharge fees," industrial and commercial entities emitting pollutants are charged a fee, which doubles if emissions exceed national standards. Consequently, a uniform pollution levy rate applies nationwide, meaning a higher provincial pollutant charge indicates greater total emissions within that province.

3.2 <u>Patent equation</u>

After deriving unbiased R&D intensity predictions ($\widehat{RDINT}_{i,t}$) from the Heckman model, this variable was used as a proxy for R&D input in the patent equation, focusing on patent applications as the dependent variable. I primarily used the negative binomial (NB2) model, with the Poisson model as a baseline, due to the uneven distribution of patent count data. The NB2 model was chosen over the Poisson model, which often fails to account for real-world data overdispersion, as it allows conditional variance to be a quadratic function of the mean (Marin, 2014).

$$PAT_{i,t} \left( \frac{ECO_{i,t}}{NECO_{i,t}} \right) = \lambda_0 \widehat{RDINT}_{i,t} + \lambda_1 lnFSTK_{i,t} + \lambda_2 lnPTL_{i,t} + \lambda_3 lnEMP_{i,t} + \delta_t + \varepsilon_{i,t}$$

(5)

Both NB2 and poisson models use individual fixed effects, removing time-invariant characteristics of each firm. Year dummies ($\delta$) are included as time fixed effects, along with Firm size ($EMP$), firm patent stock ($FSTK$) and provincial technological level ($PTL$) as controls. Total patent applications ($PAT$) are predicted in the classical CDM model, with green ($ECO$) and non-green ($NECO$) patents analyzed separately in the extended CDM model.

To refine patent count predictions, logarithmic patent predictions undergo exponential transformation into count form, adjusted using each firm's average patent applications from 2010 to 2018. This step accounts for time-invariant unobservable factors related to firm's patent preference, enhancing prediction accuracy without adding time-varying variables to the subsequent productivity equation. A small value (0.001) is added to all predictions for valid logarithmic transformation for cases of zero counts. The predicted patent counts, divided by the number of employees, yield the predicted patent intensity for the productivity equation. For entities with zero predicted patents, log(0.001) is used without dividing by employees, avoiding variance issues in zero patent intensity.

### 3.3 Productivity equation

The final step of the CDM model assesses the effect of predicted patent intensity on productivity. This involves using the logarithm of value added per employee for labor productivity, logged physical assets per employee for capital intensity, and predicted patent applications per employee for patent intensity. Additionally, logged employee numbers test for constant returns to scale. In the extended CDM model (Equation 7), total patent intensity is replaced by green and non-green patent intensities that are obtained from separate estimations of the patent equation.

$$\log\left(\frac{VA}{L}\right)_{i,t+1} = \beta_0 + \beta_1 PATINT_{i,t} + \beta_2 \log(L)_{i,t} + \beta_3 \log\left(\frac{K}{L}\right)_{i,t} + \delta_t + \varepsilon_{i,t} \quad (6)$$

$$\log\left(\frac{VA}{L}\right)_{i,t+1} = \beta_0 + \beta_1 NECOINT_{i,t} + \beta_2 ECOINT_{i,t} + \beta_3 \log(L)_{i,t} + \beta_4 \log\left(\frac{K}{L}\right)_{i,t} + \delta_t + \varepsilon_{i,t}$$

(7)

To sum up, predicted values of R&D intensity and patent intensity address endogeneity, simultaneity and omitted variable issues in the three-step structural model. Due to potential bias in standard error estimates in sequential models, bootstrapped standard errors are used for both the patent and productivity equations, following Marin and Lotti (2017).

#### 3.3.1 Non-linear relationship estimation on producitivity equation

The study then explores the non-linear relationship between patent intensity and productivity, using Unconditional Quantile Regression (UQR). UQR allows for examining the impact of predicted patent intensity at different points of the productivity distribution without relying on specific conditional assumtions that are required by Conditional Quantile Regression (CQR), preferable when conditional relationships are uncertain and complex.

$$\int IF\{y, v(F_Y)\} dF_Y = 0 \quad (8)$$

$$\int RIF\{y, v(F_Y)\} dF_Y = v(F_Y) \quad (9)$$

$$v(F_y) \sim N\left\{v(F_Y), \frac{\sigma_{IF}^2}{N}\right\} \quad (10)$$

$$\sigma_{IF}^2 = \int IF\{y, v(F_Y)\}^2 dF_Y \quad (11)$$

UQR analyzes how changes in explanatory variables, X, affect the unconditional distribution of the response variable, Y (Rios-Avila, 2020). It is based on the idea of recentered influence functions (RIF) for robust estimation against outliers and understanding the distribution structure. Influence functions (IF) utilizes Gateaux directional derivatives to estimate partial effects, revealing how small changes in distribution affect mean values. RIF assess the impact of individual observations on the mean, with IFs having an expected value of 0 (Equation 8), making RIFs' expected value equal to the distributional statistic itself (Equation 9). It is possible to estimate the asymptotic variance of any statistic by estimating the variance of the IF or RIF using sample data (Equation 11). The advantage of using RIF is it can use simple averages to recover the underlying distributional statistics and make it easier to interpretation of regression without computing counterfactual distributions.

$$RIF(Y_i, q_\tau, F_Y) = q_\tau + IF(Y_i, q_\tau, F_y) = q_\tau + \frac{\tau - \mathbb{I}\{Y_i \leq q_\tau\}}{f_y(q_\tau)} \quad (12)$$

IF is the influence of an individual observation i on the $\tau^{th}$ quantile ($q_\tau$) of unconditional distribution of $Y_i$ (firms' productivity in this case), $F_Y$ is the cumulative distribution function of $Y_i$. $\mathbb{I}\{Y_i \leq q_\tau\}$ takes 1 if a firm's productivity below or equal to $\tau^{th}\ quantile$. $f_y(q_\tau)$ is the marginal density of $Y_i$ at quantile $\tau$ that is estimated based on kernal density distribution. To sum up, RIF equals to original distributional statistic at quantile $q_\tau$ plus the marginal effects of $Y_i$ made on the quantile $\tau$ of the distribution.

$$\widehat{\beta_\tau} = (\sum_{i=1}^N X_i \cdot X_i')^{-1} \sum_{i=1}^N X_i \widehat{RIF}(Y_i, \hat{q}_\tau, F_Y) \quad (13)$$

The coefficient matrix is estimated as the Equation 13, in which it computes sample quantile of marginal distribution and acquire density estimate by kernel density to obtain RIF. The coefficient represents the marginal effect estimates of an infinitesimal location shift in the distribution of covariates X on $\tau^{th}\ quantile$ of unconditional distribution of firms' productivity, ceteris paribus. The study employs individual-time fixed effects RIF regression to examine the heterogeneity in the effects of green and non-green innovations across productivity distribution in the Extended CDM model[3].

### 3.3.2   Robustness checks for non-linear relationship estimations

#### 3.3.2.1 RIF regression with treatment effects

To check the robustness of non-linear relationship estimates in the productivity equation, a dummy variable for green patent ownership is used as a treatment variable in RIF regressions with generalized influence treatment effect, identifying effects of GI investments. However, standard RIF is limited to local approximations, especially for categorical variables. To meet the confoundedness assumption, it requires to overcome potential correlations between GI and control variables. The model incorporates Firpo and Pinto's (2016) Inverse Probability Weighting (IPW) method that combines parametric and non-parametric approaches. IPW estimates the overall treatment group gap directly on the dependent variable's distribution, rather than conditioning on explanatory variables, crucial for causal inference from observational data (Bui and Imai, 2018).

---

[3] rifhdreg with abs(id year) command allows controlling for high-dimensional fixed effects, which is equivalent to rifhdfe fixed effects settings and estimate within-firm effects when controlling individual-time effests. Moreover, it sets to report robust standard errors rather than default OLS asympototic standard errors.

$$\hat{F}_{Y_k|T=k} = \int_{i \epsilon k} F_{Y_k|x} \omega_k(x) dF_{x|t=k} \ for \ k = 0,1 \qquad (14)$$

This approach balances unconditional outcome distributions between treated and control groups by applying the weighting factor $\omega_k(x)$ to the observed data, using a probit model for average distributional treatment effects and presuming exogeneity. This assumes independence between treatment and error term, supported by the use of predicted patent intensity in the productivity equation, which removes unobservable variables from prior steps of CDM model.

Treatment effects of are determined by differences in combined RIFs between firms with and without green patents, $\nu\left(\hat{F}_{Y_1}\right)$ and $\nu\left(\hat{F}_{Y_0}\right)$, based on reweighted cumulative distribution functions, controlling for individual and time fixed effects in RIF regression[4].

$$T \times RIF\left\{y, \nu\left(\hat{F}_{Y_1}\right)\right\} + (1-T) \times RIF\left\{y, \nu\left(\hat{F}_{Y_0}\right)\right\} = \beta_0 + \beta_1 T + \varepsilon \qquad (15)$$

---

[4] To estimate the treatment effects of having green patents on firms' productivity, the command "rifhdreg y x i.T, over(T) rif (q(τ)) rwprobit(x) abs(id year) " is used. The treatment T is treated as a categorical variable, allowing for estimation of productivity at the τ-th quantile for all observations (individual-year) with and without green patents.

3.3.2.2 Conditional quantile regression for panel data

Conditional Quantile Regression (CQR), while common in quantile analysis, has limitations due to its reliance on numerous specific covariates, potentially reducing reliability across different quantiles (Bui and Imai, 2018). However, in our research, CQR is effective in CDM frameworks, using patent predictions derived from earlier stages based on limited covariates, then applied in the productivity equation to mitigate endogeneity issues caused by unobservable variables. CQR, enhanced by bootstrap standard errors, serves as a robust check against UQR results in both CDM and extended CDM models, ensuring reliable estimations.

3.4     Data description

Data on listed firms' total and green patent applications focus on inventions and utility models, excluding design patents due to their limited productivity impact[5]. This study computes non-green innovations by deducting green patent applications from the total. All patent data is calculated using a three-year rolling average to ensure a robust analysis of patent activity trends. To maintain consistency, the study excludes data from the financial sector due to their unique financial statements and innovation patterns. Also omitted are firms from Tibetan provinces, where environmental regulation data (PPC and IPT) is missing[6]. This step minimizes potential biases, considering the minimal number of affected firms. Table A2 provide a comprehensive list and detailed description of all variables used in the model.

---

[5] The data of total and green patents for all listed firms were meticulously sourced from the China Research Data Services Platform (CNRDS), WIND Financial Terminal database, and the China Stock Market and Accounting Research Database (CSMAR).

[6] The sample, spanning all industries except the financial sector, comprises 1,747 firms yielding 15,723 observations from 2010 to 2018 for R&D and patent equation predictions. After excluding firms with missing productivity-related data, the final dataset for individual fixed effect analysis consists of 11,727 observations across 1,303 firms over a nine-year period, ensuring a strong balanced panel.

# 4. RESULTS AND DISCUSSION

## 4.1 R&D equation

*Table 1 R&D equation*

| | Model 1 OLS | Model 2 Heckman | Model 3 OLS | Model 4 Heckman | Model 5 OLS | Model 6 Heckman | Model 7 OLS | Model 8 Heckman | Model 9 OLS | Model 10 Heckman | Model 11 OLS | Model 12 Heckman | Model 13 OLS | Model 14 Heckman |
|---|---|---|---|---|---|---|---|---|---|---|---|---|---|---|
| | Full sample | | High-pollution | | Non-pollution | | High-tech | | Low-tech | | Manufacturing | | Non-manufacturing | |
| Step 2: R&D investment | | | | | | | | | | | | | | |
| lnPPF(-1) | -0.0569 (0.046) | -0.0789*** (0.014) | -0.0353 (0.043) | -0.0519* (0.026) | -0.0475 (0.057) | -0.0692*** (0.016) | -0.120** (0.040) | -0.114*** (0.016) | 0.00356 (0.059) | -0.0329 (0.026) | -0.0839 (0.045) | -0.0902*** (0.020) | -0.135* (0.052) | -0.138*** (0.026) |
| LEV | -1.243*** (0.137) | -0.876*** (0.076) | -1.723*** (0.168) | -1.285*** (0.126) | -1.010*** (0.192) | -0.621*** (0.095) | -1.273*** (0.147) | -0.806*** (0.092) | -1.109*** (0.300) | -0.741*** (0.134) | -1.201*** (0.123) | -0.601*** (0.109) | -1.186*** (0.256) | -0.824*** (0.150) |
| lnEMP | -0.144*** (0.025) | -0.213*** (0.014) | -0.113* (0.054) | -0.233*** (0.026) | -0.124*** (0.031) | -0.189*** (0.016) | -0.0579 (0.032) | -0.119*** (0.016) | -0.248** (0.070) | -0.352*** (0.025) | -0.0825* (0.036) | -0.160*** (0.019) | -0.298** (0.084) | -0.387*** (0.027) |
| lnCAPINT | -0.0857* (0.036) | -0.0795*** (0.013) | 0.0336 (0.073) | 0.155*** (0.031) | -0.0435 (0.045) | -0.0392* (0.017) | -0.0403 (0.027) | -0.0410* (0.017) | -0.0803 (0.055) | -0.0756*** (0.022) | -0.00281 (0.041) | 0.013 (0.023) | -0.192*** (0.045) | -0.170*** (0.022) |
| lnPCINT | 0.368*** (0.050) | 0.379*** (0.015) | 0.414*** (0.047) | 0.327*** (0.032) | 0.313*** (0.060) | 0.337*** (0.019) | 0.367*** (0.045) | 0.374*** (0.019) | 0.411*** (0.065) | 0.403*** (0.026) | 0.465*** (0.044) | 0.441*** (0.025) | 0.238*** (0.052) | 0.243*** (0.027) |
| CR4 | -1.613*** (0.170) | -1.283*** (0.085) | -1.782*** (0.282) | -0.982*** (0.170) | -2.020*** (0.317) | -1.784*** (0.107) | -0.597** (0.188) | -0.497*** (0.099) | 0.0632 (0.285) | -0.209 (0.181) | -0.834*** (0.167) | -0.584*** (0.126) | -1.624*** (0.414) | -1.536*** (0.150) |
| SOE | -0.301*** (0.082) | -0.176*** (0.032) | -0.418*** (0.098) | -0.253*** (0.056) | -0.215* (0.098) | -0.0761 (0.039) | -0.142 (0.075) | -0.144*** (0.035) | -0.510** (0.140) | -0.282*** (0.063) | -0.182* (0.082) | -0.141*** (0.041) | -0.414** (0.122) | -0.184* (0.077) |
| _cons | 1.531** (0.541) | 1.743*** (0.132) | 0.468 (0.474) | 0.841*** (0.243) | 1.785** (0.624) | 1.935*** (0.161) | 1.424** (0.417) | 1.688*** (0.149) | -0.338 (0.523) | 0.618* (0.313) | 0.323 (0.400) | 0.870*** (0.200) | 3.063*** (0.575) | 3.500*** (0.261) |
| Year FE | Y | Y | Y | Y | Y | Y | Y | Y | Y | Y | Y | Y | Y | Y |
| Region FE | Y | Y | Y | Y | Y | Y | Y | Y | Y | Y | Y | Y | Y | Y |
| Step 1: R&D dummy | | | | | | | | | | | | | | |
| AGE | | -0.0708*** (0.002) | | -0.0764*** (0.005) | | -0.0704*** (0.003) | | -0.0854*** (0.005) | | -0.0605*** (0.003) | | -0.0885*** (0.004) | | -0.0649*** (0.003) |
| EPD | | 0.143*** (0.035) | | 0.214** (0.067) | | 0.126** (0.042) | | -0.0823 (0.065) | | 0.168*** (0.047) | | 0.057 (0.059) | | 0.0779 (0.050) |
| lnASSET | | -0.409*** (0.018) | | -0.325*** (0.045) | | -0.391*** (0.021) | | -0.165*** (0.043) | | -0.319*** (0.022) | | -0.168*** (0.039) | | -0.224*** (0.023) |
| lnPPF(-1) | | 0.0312* (0.012) | | 0.0223 (0.026) | | 0.0302* (0.014) | | -0.0946*** (0.025) | | 0.0796*** (0.016) | | -0.0281 (0.023) | | 0.00409 (0.017) |
| LEV | | -0.696*** (0.057) | | -0.587*** (0.107) | | -0.705*** (0.068) | | -0.998*** (0.103) | | -0.394*** (0.076) | | -0.841*** (0.092) | | -0.451*** (0.081) |
| lnEMP | | 0.551*** (0.016) | | 0.543*** (0.042) | | 0.505*** (0.018) | | 0.390*** (0.038) | | 0.457*** (0.020) | | 0.368*** (0.036) | | 0.355*** (0.020) |
| lnCAPINT | | 0.0580*** (0.011) | | -0.0758* (0.031) | | 0.0361** (0.013) | | 0.0059 (0.025) | | 0.0704*** (0.014) | | -0.0265 (0.025) | | 0.0075 (0.013) |
| lnPCINT | | 0.134*** (0.015) | | 0.221*** (0.029) | | 0.116*** (0.017) | | 0.0522 (0.029) | | 0.150*** (0.019) | | 0.123*** (0.027) | | 0.0751*** (0.019) |
| CR4 | | -1.142*** (0.073) | | -1.870*** (0.143) | | -0.768*** (0.091) | | -0.405** (0.138) | | 0.305** (0.109) | | -0.759*** (0.133) | | -0.349*** (0.097) |
| SOE | | -0.212*** (0.027) | | -0.162** (0.057) | | -0.226*** (0.032) | | 0.261*** (0.054) | | -0.294*** (0.036) | | 0.171*** (0.046) | | -0.311*** (0.039) |
| _cons | | 3.095*** (0.136) | | 2.743*** (0.294) | | 3.134*** (0.159) | | 2.985*** (0.287) | | 0.725*** (0.187) | | 2.293*** (0.263) | | 1.629*** (0.180) |
| Year FE | | Y | | Y | | Y | | Y | | Y | | Y | | Y |
| Region FE | | Y | | Y | | Y | | Y | | Y | | Y | | Y |
| /mills | | | | | | | | | | | | | | |
| lambda | | -0.752*** (0.069) | | -1.371*** (0.134) | | -0.732*** (0.081) | | -1.321*** (0.121) | | -0.819*** (0.114) | | -1.660*** (0.138) | | -0.771*** (0.124) |
| rho | | -0.541 | | -0.921 | | -1.000 | | -0.532 | | -0.530 | | -1.000 | | -0.486 |
| sigma | | 1.390 | | 1.488 | | 1.321 | | 1.378 | | 1.544 | | 1.660 | | 1.584 |
| Wald chi2 | | 2454.2 [0.000] | | 939.2 [0.000] | | 1511.3 [0.000] | | 1252 [0.000] | | 967.8 [0.000] | | 887.2 [0.000] | | 1148.7 [0.000] |
| F-value | 183.3 [0.000] | | 170.9 [0.000] | | 99.57 [0.000] | | 191.7 [0.000] | | 66.09 [0.000] | | 165.2 [0.000] | | 136 [0.000] | |
| adj. R-sq | 0.221 | | 0.286 | | 0.197 | | 0.233 | | 0.216 | | 0.23 | | 0.302 | |
| N | 11376 | 15639 | 3830 | 4826 | 7546 | 10813 | 7577 | 8322 | 3799 | 7317 | 8195 | 9187 | 3181 | 6452 |

Standard errors in parenthesis
* p<0.05, ** p<0.01, *** p<0.001

In evaluating the R&D equation, the Heckman selection model, compared with basic OLS, showed no significant multicollinearity (mean VIF = 1.84, see Table B1). The Heckman model's significant lambda (IMR variable) across all samples indicates sample selection bias, effectively addressed by this model (part 3 in Table 1). The first Heckman step revealed environmental performance disclosure (EPD) positively correlated with R&D expenditure disclosure in the full and certain sub-samples, while longer stock market presence (AGE) and larger assets (lnASSET) correlated negatively across all samples, validating their roles as effective exclusion restrictions (part 2 in Table 1).

In the second Heckman step (part 2 in Table 1), environmental regulations (PPF) negatively impacted R&D intensity in all samples but low-tech industries[7], indicating a crowding-out effect. In the full sample, a 1% rise in pollution charges correlates with a 0.08% decrease in R&D spending, where a provincial increase of 7.7 million in charges typically reduces a firm's R&D investment by 0.4 million RMB. Moreover, non-polluting, high-tech, and non-manufacturing firms showed greater sensitivity (higher elasticity) to environmental costs compared to their counterparts.

Other negative influences on R&D investment included financial risk (LEV), employee costs (EMP), production costs (PCINT) and market concentration (CR4). State-owned enterprises (SOE), despite governmental support, were less efficient in technological innovation. Capital intensity (CAPINT) shows a positive relationship with R&D intensity across most samples, except in high-pollution and manufacturing firms. Bartoloni (2013) highlights the crucial impact of capital structure on R&D investment. In capital-intensive sectors (e.g.manufacturing), firms often resort to financing and increasing debt for expansion, leading to a reduced focus on innovation as their total capital, including debt, grows.

---

[7] Low-tech firms, with typically smaller R&D budgets, display negligible changes in R&D investment in response to the environmental policies.

### 4.1.1 Sensitivity analysis with alternative environmental regulation proxies

In assessing the impact of environmental policies on R&D investments, this study applies two alternative proxies: a two-year lagged provincial pollutant charge (PPC) and a one-year lagged industrial pollution control investment (IPT). The results consistently indicate a significant crowding-out effects of environmental costs on R&D investments in most samples, both in the short and long term, contrasting Yuan and Zhang's (2017) findings of a positive long-term impact.

*Table 2 R&D equation with alternative environmental regulation proxies*

|  | M1 | M2 | M3 | M4 | M5 | M6 | M7 | M8 | M9 | M10 | M11 | M12 | M13 | M14 |
|---|---|---|---|---|---|---|---|---|---|---|---|---|---|---|
|  | lnPPC(-2) | nIPT(-1) | lnPPC(-2) | lnIPT(-1) | lnPPC(-2) | lnIPT(-1) | lnPPC(-2) | lnIPT(-1) | lnPPC(-2) | lnIPT(-1) | lnPPC(-2) | lnIPT(-1) | lnPPC(-2) | lnIPT(-1) |
|  | Full sample | | High-pollution | | Non-pollution | | High-tech | | Low-tech | | Manufacuring | | Non-manufacturing | |
| | | | | | | | Step 2: R&D investment | | | | | | | |
| lnPPF(-2) | -0.0771*** | | -0.0399 | | -0.115*** | | -0.0717*** | | -0.0307 | | -0.0900*** | | -0.138*** | |
|  | (0.013) | | (0.025) | | (0.015) | | (0.016) | | (0.025) | | (0.020) | | (0.025) | |
| lnIPT(-1) | | -0.0716*** | | -0.048 | | -0.0972*** | | -0.0667*** | | -0.0595 | | -0.0672** | | -0.166*** |
|  | | (0.016) | | (0.029) | | (0.018) | | (0.019) | | (0.031) | | (0.023) | | (0.032) |
| LEV | -0.875*** | -1.126*** | -1.287*** | -1.294*** | -0.802*** | -0.832*** | -0.619*** | -0.629*** | -0.739*** | -0.744*** | -0.598*** | -0.620*** | -0.828*** | -0.821*** |
|  | (0.076) | (0.075) | (0.126) | (0.126) | (0.093) | (0.090) | (0.095) | (0.095) | (0.134) | (0.133) | (0.110) | (0.107) | (0.149) | (0.150) |
| lnEMP | -0.214*** | -0.194*** | -0.233*** | -0.231*** | -0.120*** | -0.119*** | -0.190*** | -0.189*** | -0.352*** | -0.351*** | -0.160*** | -0.159*** | -0.389*** | -0.386*** |
|  | (0.014) | (0.013) | (0.026) | (0.026) | (0.016) | (0.015) | (0.016) | (0.016) | (0.025) | (0.025) | (0.019) | (0.019) | (0.027) | (0.027) |
| lnCAPINT | -0.0796*** | -0.0809*** | 0.154*** | 0.156*** | -0.0405* | -0.0441** | -0.0389* | -0.0403* | -0.0754*** | -0.0740*** | 0.0131 | 0.0117 | -0.171*** | -0.170*** |
|  | (0.013) | (0.013) | (0.031) | (0.031) | (0.017) | (0.017) | (0.017) | (0.017) | (0.022) | (0.022) | (0.023) | (0.023) | (0.022) | (0.022) |
| lnPCINT | 0.379*** | 0.394*** | 0.326*** | 0.327*** | 0.374*** | 0.371*** | 0.337*** | 0.335*** | 0.404*** | 0.401*** | 0.441*** | 0.440*** | 0.243*** | 0.241*** |
|  | (0.015) | (0.015) | (0.032) | (0.032) | (0.020) | (0.019) | (0.019) | (0.019) | (0.026) | (0.026) | (0.025) | (0.024) | (0.027) | (0.027) |
| CR4 | -1.281*** | -1.312*** | -0.988*** | -0.990*** | -0.494*** | -0.505*** | -1.781*** | -1.788*** | -0.211 | -0.211 | -0.580*** | -0.597*** | -1.533*** | -1.531*** |
|  | (0.085) | (0.084) | (0.170) | (0.169) | (0.100) | (0.097) | (0.107) | (0.106) | (0.181) | (0.180) | (0.126) | (0.124) | (0.150) | (0.150) |
| SOE | -0.177*** | -0.164*** | -0.249*** | -0.248*** | -0.147*** | -0.122*** | -0.0776* | -0.0685 | -0.280*** | -0.292*** | -0.143*** | -0.126** | -0.186* | -0.179* |
|  | (0.032) | (0.032) | (0.056) | (0.056) | (0.035) | (0.034) | (0.039) | (0.039) | (0.063) | (0.063) | (0.041) | (0.040) | (0.077) | (0.077) |
| _cons | 1.739*** | 1.742*** | 0.775** | 0.851** | 1.698*** | 1.698*** | 1.955*** | 1.986*** | 0.610* | 0.832* | 0.875*** | 0.795*** | 3.520*** | 3.813*** |
|  | (0.131) | (0.151) | (0.241) | (0.272) | (0.149) | (0.167) | (0.160) | (0.185) | (0.311) | (0.344) | (0.199) | (0.219) | (0.260) | (0.304) |
| Year FE | Y | Y | Y | Y | Y | Y | Y | Y | Y | Y | Y | Y | Y | Y |
| Region FE | Y | Y | Y | Y | Y | Y | Y | Y | Y | Y | Y | Y | Y | Y |
| Wald chi2 | 2504.110 | 2550.630 | 937.470 | 939.440 | 1240.500 | 1289.610 | 1513.540 | 1505.100 | 966.850 | 972.040 | 882.310 | 904.880 | 1152.440 | 1147.070 |
|  | [0.000] | [0.000] | [0.000] | [0.000] | [0.000] | [0.000] | [0.000] | [0.000] | [0.000] | [0.000] | [0.000] | [0.000] | [0.000] | [0.000] |
| N | 15639 | 15552 | 4826 | 4826 | 8322 | 8322 | 10813 | 10813 | 7317 | 7317 | 9187 | 9187 | 6452 | 6452 |

Standard errors in parenthesis
* $p<0.05$, ** $p<0.01$, *** $p<0.001$

However, high-polluting firms, compelled by strict environmental regulations, demonstrate a necessity to invest in GI, showing less sensitivity to increased environmental costs in both alternative models than non-heavy polluting firms (M3-M6 in Table 2). This disparity is partly due to cost transference within supply chains, with high-polluting firms passing additional environmental expenses onto other businesses. Moreover, the prevailing trend towards environmental responsibility and evolving consumer expectations encourages many low-polluting companies to voluntarily augment their environmental investments, thus diverting resources from R&D activities and exacerbating the indirect impact of environmental policies on these firms' innovation investments.

### 4.2 Patent equation

After employing the Heckman model to predict R&D intensity, this value was used in both the classical and extended CDM models to estimate patent outputs. Table 3 and Table 4 showcase these results, comparing the Poisson and negative binomial models (NB2), which both utilize individual-time double fixed effects and bootstrapped standard errors.

*Table 3 Results of patent equation-CDM*

|  | Model 15 Poisson | Model 16 NB2 | Model 17 Poisson | Model 18 NB2 | Model 19 Poisson | Model 20 NB2 | Model 21 Poisson | Model 22 NB2 | Model 23 Poisson | Model 24 NB2 | Model 25 Poisson | Model 26 NB2 | Model 27 Poisson | Model 28 NB2 |
|---|---|---|---|---|---|---|---|---|---|---|---|---|---|---|
|  | Full-sample | | High-pollution | | Non-pollution | | High-tech | | Low-tech | | Manufacturing | | Non-manufacturing | |
| $\widehat{RDINT}$ | 0.410** | 0.259*** | 0.596*** | 0.361*** | 0.214 | 0.209** | 0.567*** | 0.198*** | 0.429 | 0.254* | 0.569*** | 0.237*** | 0.206 | 0.187** |
|  | (0.131) | (0.058) | (0.144) | (0.075) | (0.154) | (0.081) | (0.133) | (0.053) | (0.307) | (0.102) | (0.113) | (0.072) | (0.281) | (0.066) |
| lnFSTK | 0.00195 | 0.00246 | -0.0014 | 0.00553 | 0.00274 | 0.00217 | 0.00155 | 0.0013 | 0.000856 | 0.00286* | 0.00146 | 0.00331 | -0.00063 | 0.00113 |
|  | (0.002) | (0.001) | (0.006) | (0.003) | (0.002) | (0.001) | (0.002) | (0.002) | (0.002) | (0.001) | (0.002) | (0.003) | (0.002) | (0.001) |
| lnPLT | 0.123 | 0.0788* | 0.319 | 0.107 | 0.0286 | 0.0691 | 0.178 | 0.0855 | -0.0294 | 0.0887 | 0.162 | 0.153** | -0.0181 | 0.102 |
|  | (0.135) | (0.038) | (0.182) | (0.079) | (0.150) | (0.055) | (0.144) | (0.063) | (0.182) | (0.058) | (0.156) | (0.051) | (0.233) | (0.063) |
| lnEMP | 0.420*** | 0.337*** | 0.533*** | 0.339*** | 0.334*** | 0.330*** | 0.418*** | 0.327*** | 0.535** | 0.351*** | 0.447*** | 0.339*** | 0.462** | 0.326*** |
|  | (0.096) | (0.033) | (0.110) | (0.054) | (0.095) | (0.036) | (0.097) | (0.032) | (0.166) | (0.060) | (0.087) | (0.035) | (0.157) | (0.050) |
| _cons |  | -0.730* |  | -0.958* |  | -0.61 |  | -0.396 |  | -1.086* |  | -0.934* |  | -1.240* |
|  |  | (0.296) |  | (0.477) |  | (0.448) |  | (0.422) |  | (0.451) |  | (0.377) |  | (0.485) |
| Individual FE | Y | Y | Y | Y | Y | Y | Y | Y | Y | Y | Y | Y | Y | Y |
| Time FE | Y | Y | Y | Y | Y | Y | Y | Y | Y | Y | Y | Y | Y | Y |
| Wald chi2 | 948.44 | 5007.8 | 694.17 | 776 | 911.23 | 2702.4 | 690.26 | 3502.7 | 442.82 | 1035.8 | 1003.53 | 3362.9 | 481.05 | 755.7 |
|  | [0.000] | [0.000] | [0.000] | [0.000] | [0.000] | [0.000] | [0.000] | [0.000] | [0.000] | [0.000] | [0.000] | [0.000] | [0.000] | [0.000] |
| Log likelihood | -88061.8 | -41633.2 | -23512.4 | -13222.5 | -63628.1 | -28375.6 | -55225.6 | -26176.6 | -31608.7 | -15217.6 | -59116.4 | -28602.3 | -27070.4 | -12714.8 |
| N | 10818 | 13890 | 3771 | 4531 | 7047 | 9359 | 5643 | 7763 | 5175 | 6127 | 6507 | 8613 | 4311 | 5277 |

Standard errors in parenthesis
* p<0.05, ** p<0.01, *** p<0.001

Key findings from Table 3 indicate that increased R&D investment (RDINT) significantly boosts patent applications across all samples, with high-polluting firms exhibiting a notably higher efficiency in converting R&D into patents than the non-polluting firms. This trend is also evident among manufacturing and low-tech firms and their counterparts. However, the influence of regional technology levels (PTL) and firm-level patent stock (FSTK) on R&D output is less significant and has smaller effects, underscoring the primary role of R&D expenditure and human resources (EMP) in driving patent outcomes.

In the extended CDM model (Table 4), R&D intensity positively impacts both green and non-green innovations. High-polluting and manufacturing firms show more efficient conversion of R&D to green patents (rates: 0.628 and 0.314) than to non-green (0.401 and 0.245), indicating a lower opportunity costs of GI in these sectors. In contrast, in the remaining samples with lower pollution issues, GI exhibits a lower R&D conversion efficiency than non-green innovations, indicating GI has crowding-out effects on other innovations.

The findings also highlight the crucial role of firm size and human resources in enhancing R&D efficiency and generating both green and non-green patents. Moreover, the regional innovation level continues to significantly influence non-green patent generation in manufacturing firms, reflecting the importance of regional capabilities in their innovation processes (Model 39).

*Table 4 Results of patent equation- Extended CDM*

|  | Model 29 | Model 30 | Model 31 | Model 32 | Model 33 | Model 34 | Model 35 | Model 36 | Model 37 | Model 38 | Model 39 | Model 40 | Model 41 | Model 42 |
|---|---|---|---|---|---|---|---|---|---|---|---|---|---|---|
|  | NECO | ECO | NECO | ECO | NECO | ECO | NECO | ECO | NECO | ECO | NECO | ECO | NECO | ECO |
|  | Full sample | | High-pollution | | Non-pollution | | High-tech | | Low-tech | | Manufacuring | | Non-manufacturing | |
| $\widehat{RDINT}$ | 0.240*** | 0.238** | 0.401*** | 0.628*** | 0.184 | 0.238 | 0.191*** | 0.165 | 0.176* | 0.116 | 0.245** | 0.314*** | 0.162* | 0.0994 |
|  | (0.060) | (0.074) | (0.084) | (0.108) | (0.114) | (0.145) | (0.047) | (0.087) | (0.071) | (0.124) | (0.079) | (0.092) | (0.070) | (0.103) |
| lnFSTK | 0.00201 | 0.00109 | 0.00587 | 0.00185 | 0.00270* | 0.00193 | 0.00069 | 0.00088 | 0.00173 | 0.00127 | 0.00309 | 0.00186 | 0.00097 | -0.00004 |
|  | (0.001) | (0.001) | (0.004) | (0.004) | (0.001) | (0.002) | (0.002) | (0.002) | (0.001) | (0.002) | (0.003) | (0.004) | (0.001) | (0.002) |
| lnPLT | 0.0677 | 0.0505 | 0.0923 | -0.00589 | 0.0907 | 0.0736 | 0.079 | 0.0702 | 0.0565 | 0.072 | 0.139** | 0.0672 | 0.0814 | 0.16 |
|  | (0.045) | (0.079) | (0.075) | (0.111) | (0.069) | (0.103) | (0.062) | (0.079) | (0.072) | (0.100) | (0.050) | (0.082) | (0.068) | (0.111) |
| lnEMP | 0.328*** | 0.321*** | 0.340*** | 0.482*** | 0.302*** | 0.312*** | 0.345*** | 0.315*** | 0.326*** | 0.278*** | 0.353*** | 0.363*** | 0.290*** | 0.253*** |
|  | (0.030) | (0.040) | (0.060) | (0.075) | (0.047) | (0.069) | (0.033) | (0.054) | (0.035) | (0.053) | (0.040) | (0.048) | (0.038) | (0.057) |
| Individual FE | Y | Y | Y | Y | Y | Y | Y | Y | Y | Y | Y | Y | Y | Y |
| Time FE | Y | Y | Y | Y | Y | Y | Y | Y | Y | Y | Y | Y | Y | Y |
| Wald chi2 | 4208.1 | 2800.1 | 1581.3 | 878.5 | 1252.4 | 1104.5 | 3359.3 | 1673.3 | 1472 | 2380.8 | 3266.3 | 1423.8 | 735.1 | 829.1 |
|  | [0.000] | [0.000] | [0.000] | [0.000] | [0.000] | [0.000] | [0.000] | [0.000] | [0.000] | [0.000] | [0.000] | [0.000] | [0.000] | [0.000] |
| Log likelihood | -39867.8 | -19099.7 | -12645.9 | -5828 | -14361.2 | -6849.4 | -25298.5 | -12174.4 | -27192.9 | -13240.2 | -27701.1 | -12784 | -11882.4 | -6240.1 |
| N | 13801 | 12627 | 4522 | 4226 | 6038 | 5292 | 7763 | 7335 | 9279 | 8401 | 8613 | 8053 | 5188 | 4574 |

Standard errors in parenthesis
* p<0.05, ** p<0.01, *** p<0.001

## 4.3 Productivity equation

The study employs the Mundlak test over the Hausman test for the productivity equation due to its reliability under issues including heteroskedasticity or serial correlation in error terms (Mundlak, 1978). The Mundlak test, incorporating panel-level means in the random effects model, significantly rejects the null hypothesis at a 99.9% confidence level (Table B7)[8]. This suggests a correlation between unobservable factors in the error term and control variables, indicating the suitability of a panel fixed effects model. The CDM and extended CDM models use individual and time dual fixed effects, with bootstrapping standard errors to address these unobservable factors.

To address the discrepancy between predicted and actual patent counts, and mitigate endogeneity arising from omitted variables like patenting preferences, the model modifies the predicted patent counts[9]. This is achieved by scaling them with each firm's actual average patent number. This adjustment method effectively neutralizes potential biases from unobserved time-invariant factors associated with patenting, while not affecting the variability in firms' productivity, thereby enhancing the accuracy of the productivity equation estimates[10].

Productivity equation models reveal that predicted patent intensity significantly boosts productivity, with a 1% increase in patent intensity leading to a 0.42% increase in productivity in the full sample (Model 43 in Table 5). Capital intensity also positively impacts productivity, while firm size shows negligible effects.

*Table 5 Results of productivity equation- CDM*

---

[8] Table A8 also shows Mundlak test results for Extended CDM model. The results also indicate random effects model is biased and fixed effects model should be used.
[9] The exponential transformed green patent and non-green patent applications compared the realized patent value are presneted in Table B5 and Table B6, respectively
[10] In Marin (2014), the absence of adjustments for potential discrepancies between predicted and actual patent counts may have contributed to their significantly different findings, particularly regarding the crowding-out effects in high-polluting firms.

|  | Model 43 | Model 44 | Model 45 | Model 46 | Model 47 | Model 48 | Model 49 |
| --- | --- | --- | --- | --- | --- | --- | --- |
|  | Full-sample | High-pollution | Non-pollution | High-tech | Low-tech | Manufacturing | Non-manufacturing |
| $\widehat{\ln PATINT}$ | 0.419*** | 0.469*** | 0.338*** | 0.387*** | 0.423*** | 0.391*** | 0.502*** |
|  | (0.035) | (0.058) | (0.069) | (0.047) | (0.052) | (0.038) | (0.049) |
| lnCAPINT | 0.139*** | 0.158*** | 0.173*** | 0.111** | 0.140*** | 0.183*** | 0.147*** |
|  | (0.025) | (0.032) | (0.034) | (0.035) | (0.034) | (0.038) | (0.037) |
| lnEMP | 0.0292 | 0.0228 | 0.043 | 0.0125 | 0.0349 | 0.0489 | 0.0408 |
|  | (0.023) | (0.045) | (0.030) | (0.042) | (0.034) | (0.025) | (0.029) |
| _cons | 5.880*** | 5.599*** | 5.000*** | 5.451*** | 6.801*** | 5.037*** | 6.201*** |
|  | (0.153) | (0.241) | (0.253) | (0.261) | (0.264) | (0.253) | (0.224) |
| Individual FE | Y | Y | Y | Y | Y | Y | Y |
| Time FE | Y | Y | Y | Y | Y | Y | Y |
| adj. R-sq | 0.144 | 0.25 | 0.128 | 0.149 | 0.115 | 0.173 | 0.149 |
| Wald chi2 | 695.7 | 431.6 | 328.8 | 468.6 | 497.4 | 704.9 | 409.8 |
|  | [0.000] | [0.000] | [0.000] | [0.000] | [0.000] | [0.000] | [0.000] |
| Log likelihood | -6635.4 | -1611 | -1582.8 | -2542.4 | -3210.7 | -2686.3 | -2409.1 |
| N | 11727 | 3888 | 3573 | 5670 | 4590 | 6507 | 3762 |

Standard errors in parenthesis
* p<0.05, ** p<0.01, *** p<0.001

*Table 6 Results of productivity equation- Extended CDM*

|  | Model 50 | Model 51 | Model 52 | Model 53 | Model 54 | Model 55 | Model 56 |
| --- | --- | --- | --- | --- | --- | --- | --- |
|  | Full-sample | High-pollution | Non-pollution | High-tech | Low-tech | Manufacturing | Non-manufacturing |
| $\widehat{\ln NECOINT}$ | 0.471*** | 0.253** | 0.527*** | 0.548*** | 0.407*** | 0.301*** | 0.534*** |
|  | (0.056) | (0.087) | (0.046) | (0.083) | (0.058) | (0.075) | (0.064) |
| $\widehat{\ln ECOINT}$ | -0.0739 | 0.198** | -0.144** | -0.156* | -0.0103 | 0.0976 | -0.114* |
|  | (0.058) | (0.076) | (0.053) | (0.077) | (0.052) | (0.068) | (0.053) |
| lnCAPINT | 0.130*** | 0.170*** | 0.120*** | 0.102*** | 0.149*** | 0.172*** | 0.109*** |
|  | (0.019) | (0.034) | (0.020) | (0.025) | (0.025) | (0.029) | (0.025) |
| lnEMP | 0.00553 | -0.0132 | 0.00491 | 0.0309 | -0.0232 | 0.0628** | -0.0258 |
|  | (0.023) | (0.048) | (0.028) | (0.032) | (0.025) | (0.024) | (0.033) |
| _cons | 5.720*** | 6.063*** | 5.656*** | 5.006*** | 6.458*** | 5.348*** | 6.636*** |
|  | (0.224) | (0.307) | (0.176) | (0.252) | (0.207) | (0.275) | (0.208) |
| Individual FE | Y | Y | Y | Y | Y | Y | Y |
| Time FE | Y | Y | Y | Y | Y | Y | Y |
| adj. R-sq | 0.131 | 0.229 | 0.106 | 0.163 | 0.108 | 0.167 | 0.115 |
| Wald chi2 | 1614.6 | 583.5 | 1066.7 | 1029.9 | 546.2 | 1712.3 | 410 |
|  | [0.000] | [0.000] | [0.000] | [0.000] | [0.000] | [0.000] | [0.000] |
| Log Likelihood | -7834.9 | -1790.8 | -5867 | -3002 | -4579.5 | -3172.3 | -4227.5 |
| N | 14721 | 4594 | 10127 | 7714 | 7007 | 8583 | 6138 |

Standard errors in parenthesis
* p<0.05, ** p<0.01, *** p<0.001

In the Extended CDM model, results show that non-green innovations primarily drive firms' productivity across most samples, where GI have negligible or negative impacts on productivity. However, the influence of green and non-green innovations (0.2% and 0.25%) is comparably beneficial to high-polluting firms' productivity (Model 51 in Table 6). The higher efficiency in R&D input-output conversion for GI (refer to Model 31-32 in Table 4), coupled with their productivity effects comparable to those of non-green innovations, indicates an absence of significant crowding-out effects of GI in these firms. This suggests that high-polluting firms can effectively balance the environmental protection costs with gains from green innovations.

### 4.3.1 Non-linear relationship estimations on productivity equation

4.3.1.1 Unconditional quantile regression

The study explores non-linear effects of innovations on productivity using unconditional quantile regression (UQR) in both CDM and Extended CDM models. In full-sample, the U-shaped relationship between firms' patent intensity (PATINT) and productivity level is evident, particularly for non-green patents (NECOINT). Green innovations (ECOINT), however, do not significantly impact productivity (see **Error! Not a valid bookmark self-reference.** and

Table *8*).

*Figure 1 Non-linear estimations based on UQR, comparing CDM and Extended CDM models in full sample*

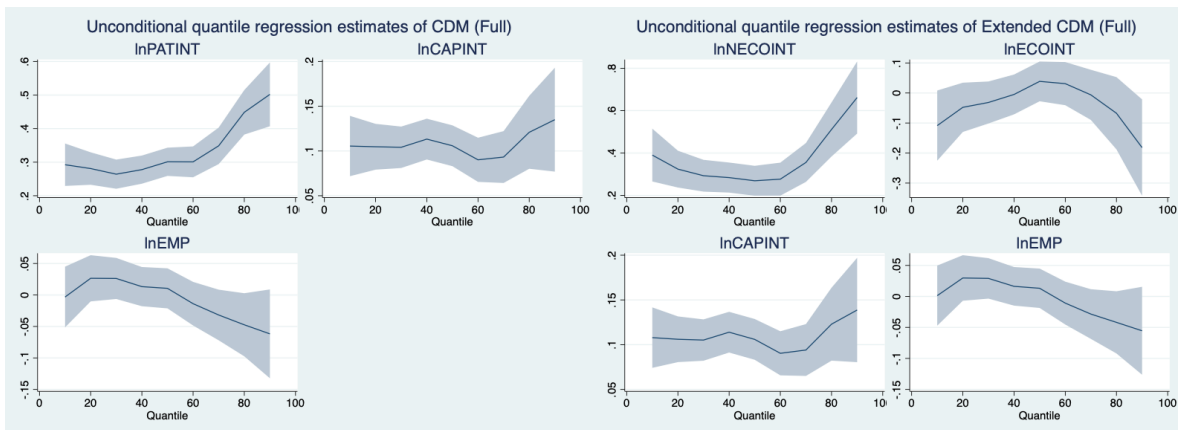

*Table 7 Results of UQR on CDM models (Full sample)*

| Full sample UQR | Model 57 Q10 | Model 58 Q20 | Model 59 Q30 | Model 60 Q40 | Model 61 Q50 | Model 62 Q60 | Model 63 Q70 | Model 64 Q80 | Model 65 Q90 |
|---|---|---|---|---|---|---|---|---|---|
| ln$\widehat{PATINT}$ | 0.292*** | 0.281*** | 0.264*** | 0.278*** | 0.301*** | 0.301*** | 0.349*** | 0.449*** | 0.502*** |
|  | (9.072) | (11.399) | (11.935) | (12.990) | (14.048) | (12.844) | (12.577) | (13.266) | (10.351) |
| lnCAPINT | 0.106*** | 0.105*** | 0.104*** | 0.113*** | 0.106*** | 0.090*** | 0.093*** | 0.121*** | 0.135*** |
|  | (6.141) | (8.028) | (8.824) | (9.746) | (9.102) | (7.166) | (6.313) | (5.819) | (4.560) |
| lnEMP | -0.003 | 0.027 | 0.026 | 0.013 | 0.011 | -0.014 | -0.032 | -0.047+ | -0.062+ |
|  | (-0.131) | (1.417) | (1.578) | (0.837) | (0.650) | (-0.783) | (-1.550) | (-1.853) | (-1.716) |
| _cons | 4.951*** | 5.282*** | 5.549*** | 5.703*** | 5.932*** | 6.264*** | 6.458*** | 6.541*** | 6.936*** |
|  | (42.436) | (58.269) | (67.662) | (70.893) | (73.963) | (71.358) | (62.268) | (47.360) | (35.751) |
| Individual FE | Y | Y | Y | Y | Y | Y | Y | Y | Y |
| Time FE | Y | Y | Y | Y | Y | Y | Y | Y | Y |
| adj. R-sq | 0.612 | 0.612 | 0.612 | 0.612 | 0.612 | 0.612 | 0.612 | 0.612 | 0.612 |
| F-Stat | 56.67 | 56.67 | 56.67 | 56.67 | 56.67 | 56.67 | 56.67 | 56.67 | 56.67 |
| Prob > F | [0.000] | [0.000] | [0.000] | [0.000] | [0.000] | [0.000] | [0.000] | [0.000] | [0.000] |
| Log likelihood | -23266 | -23266 | -23266 | -23266 | -23266 | -23266 | -23266 | -23266 | -23266 |
| N | 14,721 | 14,721 | 14,721 | 14,721 | 14,721 | 14,721 | 14,721 | 14,721 | 14,721 |

Standard errors in parenthesis
*** p<0.001, ** p<0.01, * p<0.05, + p<0.1

*Table 8 Results of UQR on Extended CDM models (Full sample)*

| Full sample UQR | Model 66 Q10 | Model 67 Q20 | Model 68 Q30 | Model 69 Q40 | Model 70 Q50 | Model 71 Q60 | Model 72 Q70 | Model 73 Q80 | Model 74 Q90 |
|---|---|---|---|---|---|---|---|---|---|
| ln$\widehat{NECOINT}$ | 0.390*** | 0.324*** | 0.293*** | 0.284*** | 0.269*** | 0.277*** | 0.356*** | 0.511*** | 0.662*** |
|  | (6.147) | (7.317) | (7.668) | (7.857) | (7.475) | (6.967) | (7.609) | (7.862) | (7.633) |
| ln$\widehat{ECOINT}$ | -0.109+ | -0.048 | -0.032 | -0.005 | 0.039 | 0.031 | -0.006 | -0.067 | -0.182* |
|  | (-1.823) | (-1.142) | (-0.886) | (-0.144) | (1.146) | (0.850) | (-0.146) | (-1.100) | (-2.221) |
| lnCAPINT | 0.108*** | 0.106*** | 0.105*** | 0.114*** | 0.106*** | 0.090*** | 0.094*** | 0.123*** | 0.139*** |
|  | (6.262) | (8.144) | (8.929) | (9.824) | (9.116) | (7.190) | (6.366) | (5.913) | (4.665) |
| lnEMP | 0.001 | 0.03 | 0.029+ | 0.016 | 0.013 | -0.011 | -0.028 | -0.042 | -0.055 |
|  | (0.051) | (1.588) | (1.748) | (1.024) | (0.812) | (-0.611) | (-1.393) | (-1.637) | (-1.532) |
| _cons | 4.738*** | 5.214*** | 5.517*** | 5.735*** | 6.073*** | 6.386*** | 6.501*** | 6.451*** | 6.587*** |
|  | (26.117) | (40.501) | (49.007) | (53.151) | (56.299) | (53.665) | (45.739) | (32.079) | (23.381) |
| Individual FE | Y | Y | Y | Y | Y | Y | Y | Y | Y |
| Time FE | Y | Y | Y | Y | Y | Y | Y | Y | Y |
| adj. R-sq | 0.613 | 0.613 | 0.613 | 0.613 | 0.613 | 0.613 | 0.613 | 0.613 | 0.613 |
| F-Stat | 43 | 43 | 43 | 43 | 43 | 43 | 43 | 43 | 43 |
| Prob > F | [0.000] | [0.000] | [0.000] | [0.000] | [0.000] | [0.000] | [0.000] | [0.000] | [0.000] |
| Log likelihood | -23261 | -23261 | -23261 | -23261 | -23261 | -23261 | -23261 | -23261 | -23261 |
| N | 14,721 | 14,721 | 14,721 | 14,721 | 14,721 | 14,721 | 14,721 | 14,721 | 14,721 |

Standard errors in parenthesis
*** p<0.001, ** p<0.01, * p<0.05, + p<0.1

While capital intensity (CAPTINT) exhibits a positive correlation with productivity, particularly in the higher productivity quantiles, labor input (EMP) does not demonstrate a substantial impact. This suggests a predominant reliance on non-green innovation and capital across all productivity levels, as opposed to GI and human resources, for listed firms in China in overall. In the ensuing comparative analysis, it scrutinizes sub-samples to evaluate whether the factors influencing productivity, with a specific emphasis on the crowding-out effects of GI, manifest differently in varied industrial landscapes.

*Figure 2 Non-linear estimations based on UQR of CDM model, high-pollution v.s. non-pollution sample*

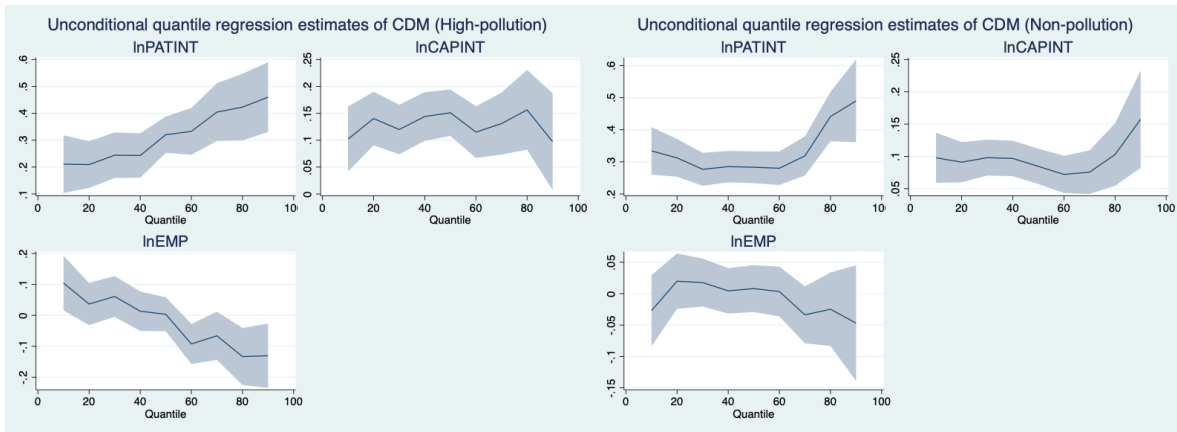

*Figure 3 Non-linear estimations based on UQR of Extended CDM model, high-pollution v.s. non-pollution sample*

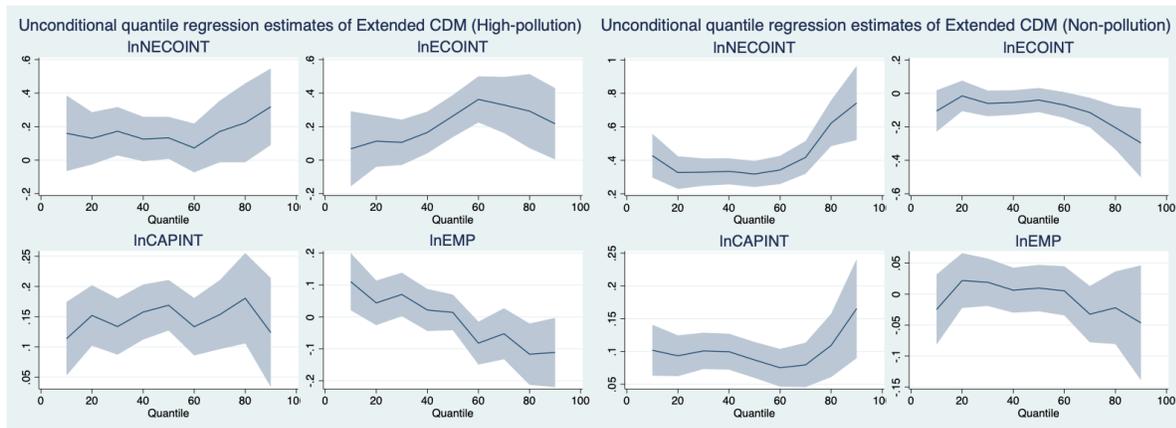

High-polluting firms show a linear increase in economic gains with patent intensity across the productivity distribution, whereas non-polluting firms experience innovation returns stagnate until higher productivity levels (Figure 2). These results from CDM models suggest high-polluting firms find innovation more rewarding. The Extended CDM models show an inverse U-shaped relationship between green patent intensity and productivity in high-polluting and non-polluting samples, while non-green patents exhibit a U-shaped pattern (Figure 3).

*Table 9 Results of UQR on Extended CDM models (High-pollution sample)*

| High-pollution UQR | Model 93 Q10 | Model 94 Q20 | Model 95 Q30 | Model 96 Q40 | Model 97 Q50 | Model 98 Q60 | Model 99 Q70 | Model 100 Q80 | Model 101 Q90 |
|---|---|---|---|---|---|---|---|---|---|
| $\widehat{lnNECOINT}$ | 0.16 | 0.13 | 0.172* | 0.126+ | 0.133* | 0.073 | 0.171+ | 0.223+ | 0.319** |
|  | (1.392) | (1.638) | (2.340) | (1.861) | (2.080) | (0.978) | (1.825) | (1.856) | (2.741) |
| $\widehat{lnECOINT}$ | 0.068 | 0.113 | 0.106 | 0.165** | 0.263*** | 0.362*** | 0.329*** | 0.293** | 0.217* |
|  | (0.591) | (1.453) | (1.542) | (2.605) | (4.106) | (5.159) | (3.856) | (2.597) | (1.996) |
| lnCAPINT | 0.114*** | 0.152*** | 0.134*** | 0.158*** | 0.169*** | 0.133*** | 0.153*** | 0.181*** | 0.124** |
|  | (3.661) | (5.963) | (5.628) | (6.784) | (7.928) | (5.502) | (5.272) | (4.735) | (2.691) |
| lnEMP | 0.110* | 0.044 | 0.070* | 0.022 | 0.014 | -0.082* | -0.053 | -0.117* | -0.112* |
|  | (2.411) | (1.229) | (2.021) | (0.641) | (0.487) | (-2.403) | (-1.296) | (-2.388) | (-2.019) |
| _cons | 5.091*** | 5.317*** | 5.567*** | 5.787*** | 6.035*** | 6.797*** | 6.746*** | 6.861*** | 7.413*** |
|  | (14.935) | (20.148) | (22.776) | (25.251) | (28.825) | (28.605) | (22.582) | (17.334) | (17.009) |
| Individual FE | Y | Y | Y | Y | Y | Y | Y | Y | Y |
| Time FE | Y | Y | Y | Y | Y | Y | Y | Y | Y |
| adj. R-sq | 0.555 | 0.555 | 0.555 | 0.555 | 0.555 | 0.555 | 0.555 | 0.555 | 0.555 |
| F-Stat | 25.15 | 25.15 | 25.15 | 25.15 | 25.15 | 25.15 | 25.15 | 25.15 | 25.15 |
| Prob > F | [0.000] | [0.000] | [0.000] | [0.000] | [0.000] | [0.000] | [0.000] | [0.000] | [0.000] |
| Log likelihood | -6158 | -6158 | -6158 | -6158 | -6158 | -6158 | -6158 | -6158 | -6158 |
| N | 4,594 | 4,594 | 4,594 | 4,594 | 4,594 | 4,594 | 4,594 | 4,594 | 4,594 |

Standard errors in parenthesis
*** p<0.001, ** p<0.01, * p<0.05, + p<0.1

*Table 10 Results of UQR on Extended CDM models (Non-pollution sample)*

| Non-pollution UQR | Model 102 Q10 | Model 103 Q20 | Model 104 Q30 | Model 105 Q40 | Model 106 Q50 | Model 107 Q60 | Model 108 Q70 | Model 109 Q80 | Model 110 Q90 |
|---|---|---|---|---|---|---|---|---|---|
| $\widehat{lnNECOINT}$ | 0.428*** | 0.327*** | 0.329*** | 0.334*** | 0.318*** | 0.342*** | 0.418*** | 0.622*** | 0.743*** |
|  | (6.406) | (6.554) | (7.902) | (8.424) | (8.028) | (7.961) | (8.365) | (8.895) | (6.568) |
| $\widehat{lnECOINT}$ | -0.105+ | -0.015 | -0.06 | -0.055 | -0.04 | -0.070+ | -0.115* | -0.205** | -0.296** |
|  | (-1.670) | (-0.328) | (-1.532) | (-1.469) | (-1.084) | (-1.770) | (-2.547) | (-3.070) | (-2.818) |
| lnCAPINT | 0.102*** | 0.093*** | 0.101*** | 0.100*** | 0.087*** | 0.075*** | 0.080*** | 0.109*** | 0.165*** |
|  | (5.135) | (5.880) | (7.137) | (7.130) | (6.234) | (5.129) | (4.610) | (4.433) | (4.290) |
| lnEMP | -0.025 | 0.022 | 0.019 | 0.006 | 0.01 | 0.005 | -0.032 | -0.022 | -0.046 |
|  | (-0.859) | (0.970) | (0.979) | (0.342) | (0.505) | (0.261) | (-1.396) | (-0.738) | (-0.980) |
| _cons | 4.770*** | 5.346*** | 5.496*** | 5.736*** | 6.068*** | 6.287*** | 6.403*** | 6.265*** | 6.329*** |
|  | (25.130) | (36.765) | (44.609) | (48.437) | (50.653) | (48.336) | (41.197) | (28.303) | (17.890) |
| Individual FE | Y | Y | Y | Y | Y | Y | Y | Y | Y |
| Time FE | Y | Y | Y | Y | Y | Y | Y | Y | Y |
| adj. R-sq | 0.641 | 0.641 | 0.641 | 0.641 | 0.641 | 0.641 | 0.641 | 0.641 | 0.641 |
| F-Stat | 24.64 | 24.64 | 24.64 | 24.64 | 24.64 | 24.64 | 24.64 | 24.64 | 24.64 |
| Prob > F | [0.000] | [0.000] | [0.000] | [0.000] | [0.000] | [0.000] | [0.000] | [0.000] | [0.000] |
| Log likelihood | -17136 | -17136 | -17136 | -17136 | -17136 | -17136 | -17136 | -17136 | -17136 |
| N | 10,127 | 10,127 | 10,127 | 10,127 | 10,127 | 10,127 | 10,127 | 10,127 | 10,127 |

Standard errors in parenthesis
*** p<0.001, ** p<0.01, * p<0.05, + p<0.1

Contrary to the non-pollution sample where negative coefficients of GI suggest crowding-out effects (Table 10), UQR in the high-pollution sample reveals a notably robust impact of GI on productivity, surpassing non-green innovations within the 40th to 80th productivity quantiles (Table 9). This finding underscores the absence of crowding-out effects and highlights the beneficial role of GI in enhancing the performance of median-high productive, high-polluting firms. However, at the 90th productivity quantile, GI's productivity returns diminish, showing lower benefits compared to non-green innovations, indicating crowding-out effects.

In addition to innovations, the impact of different resources on productivity varies across industrial types and productivity levels. Both CDM and extended CDM models show that capital intensity generally boosts productivity more in high-polluting firms (Table C1-Table C2 and Table 9-Table 10). In contrast, non-polluting firms rely more on capital at high productivity (Figure 3). While employee size do not significantly benefit non-polluting firms, they positively impact the productivity of high-polluting firms at lower levels but yield negative returns at higher productivity levels.

*Figure 4 Non-linear estimations based on UQR of CDM model, high-tech v.s. low-tech sample*

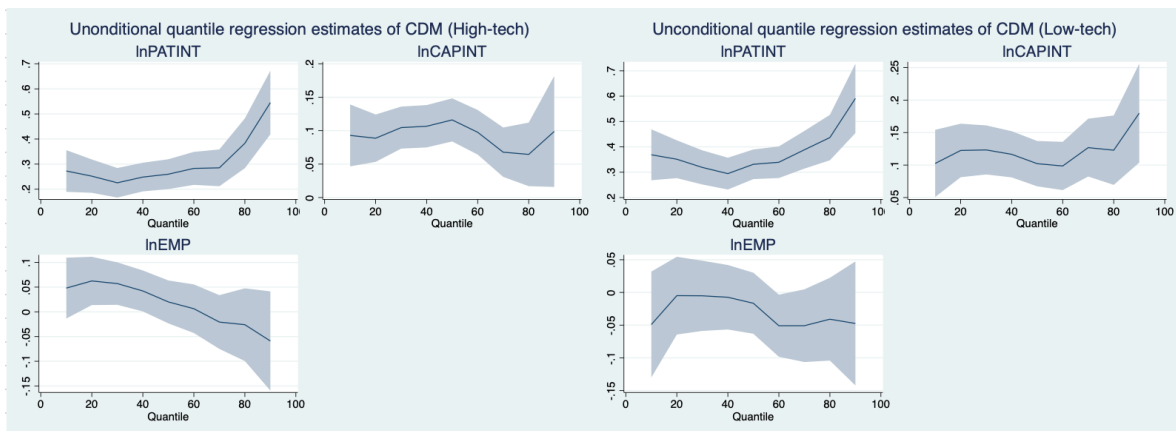

*Figure 5 Non-linear estimations based on UQR of Extended CDM model, high-tech v.s. low-tech sample*

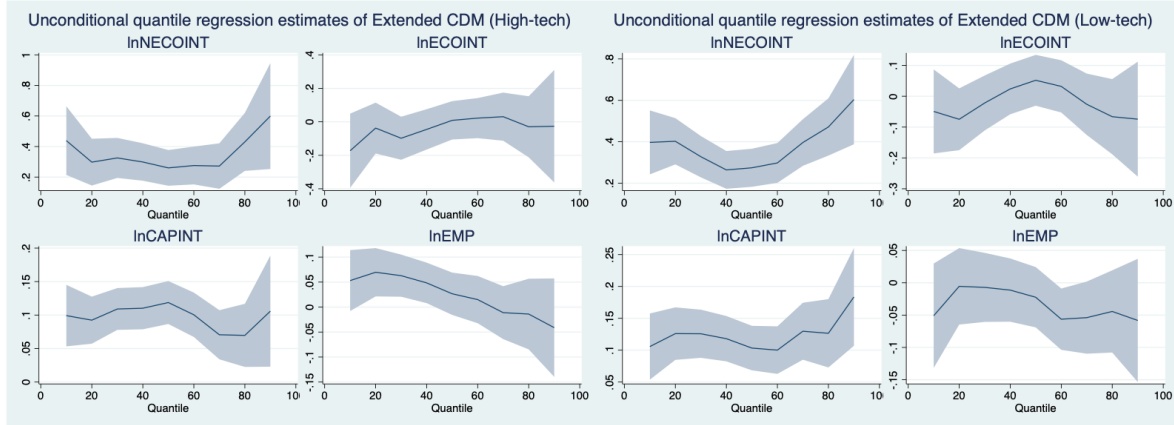

*Table 11 Results of UQR on Extended CDM models (High-tech sample)*

| High-tech UQR | Model 129 Q10 | Model 130 Q20 | Model 131 Q30 | Model 132 Q40 | Model 133 Q50 | Model 134 Q60 | Model 135 Q70 | Model 136 Q80 | Model 137 Q90 |
|---|---|---|---|---|---|---|---|---|---|
| $\widehat{\ln NECOINT}$ | 0.439*** | 0.298*** | 0.325*** | 0.299*** | 0.260*** | 0.276*** | 0.272*** | 0.430*** | 0.600*** |
|  | (3.829) | (3.817) | (4.876) | (4.783) | (4.362) | (4.373) | (3.587) | (4.448) | (3.393) |
| $\widehat{\ln ECOINT}$ | -0.173 | -0.037 | -0.098 | -0.045 | 0.009 | 0.022 | 0.031 | -0.03 | -0.026 |
|  | (-1.528) | (-0.481) | (-1.497) | (-0.735) | (0.153) | (0.364) | (0.419) | (-0.319) | (-0.153) |
| lnCAPINT | 0.099*** | 0.092*** | 0.109*** | 0.110*** | 0.119*** | 0.100*** | 0.071*** | 0.070** | 0.106* |
|  | (4.236) | (5.160) | (6.857) | (6.871) | (7.259) | (5.943) | (3.777) | (2.900) | (2.503) |
| lnEMP | 0.053+ | 0.070** | 0.063** | 0.048* | 0.027 | 0.015 | -0.011 | -0.014 | -0.041 |
|  | (1.707) | (2.828) | (2.925) | (2.343) | (1.233) | (0.621) | (-0.415) | (-0.388) | (-0.820) |
| _cons | 4.270*** | 4.962*** | 4.993*** | 5.265*** | 5.532*** | 5.795*** | 6.211*** | 6.104*** | 6.001*** |
|  | (12.496) | (20.886) | (24.076) | (26.533) | (28.556) | (28.286) | (25.848) | (19.862) | (10.818) |
| Individual FE | Y | Y | Y | Y | Y | Y | Y | Y | Y |
| Time FE | Y | Y | Y | Y | Y | Y | Y | Y | Y |
| adj. R-sq | 0.591 | 0.591 | 0.591 | 0.591 | 0.591 | 0.591 | 0.591 | 0.591 | 0.591 |
| F-Stat | 29.02 | 29.02 | 29.02 | 29.02 | 29.02 | 29.02 | 29.02 | 29.02 | 29.02 |
| Prob > F | [0.000] | [0.000] | [0.000] | [0.000] | [0.000] | [0.000] | [0.000] | [0.000] | [0.000] |
| Log likelihood | -11270 | -11270 | -11270 | -11270 | -11270 | -11270 | -11270 | -11270 | -11270 |
| N | 7,714 | 7,714 | 7,714 | 7,714 | 7,714 | 7,714 | 7,714 | 7,714 | 7,714 |

Standard errors in parenthesis
*** p<0.001, ** p<0.01, * p<0.05, + p<0.1

The U-shaped relationship between patent intensity and productivity is consistent in both high-tech and low-tech samples, with low-tech firms showing greater productivity growth per innovation (Figure 4 and Table C3-Table C4). The "Smiling Curve" theory in innovation explains this phenomenon, highlighting higher returns at early and late technology development stages.

Differences in crowding-out effects of GI between high-pollution and non-pollution samples aren't mirrored in high-tech vs. low-tech firms (Figure 5). In the extended CDM model, both firm types show similar patterns in non-green and green innovation's impact on productivity, and neither benefits significantly from GI (

Table *11*and

Table *12*). This similarity suggests pollution intensity, not technological level, primarily drives GI's economic returns, making high-tech and low-tech firms suitable for a placebo test to confirm pollution's influence on GI activities.

*Table 12 Results of UQR on Extended CDM models (Low-tech sample)*

| Low-tech UQR | Model 138 Q10 | Model 139 Q20 | Model 140 Q30 | Model 141 Q40 | Model 142 Q50 | Model 143 Q60 | Model 144 Q70 | Model 145 Q80 | Model 146 Q90 |
|---|---|---|---|---|---|---|---|---|---|
| $\widehat{lnNECOINT}$ | 0.397*** | 0.402*** | 0.328*** | 0.264*** | 0.274*** | 0.298*** | 0.396*** | 0.471*** | 0.604*** |
|  | (5.041) | (7.062) | (6.443) | (5.677) | (5.854) | (6.074) | (6.908) | (6.689) | (5.483) |
| $\widehat{lnECOINT}$ | -0.049 | -0.075 | -0.022 | 0.024 | 0.052 | 0.032 | -0.026 | -0.066 | -0.074 |
|  | (-0.710) | (-1.462) | (-0.480) | (0.566) | (1.233) | (0.742) | (-0.513) | (-1.067) | (-0.782) |
| lnCAPINT | 0.106*** | 0.126*** | 0.126*** | 0.118*** | 0.103*** | 0.100*** | 0.130*** | 0.126*** | 0.183*** |
|  | (3.993) | (5.997) | (6.531) | (6.512) | (5.800) | (5.266) | (5.695) | (4.627) | (4.696) |
| lnEMP | -0.051 | -0.005 | -0.007 | -0.011 | -0.022 | -0.056* | -0.054+ | -0.044 | -0.058 |
|  | (-1.237) | (-0.181) | (-0.260) | (-0.453) | (-0.939) | (-2.329) | (-1.895) | (-1.367) | (-1.198) |
| _cons | 5.298*** | 5.496*** | 5.900*** | 6.296*** | 6.725*** | 6.980*** | 6.993*** | 7.330*** | 7.576*** |
|  | (23.496) | (32.203) | (38.542) | (44.773) | (47.522) | (45.418) | (38.056) | (33.323) | (23.285) |
| Individual FE | Y | Y | Y | Y | Y | Y | Y | Y | Y |
| Time FE | Y | Y | Y | Y | Y | Y | Y | Y | Y |
| adj. R-sq | 0.644 | 0.644 | 0.644 | 0.644 | 0.644 | 0.644 | 0.644 | 0.644 | 0.644 |
| F-Stat | 23.74 | 23.74 | 23.74 | 23.74 | 23.74 | 23.74 | 23.74 | 23.74 | 23.74 |
| Prob > F | [0.000] | [0.000] | [0.000] | [0.000] | [0.000] | [0.000] | [0.000] | [0.000] | [0.000] |
| Log likelihood | -11609 | -11609 | -11609 | -11609 | -11609 | -11609 | -11609 | -11609 | -11609 |
| N | 7,007 | 7,007 | 7,007 | 7,007 | 7,007 | 7,007 | 7,007 | 7,007 | 7,007 |

Standard errors in parenthesis
*** p<0.001, ** p<0.01, * p<0.05, + p<0.1

Overall, non-manufacturing firms exhibit higher innovation-driven economic returns across all productivity quantiles compared to manufacturing firms (**Error! Not a valid bookmark self-reference.**). Notably, in manufacturing firms, the relationship between patent intensity and productivity is positively and linearly related, indicating a constant growth in productivity with a 1% increase in innovations at any level of productivity among these firms.

While non-manufacturing firms are negatively impacted by GI, for manufacturing firms, GI impacts productivity similarly to non-green innovations, showing no significant crowding-out effects of GI (

Figure 7). This is likely due to manufacturing firms' scale-induced productivity gains from GI in response to pollution-related resource allocation inefficiencies. Green patent intensity have positive and linearly increasing economic returns along the productivity distribution, differing from the U-shaped pattern of non-green innovation. Additionally, labor substantially enhances productivity in labor-intensive manufacturing firms (

Table 13), while human resources yield minimal or even burdensome effects in other sectors.

*Figure 6 Non-linear estimations based on UQR of CDM model, manufacturing v.s. non-manufacturing sample*

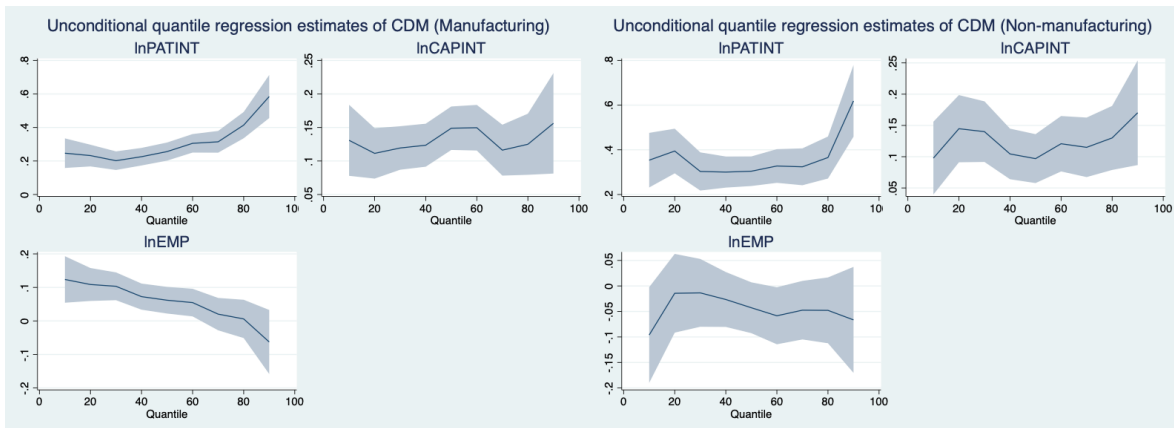

*Figure 7 Non-linear estimations based on UQR of Extended CDM model, manufacturing v.s. non-manufacturing sample*

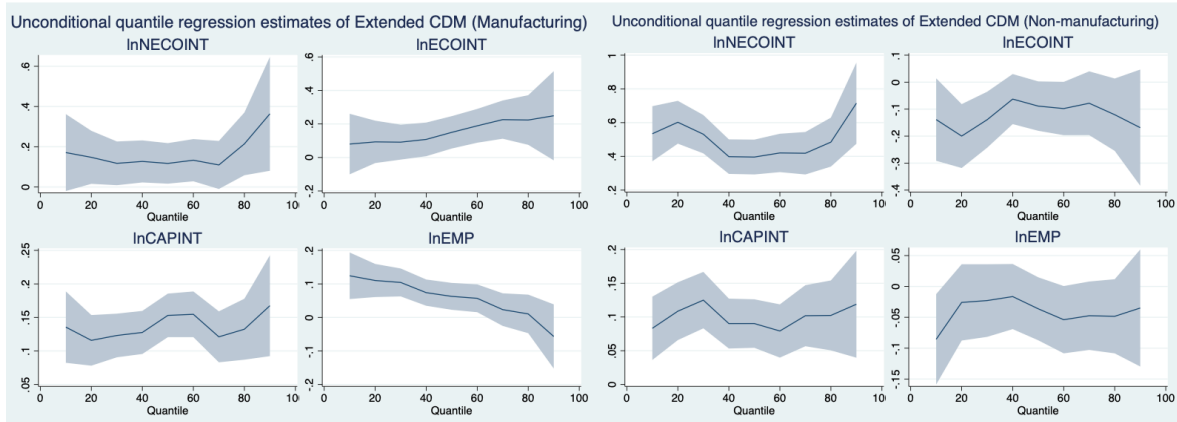

*Table 13 Results of UQR on Extended CDM models (Manufacturing sample)*

| Manufacturing UQR | Model 165 Q10 | Model 166 Q20 | Model 167 Q30 | Model 168 Q40 | Model 169 Q50 | Model 170 Q60 | Model 171 Q70 | Model 172 Q80 | Model 173 Q90 |
|---|---|---|---|---|---|---|---|---|---|
| lnNECOINT | 0.171+ | 0.147* | 0.118* | 0.127* | 0.117* | 0.133* | 0.110+ | 0.214** | 0.364* |
|  | (1.757) | (2.195) | (2.122) | (2.385) | (2.283) | (2.494) | (1.799) | (2.696) | (2.520) |
| lnECOINT | 0.08 | 0.093 | 0.092+ | 0.108* | 0.150** | 0.189*** | 0.226*** | 0.224** | 0.249+ |
|  | (0.871) | (1.444) | (1.722) | (2.110) | (3.048) | (3.676) | (3.885) | (2.961) | (1.835) |
| lnCAPINT | 0.136*** | 0.116*** | 0.123*** | 0.128*** | 0.153*** | 0.155*** | 0.121*** | 0.132*** | 0.167*** |
|  | (4.993) | (6.013) | (7.417) | (7.785) | (9.243) | (8.901) | (6.244) | (5.711) | (4.366) |
| lnEMP | 0.124*** | 0.110*** | 0.104*** | 0.074*** | 0.063** | 0.057** | 0.023 | 0.01 | -0.057 |
|  | (3.511) | (4.379) | (4.916) | (3.705) | (3.078) | (2.700) | (0.939) | (0.357) | (-1.163) |
| _cons | 4.738*** | 5.219*** | 5.450*** | 5.617*** | 5.713*** | 5.890*** | 6.396*** | 6.391*** | 6.390*** |
|  | (14.565) | (23.480) | (29.096) | (30.752) | (31.909) | (31.554) | (30.124) | (24.075) | (13.342) |
| Individual FE | Y | Y | Y | Y | Y | Y | Y | Y | Y |
| Time FE | Y | Y | Y | Y | Y | Y | Y | Y | Y |
| adj. R-sq | 0.595 | 0.595 | 0.595 | 0.595 | 0.595 | 0.595 | 0.595 | 0.595 | 0.595 |
| F-Stat | 36.71 | 36.71 | 36.71 | 36.71 | 36.71 | 36.71 | 36.71 | 36.71 | 36.71 |
| Prob > F | [0.000] | [0.000] | [0.000] | [0.000] | [0.000] | [0.000] | [0.000] | [0.000] | [0.000] |
| Log likelihood | -12192 | -12192 | -12192 | -12192 | -12192 | -12192 | -12192 | -12192 | -12192 |
| N | 8,583 | 8,583 | 8,583 | 8,583 | 8,583 | 8,583 | 8,583 | 8,583 | 8,583 |

Standard errors in parenthesis
*** p<0.001, ** p<0.01, * p<0.05, + p<0.1

*Table 14 Results of UQR on Extended CDM models (Non-manufacturing sample)*

| Non-manufacturing UQR | Model 174 Q10 | Model 175 Q20 | Model 176 Q30 | Model 177 Q40 | Model 178 Q50 | Model 179 Q60 | Model 180 Q70 | Model 181 Q80 | Model 182 Q90 |
|---|---|---|---|---|---|---|---|---|---|
| lnNECOINT | 0.534*** | 0.602*** | 0.531*** | 0.398*** | 0.396*** | 0.420*** | 0.419*** | 0.485*** | 0.715*** |
|  | (6.420) | (9.309) | (9.170) | (7.639) | (7.527) | (7.250) | (6.524) | (6.548) | (5.835) |
| lnECOINT | -0.139+ | -0.200*** | -0.139** | -0.063 | -0.089+ | -0.098+ | -0.078 | -0.121+ | -0.169 |
|  | (-1.780) | (-3.312) | (-2.633) | (-1.327) | (-1.903) | (-1.944) | (-1.298) | (-1.764) | (-1.536) |
| lnCAPINT | 0.083*** | 0.109*** | 0.125*** | 0.090*** | 0.090*** | 0.079*** | 0.102*** | 0.102*** | 0.119** |
|  | (3.478) | (4.972) | (5.862) | (4.788) | (4.926) | (3.940) | (4.415) | (3.875) | (2.942) |
| lnEMP | -0.086* | -0.026 | -0.023 | -0.016 | -0.036 | -0.054+ | -0.047+ | -0.048 | -0.035 |
|  | (-2.295) | (-0.821) | (-0.763) | (-0.610) | (-1.392) | (-1.939) | (-1.680) | (-1.579) | (-0.723) |

| _cons | 5.369*** | 5.507*** | 5.865*** | 6.455*** | 6.669*** | 7.000*** | 7.222*** | 7.536*** | 7.991*** |
|---|---|---|---|---|---|---|---|---|---|
| | (26.030) | (31.796) | (37.001) | (45.726) | (47.381) | (44.111) | (38.995) | (35.648) | (24.098) |
| Individual FE | Y | Y | Y | Y | Y | Y | Y | Y | Y |
| Time FE | Y | Y | Y | Y | Y | Y | Y | Y | Y |
| adj. R-sq | 0.617 | 0.617 | 0.617 | 0.617 | 0.617 | 0.617 | 0.617 | 0.617 | 0.617 |
| F-Stat | 21.46 | 21.46 | 21.46 | 21.46 | 21.46 | 21.46 | 21.46 | 21.46 | 21.46 |
| Prob > F | [0.000] | [0.000] | [0.000] | [0.000] | [0.000] | [0.000] | [0.000] | [0.000] | [0.000] |
| Log likelihood | -10541 | -10541 | -10541 | -10541 | -10541 | -10541 | -10541 | -10541 | -10541 |
| N | 6,138 | 6,138 | 6,138 | 6,138 | 6,138 | 6,138 | 6,138 | 6,138 | 6,138 |

Standard errors in parenthesis
*** p<0.001, ** p<0.01, * p<0.05, + p<0.1

*Figure 8 Opportunity costs of green innovation across productivity distribution*

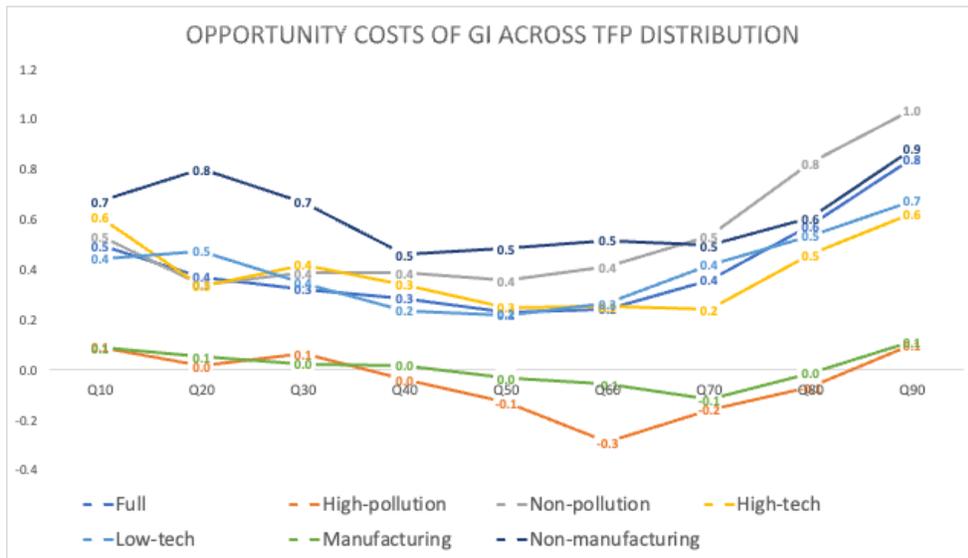

In summary, aligning with Porter's Hypothesis, the study reveals that firms with heavier pollution issues, particularly at mid to high productivity levels, substantially benefit from GI. As evidenced by the U-shaped relationship between non-green innovations and productivity in UQR analyses, high-polluting and manufacturing firms experience negative opportunity costs in GI investments within the 50th to 80th productivity quantiles (Figure 8). In contrast, other firms with lower pollution issues face much higher opportunity costs with GI, making non-green innovations more economically advantageous.

The study's insights are crucial for corporate strategy, highlighting that the relative benefits of GI vary with a firm's pollution level and productivity level. For high-polluting firms operating at moderate productivity levels, GI is a cost-effective strategy, offering higher returns than other innovations. However, at very low or high productivity levels, or in firms with minimal pollution concerns, the opportunity costs of GI outweigh its benefits, making non-green innovation a more lucrative option. These conclusions provide key strategic direction for firms in allocating resources for innovation.

4.3.2 <u>Robustness checks on non-linear relationship estimations</u>

4.3.2.1 RIF treatment effects

*Figure 9 Non-linear estimations based on RIF treatment effects (Full sample)*

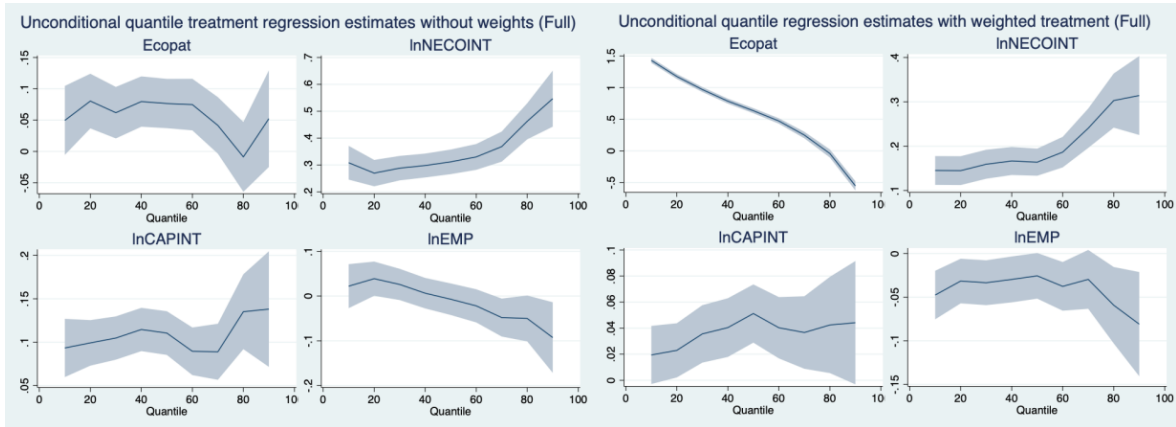

*Table 15 Results of RIF treatment effects with IPW (Full sample)*

| Full sample RIF-Weighted Treat | Model 192 Q10 | Model 193 Q20 | Model 194 Q30 | Model 195 Q40 | Model 196 Q50 | Model 197 Q60 | Model 198 Q70 | Model 199 Q80 | Model 200 Q90 |
|---|---|---|---|---|---|---|---|---|---|
| ECO | 1.431*** | 1.176*** | 0.970*** | 0.786*** | 0.635*** | 0.472*** | 0.249*** | -0.042 | -0.553*** |
|  | (70.435) | (58.908) | (48.619) | (39.494) | (31.342) | (21.539) | (9.420) | (-1.287) | (-14.787) |
| lnNECOINT | 0.145*** | 0.145*** | 0.159*** | 0.167*** | 0.164*** | 0.187*** | 0.240*** | 0.303*** | 0.314*** |
|  | (8.760) | (8.720) | (9.548) | (10.336) | (10.492) | (10.614) | (10.616) | (9.722) | (6.918) |
| lnCAPINT | 0.019+ | 0.023* | 0.036** | 0.040*** | 0.051*** | 0.040*** | 0.037** | 0.042* | 0.044+ |
|  | (1.708) | (2.165) | (3.165) | (3.519) | (4.497) | (3.360) | (2.582) | (2.253) | (1.829) |
| lnEMP | -0.048*** | -0.031* | -0.033* | -0.030* | -0.026+ | -0.038** | -0.030+ | -0.059** | -0.081** |
|  | (-3.352) | (-2.414) | (-2.551) | (-2.210) | (-1.912) | (-2.641) | (-1.720) | (-2.651) | (-2.655) |
| _cons | 6.553*** | 6.919*** | 7.208*** | 7.480*** | 7.662*** | 7.944*** | 8.520*** | 9.012*** | 9.519*** |
|  | (64.813) | (65.454) | (66.994) | (69.332) | (71.075) | (67.308) | (57.025) | (44.510) | (31.511) |
| Individual FE | Y | Y | Y | Y | Y | Y | Y | Y | Y |
| Time FE | Y | Y | Y | Y | Y | Y | Y | Y | Y |
| adj. R-sq | 0.674 | 0.674 | 0.674 | 0.674 | 0.674 | 0.674 | 0.674 | 0.674 | 0.674 |
| F-Stat | 73.35 | 73.35 | 73.35 | 73.35 | 73.35 | 73.35 | 73.35 | 73.35 | 73.35 |
| Prob > F | [0.000] | [0.000] | [0.000] | [0.000] | [0.000] | [0.000] | [0.000] | [0.000] | [0.000] |
| Log likelihood | 27402 | 27402 | 27402 | 27402 | 27402 | 27402 | 27402 | 27402 | 27402 |
| N | 14,721 | 14,721 | 14,721 | 14,721 | 14,721 | 14,721 | 14,721 | 14,721 | 14,721 |

Standard errors in parenthesis
*** p<0.001, ** p<0.01, * p<0.05, + p<0.1

The RIF treatment effects models[11], both standard and IPW-adjusted, demonstrate that China's listed firms investing in GI consistently exhibit higher productivity up to the 80th productivity quantile compared to those without GI (Figure 9, Table 15 and Table D1). While UQR analysis previously reveals some negative, albeit insignificant, marginal effects of GI, RIF treatment effects affirm the positive average economic impact of GI particularly at lower levels of productivity.

*Figure 10 Non-linear estimations based on RIF treatment effects (High-pollution sample)*

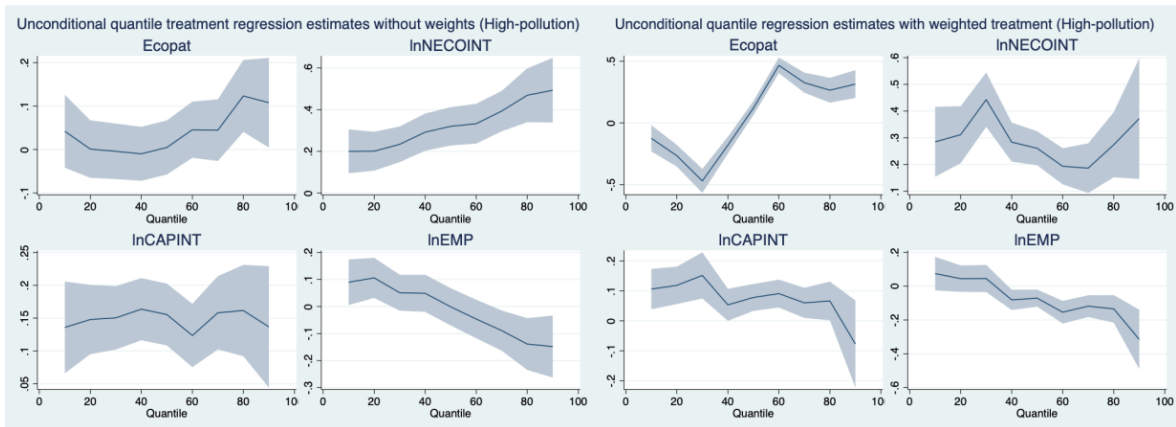

*Table 16 Results of RIF treatment effects with IPW (High-pollution sample)*

| High-pollution RIF-Weighted Treat | Model 210 Q10 | Model 211 Q20 | Model 212 Q30 | Model 213 Q40 | Model 214 Q50 | Model 215 Q60 | Model 216 Q70 | Model 217 Q80 | Model 218 Q90 |
|---|---|---|---|---|---|---|---|---|---|
| ECO | -0.125* | -0.266*** | -0.470*** | -0.181*** | 0.120*** | 0.467*** | 0.327*** | 0.266*** | 0.315*** |
|  | (-2.299) | (-5.911) | (-9.517) | (-5.190) | (4.181) | (14.578) | (7.873) | (5.168) | (5.491) |
| lnNECOINT̂ | 0.285*** | 0.311*** | 0.443*** | 0.283*** | 0.260*** | 0.193*** | 0.186*** | 0.274*** | 0.372** |
|  | (4.262) | (5.736) | (8.516) | (7.604) | (8.036) | (5.591) | (3.936) | (4.400) | (3.225) |
| lnCAPINT | 0.106** | 0.118*** | 0.152*** | 0.053* | 0.078*** | 0.091*** | 0.060* | 0.066* | -0.077 |

---

[11] A rolling three-year average number of green patent counts is used as a measure of whether a firm has invested in green innovation. If it is greater than zero, it means that the firm has had a positive average investment in green innovation over past three years. This approach helps mitigate potential endogeneity issues when estimating the average effects between the two groups, as it accounts for unobservable factors that may influence productivity differently from the level of green patent counts or the amount of R&D investment in a single year.

|  | (3.082) | (3.728) | (3.859) | (1.966) | (3.400) | (3.830) | (2.341) | (2.015) | (-1.042) |
|---|---|---|---|---|---|---|---|---|---|
| lnEMP | 0.075 | 0.044 | 0.045 | -0.081** | -0.071** | -0.154*** | -0.118*** | -0.134** | -0.315*** |
|  | (1.476) | (1.111) | (1.106) | (-2.648) | (-2.757) | (-4.509) | (-3.614) | (-3.234) | (-3.527) |
| _cons | 5.717*** | 6.122*** | 6.410*** | 7.133*** | 6.977*** | 6.838*** | 7.165*** | 7.576*** | 9.072*** |
|  | (24.387) | (28.377) | (24.231) | (38.550) | (45.062) | (41.631) | (40.106) | (33.806) | (19.199) |
| Individual FE | Y | Y | Y | Y | Y | Y | Y | Y | Y |
| Time FE | Y | Y | Y | Y | Y | Y | Y | Y | Y |
| adj. R-sq | 0.668 | 0.668 | 0.668 | 0.668 | 0.668 | 0.668 | 0.668 | 0.668 | 0.668 |
| F-Stat | 20.63 | 20.63 | 20.63 | 20.63 | 20.63 | 20.63 | 20.63 | 20.63 | 20.63 |
| Prob > F | [0.000] | [0.000] | [0.000] | [0.000] | [0.000] | [0.000] | [0.000] | [0.000] | [0.000] |
| Log likelihood | -5059 | -5059 | -5059 | -5059 | -5059 | -5059 | -5059 | -5059 | -5059 |
| N | 4,594 | 4,594 | 4,594 | 4,594 | 4,594 | 4,594 | 4,594 | 4,594 | 4,594 |

Standard errors in parenthesis
*** p<0.001, ** p<0.01, * p<0.05, + p<0.1

The RIF treatment effects models, both standard and IPW-adjusted, demonstrate that China's listed firms investing in GI consistently exhibit higher productivity up to the 80th productivity quantile compared to those without GI (Figure 9, Table 15 and Table D1). While UQR analysis previously reveals some negative, albeit insignificant, marginal effects of GI, RIF treatment effects affirm the positive average economic impact of GI particularly at lower levels of productivity.

Figure 10 indicates that in high-polluting firms, GI's impact on productivity shifts with productivity levels. Firms with lower productivity initially face negative effects, hindered by limited capabilities in efficient technology implementation. As productivity increases, these firms become more adept at leveraging new technology, both green and non-green innovations, for productivity gains (Table 16). This highlights the critical role of a firm's current productivity level in the success of GI investments.

Additionally, Table 16 reveal that the positive impact of GI moderates when reaching the 60th productivity quantile, with non-green innovation continuing to drive productivity at higher levels. This suggests that while GI contributes positively when high-polluting firms achieve median-level production efficiency, its impact wanes at the highest productivity levels, where core product innovations predominate and the gains from resource-efficient GI become more limited.

*Figure 11 Non-linear estimations based on RIF treatment effects (Non-pollution sample)*

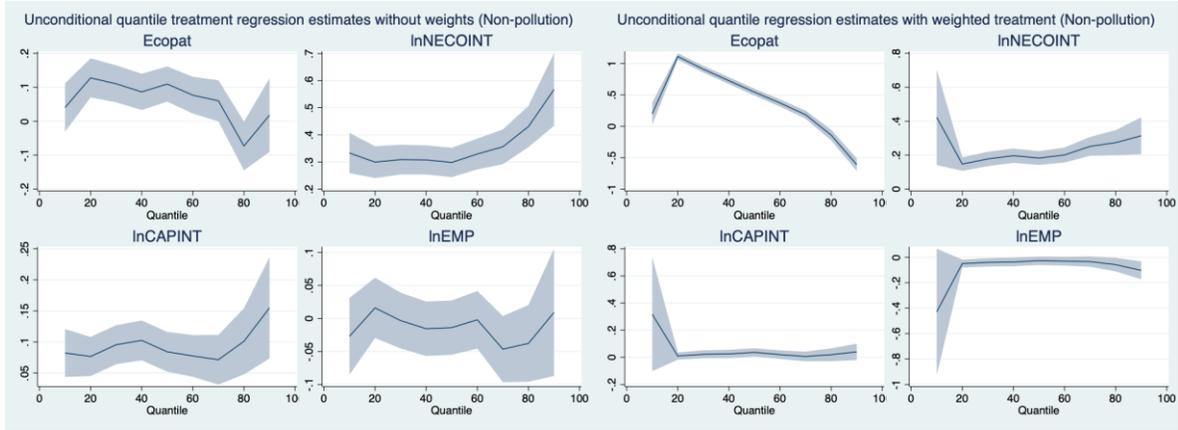

For non-polluting firms, GI yields positive economic returns for firms at lower productivity levels, peaking at the 20th productivity quantile (The RIF treatment effects models, both standard and IPW-adjusted, demonstrate that China's listed firms investing in GI consistently exhibit higher productivity up to the 80th productivity quantile compared to those without GI (Figure 9, Table 15 and Table D1). While UQR analysis previously reveals some negative, albeit insignificant, marginal effects of GI, RIF treatment effects affirm the positive average economic impact of GI particularly at lower levels of productivity.

Figure 10 indicates that in high-polluting firms, GI's impact on productivity shifts with productivity levels. Firms with lower productivity initially face negative effects, hindered by limited capabilities in efficient technology implementation. As productivity increases, these firms become more adept at leveraging new technology, both green and non-green innovations, for productivity gains (Table 16). This highlights the critical role of a firm's current productivity level in the success of GI investments.

Additionally, Table 16 reveal that the positive impact of GI moderates when reaching the 60th productivity quantile, with non-green innovation continuing to drive productivity at higher levels. This suggests that while GI contributes positively when high-polluting firms achieve median-level production efficiency, its impact wanes at the highest productivity levels, where core product innovations predominate and the gains from resource-efficient GI become more limited.

Figure 11). Yet, these benefits diminish and eventually turn negative beyond the 80th productivity quantile. This trend arises as these firms primarily employ GI to optimize resource utilization rather than to tackle pollution challenges. Consequently, they experience rapid productivity gains from GI at lower levels, but these improvements are not sustained over time due to limited room for further efficiency gains. Additionally, other samples, including non-manufacturing firms, exhibit similar indications to the full sample and non-polluting samples, generally show higher productivity impacts from GI at lower productivity levels[12].

In stark contrast, for manufacturing firms, the findings are similar to those of high-polluting firms, with GI realizes significant positive influencing on productivity at higher quantiles (Table D8, Table D9 and Figure D 1). Firms with pollution issues necessitate a greater level of productivity to effectively harness GI for productivity gains. Their reliance on GI stems from the imperative to address environmental concerns, which translates into more enduring productivity enhancements at higher levels, as they integrate green practices into their production and product offerings.

*Table 17 Results of RIF treatment effects with IPW (Non-pollution sample)*

| Non-pollution RIF-Weighted Treat | Model 228 Q10 | Model 229 Q20 | Model 230 Q30 | Model 231 Q40 | Model 232 Q50 | Model 233 Q60 | Model 234 Q70 | Model 235 Q80 | Model 236 Q90 |
| --- | --- | --- | --- | --- | --- | --- | --- | --- | --- |

---

[12] The results of RIF treatment effects models of other samples are detailed in the appendix (Figure D 1-Figure D4 and Table D4-Table D11).

| | | | | | | | | | |
|---|---|---|---|---|---|---|---|---|---|
| ECO | 0.200* | 1.111*** | 0.909*** | 0.727*** | 0.544*** | 0.371*** | 0.185*** | -0.140** | -0.612*** |
| | (2.221) | (42.774) | (33.409) | (26.205) | (19.514) | (12.699) | (5.564) | (-3.262) | (-11.855) |
| $\widehat{\text{lnNECOINT}}$ | 0.423** | 0.147*** | 0.177*** | 0.197*** | 0.183*** | 0.201*** | 0.252*** | 0.274*** | 0.314*** |
| | (2.950) | (7.282) | (8.190) | (9.143) | (8.896) | (8.895) | (8.926) | (7.267) | (5.673) |
| lnCAPINT | 0.318 | 0.01 | 0.023 | 0.025 | 0.037* | 0.02 | 0.007 | 0.019 | 0.04 |
| | (1.486) | (0.720) | (1.527) | (1.604) | (2.372) | (1.262) | (0.357) | (0.807) | (1.295) |
| lnEMP | -0.429+ | -0.050** | -0.040* | -0.037* | -0.027 | -0.031+ | -0.034 | -0.058* | -0.103** |
| | (-1.697) | (-3.144) | (-2.396) | (-2.144) | (-1.603) | (-1.724) | (-1.626) | (-2.136) | (-2.869) |
| _cons | 5.227*** | 6.848*** | 7.252*** | 7.592*** | 7.783*** | 8.138*** | 8.646*** | 9.011*** | 9.509*** |
| | (3.548) | (63.521) | (61.770) | (63.910) | (66.899) | (64.494) | (55.361) | (42.874) | (30.967) |
| Individual FE | Y | Y | Y | Y | Y | Y | Y | Y | Y |
| Time FE | Y | Y | Y | Y | Y | Y | Y | Y | Y |
| adj. R-sq | 0.697 | 0.697 | 0.697 | 0.697 | 0.697 | 0.697 | 0.697 | 0.697 | 0.697 |
| F-Stat | 53.05 | 53.05 | 53.05 | 53.05 | 53.05 | 53.05 | 53.05 | 53.05 | 53.05 |
| Prob > F | [0.000] | [0.000] | [0.000] | [0.000] | [0.000] | [0.000] | [0.000] | [0.000] | [0.000] |
| Log likelihood | 1007 | 1007 | 1007 | 1007 | 1007 | 1007 | 1007 | 1007 | 1007 |
| N | 10,127 | 10,127 | 10,127 | 10,127 | 10,127 | 10,127 | 10,127 | 10,127 | 10,127 |

Standard errors in parenthesis
*** p<0.001, ** p<0.01, * p<0.05, + p<0.1

### 4.3.2.2 Conditional quantile regression

The paper utilizes conditional quantile regressions with bootstrapping (

Table *D12*

*Table D13*) to validate results from the panel fixed effects and UQR methods. The CDM and extended CDM findings generally concur with those of the panel fixed effects model. Notably, the impact of GI on productivity is found to be nearly identical to the contribution of other innovations to productivity growth in the high-polluting industries, suggesting no significant crowding-out effect. However, negative coefficients of green patent intensity in the full, non-pollution, high-tech, low-tech, and non-manufacturing samples indicate crowding-out effects of GI in firms without environmental issues.

## 5. CONCLUSION

The article begins by estimating the classical CDM model to examine the economic effects of R&D activities in listed companies. It then introduces an extended CDM model to test for a crowding-out effect of green innovation (GI) on other innovations, by assessing the separate impacts of GI and non-green innovations on firm productivity. Additionally, unconditional quantile regression (UQR) is employed to investigate potential non-linear relationships between innovations and productivity across the productivity distribution, and to examine how crowding-out effects may vary among different types of firms at different levels of productivity. Finally, the robustness checks conduct conditional quantile regression (CQR) estimations, as well as recentered influence function (RIF) estimations that compare the treatment effects on productivity when a firm invests in GI compared to no GI investments.

The results of the first step of the CDM model, which estimates R&D intensity, indicate that environmental regulations impose additional cost burdens and reduce funds available for firms' R&D investments. The analysis corrects for firm selection bias in R&D expenditure disclosure using the Hackman model. In the second step, the bias-corrected R&D intensity is used to predict firms' patents in the CDM model and green/non-green patents in the extended CDM models. The findings suggest that the efficiency of R&D input-output conversion is low for GI in firms without pollution issues but high in firms with pollution issues, particularly high-polluting firms and manufacturing firms. This implies that crowding-out effects of GI in terms of R&D resource usage efficiency primarily occur in non-heavy polluting firms, while high-polluting firms are more effective in utilizing R&D resources to generate GI compared to other types of innovations.

The CDM model results, based on the predicted patent intensity from the previous patent equation, show that higher patent intensity significantly improves firms' productivity. Furthermore, the estimation using UQR reveals a non-linear (U-shaped) marginal effect of patent intensity on productivity throughout the productivity distribution.

Moreover, the estimations from the extended CDM models indicate that there are crowding-out effects of GI on other innovations in firms without pollution issues. GI has negative or insignificant marginal impacts on productivity, while other patents yield significantly higher productivity returns. However, the positive economic returns of green patents are comparable to other patents for high-polluting firms and manufacturing firms, suggesting that GI can offset environmental costs and potentially improve resource allocation efficiency as much as other types of innovations for firms with pollution issues. Therefore, firms with pollution issues do not exhibit significant crowding-out effects of green innovation.

The UQR estimations show that the opportunity costs of GI in high-polluting firms with mid-to-high levels of productivity are below zero, indicating that GI yields higher economic returns compared to non-green innovations. Additionally, the relationship between non-green innovation and productivity follows a U-shaped non-linear pattern, while the relationships between GI and productivity demonstrate inverse U-shaped patterns in most samples, except for manufacturing firms, where a linearly increasing relationship indicates continuous support of GI for economic performnce improvements at any productivity level.

Lastly, the estimations of recentered influence function (RIF) treatment effects indicate that, on average, having GI brings positive productivity growth below the 80th quantile of the productivity distribution. However, it becomes pure costs for firms without pollution issues at the top-level of productivity. High-polluting firms with productivity levels above the median are more likely to benefit from the reduction of pollution costs or efficiency improvements induced by GI, as there may be a capability threshold for transforming technology into productivity.

Despite the valuable insights provided by this research, there are some limitations to be considered. Firstly, the analysis is based on data from listed companies, which may not fully represent the entire population of firms. The findings may not be generalized to non-listed or smaller firms that might have different characteristics and resource constraints. Secondly, the study focuses on the impact of GI on firm productivity, neglecting other important factors that could influence productivity, such as market competition and managerial practices. Future research should consider exploring the mechanisms through which GI impacts productivity, such as the role of organizational capabilities, knowledge transfer, and technological spillovers, could deepen our understanding of the underlying processes driving the relationship. Moreover, future studies could investigate the long-term effects of GI on firm performance and sustainability, and explore the underlying mechanisms driving the relationship.

**Declaration of generative AI and AI-assisted technologies in the writing process**

During the preparation of this work the author used Chat-GPT in order to improve the language. After using this tool/service, the author reviewed and edited the content as needed and takes full responsibility for the content of the publication.

# Appendix

*Table A1 Description of variables*

| Variable | Obs | Mean | Std. Dev. | Min | Max |
|---|---|---|---|---|---|
| RD | 15,714 | 0.72 | 0.45 | 0 | 1 |
| RDINT | 11,378 | 2.77 | 1.50 | -7.20 | 13.23 |
| PAT | 15,714 | 70.87 | 333.11 | 0 | 11171 |
| ECO | 15,714 | 8.92 | 50.89 | 0 | 1484 |
| NECO | 15,714 | 61.95 | 293.54 | 0 | 10045 |
| lnVA | 15,703 | 6.93 | 0.97 | -1.08 | 11.78 |
| lnPPC | 15,714 | 6.29 | 1.00 | 3.35 | 7.83 |
| lnIPT | 15,714 | 7.59 | 0.92 | 3.57 | 9.56 |
| lnFSTK | 14,971 | -4.98 | 44.88 | -2302.59 | 269.12 |
| lnPLT | 15,714 | 8.04 | 1.06 | 5.12 | 9.75 |
| EPD | 15,714 | 0.17 | 0.37 | 0 | 1 |
| AGE | 15,714 | 11.80 | 6.36 | 0 | 28 |
| lnASSET | 15,677 | 8.40 | 1.40 | -0.74 | 14.70 |
| LEV | 15,713 | 0.48 | 0.43 | 0 | 29.70 |
| lnEMP | 15,700 | 0.81 | 1.39 | -5.81 | 6.32 |
| lnCAPINT | 15,686 | 5.66 | 1.22 | -2.78 | 12.33 |
| lnPCINT | 15,685 | 6.53 | 1.10 | -4.66 | 11.77 |
| CR4 | 15,714 | 0.37 | 0.17 | 0.11 | 1 |
| SOE | 15,714 | 0.46 | 0.50 | 0 | 1 |

*Table A2 Variable explanations*

| Type | Symbol | Variable | Definition | Source |
|---|---|---|---|---|
| Dependent variable | RD | R&D dummy | Dummy variable that whether a firm disclose its R&D expenditure information | Wind Database |
| | RDINT | R&D expenditure intensity | The natural logarithm of the amount of R&D expenditure per employee of listed firms (in millions RMB per thousand employees). | Wind Database |
| | PAT | Patent applications | Rolling average of firms' patent applications in three years. | Wind/CNRDS |
| | lnPATINT | Patent intensity | The logarithm of patent applications per employee of listed firms | Wind/CNRDS |
| | ECO | Eco-patent applications | Rolling average of firms' eco-patent applications in three years. | CNRDS |
| | lnECOINT | Eco-patent intensity | The logarithm of eco-patent applications per employee of listed firms | CNRDS |
| | NECO | Other patent applications | Rolling average of firms' patent applications other than eco-patent applications in three years (All patent application minus eco-patent applications). | Self-calculated |
| | lnNECOINT | Other patent intensity | The logarithm of other patent applications per employee of listed firms | Self-calculated |
| | lnVA | Value added | The natural logarithm of total sales revenue per employee of listed firms. | CSMAR |
| Key variables | lnPPC | Provincial pollution charge | The natural logarithm of total amount of pollutant emission charges (in millions RMB). | China Environmental Yearbook |
| | lnIPT | Industrial pollution treatment investment | The natural logarithm of total amount of industrial pollution treatment investment in each provinces (in millions RMB). | China Statistical Yearbook |
| | lnFSTK | Firm technological knowledge stock | The natural logarithm of firms' patent applications in the past 2 years | Wind/CNRDS |
| | lnPTL | Provincial technological innovation capability level | The natural logarithm of total number of patent applications of each province divided by the provincial population in the past year. | State Statistics Bureau |
| Control variable | lnEMP | Firm size | The natural logarithm of total number of employees. | Wind Database |
| | lnCAPINT | Capital intensity | The natural logarithm of Physical capital per employee (in thousands). | CSMAR |
| | lnPCINT | Production costs intensity | The natural logarithm of production costs per employee (in millions RMB). | CSMAR |
| | LEV | Leverage | Total debt divided by year-end total assets. | CSMAR |
| | SOE | State-owned enterprises | A dummy variable that is equal to 1 if the firm i is a state-owned enterprise and 0 otherwise. | Wind Database |
| | CR4 | Industrial market concentration | The market occupation ratio of the largest four firms in the industry | Self-calculated |
| Exclusion restrictions | EPD | Environmental performance disclosure | If any environmental information was disclosed under Hexun 4 disclose categories, ERD is 1. If none of environmental information was disclosed, ERD is 0. | Hexun Database |
| | AGE | IPO age | Years that a firm has been registered at IPO | Wind Database |
| | lnASSET | Total assets | The natural logarithm of firm total assets. | CSMAR |

*Table B1 Variance Inflation Factors (VIF) tests*

| Variable | VIF | 1/VIF |
|---|---|---|
| IMR | 2.31 | 0.43 |
| lnPPF(-1) | 1.09 | 0.92 |
| LEV | 1.75 | 0.57 |
| lnEMP | 1.66 | 0.60 |
| lnCAPINT | 1.26 | 0.80 |
| lnPCINT | 1.41 | 0.71 |
| CR4 | 1.21 | 0.82 |
| SOE | 1.53 | 0.65 |
| year | | |
| 2011 | 1.87 | 0.54 |
| 2012 | 2.18 | 0.46 |
| 2013 | 2.22 | 0.45 |
| 2014 | 2.16 | 0.46 |
| 2015 | 2.04 | 0.49 |
| 2016 | 2.26 | 0.44 |
| 2017 | 2.29 | 0.44 |

|  | 2018 | 2.49 | 0.40 |
|---|---|---|---|
|  | central | 1.82 | 0.55 |
|  | northeast | 1.26 | 0.79 |
|  | east | 2.10 | 0.48 |
|  | Mean VIF |  | 1.84 |

*Table B2 Full Heckman estimation of R&D equation with alternative environmental regualtion proxies*

| | M1 | M2 | M3 | M4 | M5 | M6 | M7 | M8 | M9 | M10 | M11 | M12 | M13 | M14 |
|---|---|---|---|---|---|---|---|---|---|---|---|---|---|---|
| | lnPPC(-2) | nIPT(-1) | lnPPC(-2) | lnIPT(-1) | lnPPC(-2) | lnIPT(-1) | lnPPC(-2) | lnIPT(-1) | lnPPC(-2) | lnIPT(-1) | lnPPC(-2) | lnIPT(-1) | lnPPC(-2) | lnIPT(-1) |
| | Full sample | | High-pollution | | Non-pollution | | High-tech | | Low-tech | | Manufacuring | | Non-manufacturing | |
| | | | | | | | Step 2: R&D investment | | | | | | | |
| lnPPF(-2) | -0.0771*** | | -0.0399 | | -0.115*** | | -0.0717*** | | -0.0307 | | -0.0900*** | | -0.138*** | |
| | (0.013) | | (0.025) | | (0.015) | | (0.016) | | (0.025) | | (0.020) | | (0.025) | |
| lnIPT(-1) | | -0.0716*** | | -0.048 | | -0.0972*** | | -0.0667*** | | -0.0595 | | -0.0672** | | -0.166*** |
| | | (0.016) | | (0.029) | | (0.018) | | (0.019) | | (0.031) | | (0.023) | | (0.032) |
| LEV | -0.875*** | -1.126*** | -1.287*** | -1.294*** | -0.802*** | -0.832*** | -0.619*** | -0.629*** | -0.739*** | -0.744*** | -0.598*** | -0.620*** | -0.828*** | -0.821*** |
| | (0.076) | (0.075) | (0.126) | (0.126) | (0.093) | (0.090) | (0.095) | (0.095) | (0.134) | (0.133) | (0.110) | (0.107) | (0.149) | (0.150) |
| lnEMP | -0.214*** | -0.194*** | -0.233*** | -0.231*** | -0.120*** | -0.119*** | -0.190*** | -0.189*** | -0.352*** | -0.351*** | -0.160*** | -0.159*** | -0.389*** | -0.386*** |
| | (0.014) | (0.013) | (0.026) | (0.026) | (0.016) | (0.015) | (0.016) | (0.016) | (0.025) | (0.025) | (0.019) | (0.019) | (0.027) | (0.027) |
| lnCAPINT | -0.0796*** | -0.0809*** | 0.154*** | 0.156*** | -0.0405* | -0.0441** | -0.0389* | -0.0403* | -0.0754*** | -0.0740*** | 0.0131 | 0.0117 | -0.171*** | -0.170*** |
| | (0.013) | (0.013) | (0.031) | (0.031) | (0.017) | (0.017) | (0.017) | (0.017) | (0.022) | (0.022) | (0.023) | (0.023) | (0.022) | (0.022) |
| lnPCINT | 0.379*** | 0.394*** | 0.326*** | 0.327*** | 0.374*** | 0.371*** | 0.337*** | 0.335*** | 0.404*** | 0.401*** | 0.441*** | 0.440*** | 0.243*** | 0.241*** |
| | (0.015) | (0.016) | (0.032) | (0.032) | (0.020) | (0.019) | (0.019) | (0.019) | (0.026) | (0.026) | (0.025) | (0.024) | (0.027) | (0.027) |
| CR4 | -1.281*** | -1.312*** | -0.988*** | -0.990*** | -0.494*** | -0.505*** | -1.781*** | -1.788*** | -0.211 | -0.211 | -0.580*** | -0.597*** | -1.533*** | -1.531*** |
| | (0.085) | (0.084) | (0.170) | (0.169) | (0.100) | (0.097) | (0.107) | (0.106) | (0.181) | (0.180) | (0.126) | (0.124) | (0.150) | (0.150) |
| SOE | -0.177*** | -0.164*** | -0.249*** | -0.248*** | -0.147*** | -0.122*** | -0.0776* | -0.0685 | -0.280*** | -0.292*** | -0.143*** | -0.126** | -0.186* | -0.179* |
| | (0.032) | (0.032) | (0.056) | (0.056) | (0.035) | (0.034) | (0.039) | (0.039) | (0.063) | (0.063) | (0.041) | (0.040) | (0.077) | (0.077) |
| _cons | 1.739*** | 1.742*** | 0.775** | 0.851** | 1.698*** | 1.698*** | 1.955*** | 1.986*** | 0.610* | 0.832* | 0.875*** | 0.795*** | 3.520*** | 3.813*** |
| | (0.131) | (0.151) | (0.241) | (0.272) | (0.149) | (0.167) | (0.160) | (0.185) | (0.311) | (0.344) | (0.199) | (0.219) | (0.260) | (0.304) |
| Year FE | Y | Y | Y | Y | Y | Y | Y | Y | Y | Y | Y | Y | Y | Y |
| Region FE | Y | Y | Y | Y | Y | Y | Y | Y | Y | Y | Y | Y | Y | Y |
| | | | | | | | Step 1: R&D dummy | | | | | | | |
| AGE | -0.0708*** | -0.0751*** | -0.0764*** | -0.0763*** | -0.0855*** | -0.0859*** | -0.0703*** | -0.0705*** | -0.0606*** | -0.0610*** | -0.0885*** | -0.0890*** | -0.0649*** | -0.0649*** |
| | (0.002) | (0.002) | (0.005) | (0.005) | (0.005) | (0.005) | (0.003) | (0.003) | (0.003) | (0.003) | (0.004) | (0.004) | (0.003) | (0.003) |
| EPD | 0.143*** | 0.156*** | 0.213** | 0.209** | -0.0804 | -0.0792 | 0.126** | 0.123** | 0.167*** | 0.161*** | 0.0578 | 0.0512 | 0.0781 | 0.0776 |
| | (0.035) | (0.035) | (0.067) | (0.067) | (0.065) | (0.065) | (0.042) | (0.042) | (0.047) | (0.047) | (0.059) | (0.059) | (0.050) | (0.050) |
| lnASSET | -0.409*** | -0.426*** | -0.325*** | -0.332*** | -0.165*** | -0.160*** | -0.391*** | -0.391*** | -0.319*** | -0.320*** | -0.168*** | -0.170*** | -0.224*** | -0.224*** |

|  | (1) | (2) | (3) | (4) | (5) | (6) | (7) | (8) | (9) | (10) | (11) | (12) | (13) | (14) |
|---|---|---|---|---|---|---|---|---|---|---|---|---|---|---|
|  | (0.018) | (0.018) | (0.046) | (0.045) | (0.043) | (0.043) | (0.021) | (0.021) | (0.022) | (0.022) | (0.039) | (0.039) | (0.023) | (0.023) |
| lnPPF(-2) | 0.0325** |  | 0.0209 |  | -0.0941*** |  | 0.0323* |  | 0.0787*** |  | -0.0271 |  | 0.00596 |  |
|  | (0.012) |  | (0.026) |  | (0.025) |  | (0.014) |  | (0.016) |  | (0.023) |  | (0.016) |  |
| lnIPT(-1) |  | 0.0224 |  | -0.00202 |  | -0.133*** |  | 0.0326 |  | 0.0729*** |  | -0.0763** |  | 0.00701 |
|  |  | (0.015) |  | (0.030) |  | (0.030) |  | (0.017) |  | (0.019) |  | (0.027) |  | (0.020) |
| LEV | -0.696*** | -0.587*** | -0.587*** | -0.586*** | -0.998*** | -0.991*** | -0.705*** | -0.703*** | -0.391*** | -0.390*** | -0.841*** | -0.843*** | -0.452*** | -0.452*** |
|  | (0.057) |  | (0.107) | (0.107) | (0.103) | (0.103) | (0.068) | (0.068) | (0.076) | (0.076) | (0.092) | (0.092) | (0.081) | (0.081) |
| lnEMP | 0.551*** | 0.544*** | 0.544*** | 0.549*** | 0.390*** | 0.384*** | 0.505*** | 0.506*** | 0.457*** | 0.457*** | 0.368*** | 0.371*** | 0.355*** | 0.355*** |
|  | (0.016) | (0.016) | (0.042) | (0.042) | (0.038) | (0.038) | (0.018) | (0.018) | (0.020) | (0.020) | (0.036) | (0.036) | (0.020) | (0.020) |
| lnCAPINT | 0.0578*** | 0.0684*** | -0.0755** | -0.0733* | 0.00542 | 0.00296 | 0.0358** | 0.0367** | 0.0702*** | 0.0713*** | -0.0267 | -0.0261 | 0.00738 | 0.00747 |
|  | (0.011) | (0.011) | (0.031) | (0.030) | (0.025) | (0.025) | (0.013) | (0.013) | (0.014) | (0.014) | (0.025) | (0.025) | (0.013) | (0.013) |
| lnPCINT | 0.133*** | 0.107*** | 0.221*** | 0.223*** | 0.0529 | 0.0488 | 0.116*** | 0.117*** | 0.150*** | 0.150*** | 0.123*** | 0.123*** | 0.0750*** | 0.0750*** |
|  | (0.015) | (0.014) | (0.029) | (0.029) | (0.029) | (0.029) | (0.017) | (0.017) | (0.019) | (0.019) | (0.027) | (0.027) | (0.019) | (0.019) |
| CR4 | -1.142*** | -1.133*** | -1.870*** | -1.863*** | -0.404** | -0.404** | -0.767*** | -0.768*** | 0.307** | 0.300** | -0.759*** | -0.751*** | -0.349*** | -0.350*** |
|  | (0.073) | (0.073) | (0.143) | (0.143) | (0.138) | (0.138) | (0.091) | (0.091) | (0.109) | (0.109) | (0.133) | (0.133) | (0.097) | (0.097) |
| SOE | -0.212*** | -0.225*** | -0.162** | -0.168** | 0.260*** | 0.272*** | -0.225*** | -0.228*** | -0.293*** | -0.300*** | 0.171*** | 0.172*** | -0.311*** | -0.311*** |
|  | (0.028) | (0.027) | (0.057) | (0.057) | (0.054) | (0.054) | (0.032) | (0.032) | (0.036) | (0.036) | (0.046) | (0.046) | (0.039) | (0.039) |
| _cons | 3.084*** | 2.959*** | 2.751*** | 2.918*** | 2.983*** | 3.333*** | 3.119*** | 3.089*** | 0.724*** | 0.713*** | 2.286*** | 2.662*** | 1.617*** | 1.605*** |
|  | (0.135) | (0.159) | (0.294) | (0.324) | (0.288) | (0.320) | (0.159) | (0.180) | (0.186) | (0.209) | (0.264) | (0.292) | (0.179) | (0.204) |
| Year FE | Y | Y | Y | Y | Y | Y | Y | Y | Y | Y | Y | Y | Y | Y |
| Region FE | Y | Y | Y | Y | Y | Y | Y | Y | Y | Y | Y | Y | Y | Y |
| /mills |  |  |  |  |  |  |  |  |  |  |  |  |  |  |
| lambda | -0.755*** | -0.699*** | -1.370*** | -1.361*** | -1.329*** | -1.290*** | -0.737*** | -0.724*** | -0.824*** | -0.805*** | -1.665*** | -1.633*** | -0.768*** | -0.768*** |
|  | (0.069) | (0.066) | (0.134) | (0.133) | (0.122) | (0.118) | (0.081) | (0.081) | (0.114) | (0.113) | (0.138) | (0.135) | (0.124) | (0.124) |
| rho | -0.543 | -0.506 | -0.921 | -0.916 | -1.000 | -1.000 | -0.535 | -0.526 | -0.533 | -0.522 | -1.000 | -1.000 | -0.485 | -0.485 |
| sigma | 1.391 | 1.380 | 1.488 | 1.485 | 1.329 | 1.290 | 1.378 | 1.376 | 1.546 | 1.540 | 1.665 | 1.633 | 1.583 | 1.584 |
| Wald chi2 | 2504.110 | 2550.630 | 937.470 | 939.440 | 1240.500 | 1289.610 | 1513.540 | 1505.100 | 966.850 | 972.040 | 882.310 | 904.880 | 1152.440 | 1147.070 |
|  | (0.000) | (0.000) | (0.000) | (0.000) | (0.000) | (0.000) | (0.000) | (0.000) | (0.000) | (0.000) | (0.000) | (0.000) | (0.000) | (0.000) |
| N | 15639 | 15552 | 4826 | 4826 | 8322 | 8322 | 10813 | 10813 | 7317 | 7317 | 9187 | 9187 | 6452 | 6452 |

Standard errors in parenthesis
* p<0.05, ** p<0.01, *** p<0.001

*Table B3 Summary table of predicted R&D intensity*

| Sample | R/P | Obs | Mean | Std.Dev. | Min | Max |
|---|---|---|---|---|---|---|
| Full sample | Realistic | 11,378 | 2.77 | 1.50 | -7.20 | 13.23 |
|  | Predicted | 15,639 | 2.66 | 0.85 | -15.36 | 5.97 |
| High-pollution | Realistic | 3,832 | 2.54 | 1.54 | -6.02 | 13.23 |
|  | Predicted | 4,826 | 2.37 | 1.09 | -21.06 | 4.77 |
| Non-pollution | Realistic | 7,546 | 2.89 | 1.47 | -7.20 | 12.65 |
|  | Predicted | 10,813 | 2.80 | 0.77 | -12.01 | 5.61 |
| High-tech | Realistic | 7,578 | 3.15 | 1.28 | -7.20 | 13.23 |
|  | Predicted | 8,322 | 3.08 | 0.82 | -21.31 | 4.97 |
| Low-tech | Realistic | 3,800 | 2.02 | 1.63 | -5.78 | 6.57 |
|  | Predicted | 7,317 | 1.94 | 0.90 | -11.33 | 5.72 |
| Manufacturing | Realistic | 8,196 | 2.94 | 1.34 | -7.20 | 13.23 |
|  | Predicted | 9,187 | 2.86 | 0.83 | -19.96 | 5.05 |
| Non-manufacturing | Realistic | 3,182 | 2.34 | 1.79 | -5.94 | 6.57 |
|  | Predicted | 6,452 | 2.21 | 1.06 | -11.94 | 6.27 |

*Table B4 Summary table of exponential transformed predicted patent applications*

| Sample | R/P | Obs | Mean | Std.Dev. | Min | Max |
|---|---|---|---|---|---|---|
| Full sample | Realistic | 15,714 | 70.87 | 333.11 | 0 | 11171.00 |
| | Predicted | 14,904 | 72.21 | 323.78 | 0 | 8899.83 |
| High-pollution | Realistic | 4,833 | 52.45 | 258.49 | 0 | 5843.33 |
| | Predicted | 4,648 | 53.53 | 265.00 | 0 | 6979.19 |
| Non-pollution | Realistic | 10,881 | 79.05 | 361.06 | 0 | 11171.00 |
| | Predicted | 10,256 | 80.67 | 346.25 | 0 | 8825.55 |
| High-tech | Realistic | 8,361 | 94.10 | 375.60 | 0 | 11171.00 |
| | Predicted | 7,805 | 97.56 | 361.59 | 0 | 8613.34 |
| Low-tech | Realistic | 7,353 | 44.46 | 274.64 | 0 | 5843.33 |
| | Predicted | 7,099 | 44.33 | 273.11 | 0 | 7685.63 |
| Manufacturing | Realistic | 9,207 | 89.84 | 360.99 | 0 | 11171.00 |
| | Predicted | 8,676 | 92.26 | 345.96 | 0 | 8632.77 |
| Non-manufacturing | Realistic | 6,507 | 44.04 | 287.00 | 0 | 5843.33 |
| | Predicted | 5,291 | 46.36 | 309.89 | 0 | 7723.24 |

*Table B5 Summary table of exponential transformed predicted green patent applications*

| Sample | R/P | Obs | Mean | Std.Dev. | Min | Max |
|---|---|---|---|---|---|---|
| Full sample | Realistic | 15,714 | 8.92 | 50.89 | 0 | 1484.00 |
| | Predicted | 14,904 | 9.07 | 51.71 | 0 | 1844.50 |
| High-pollution | Realistic | 4,833 | 7.25 | 51.01 | 0 | 1427.00 |
| | Predicted | 4,648 | 7.42 | 54.47 | 0 | 1790.78 |
| Non-pollution | Realistic | 10,881 | 9.66 | 50.82 | 0 | 1484.00 |
| | Predicted | 10,256 | 9.81 | 50.10 | 0 | 1727.47 |
| High-tech | Realistic | 8,361 | 10.52 | 49.35 | 0 | 1484.00 |
| | Predicted | 7,805 | 10.88 | 49.48 | 0 | 1662.32 |
| Low-tech | Realistic | 7,353 | 7.10 | 52.53 | 0 | 1427.00 |
| | Predicted | 7,099 | 7.07 | 54.90 | 0 | 1997.41 |
| Manufacturing | Realistic | 9,207 | 9.41 | 45.37 | 0 | 1484.00 |
| | Predicted | 8,676 | 9.63 | 45.46 | 0 | 1643.66 |
| Non-manufacturing | Realistic | 6,507 | 8.23 | 57.79 | 0 | 1427.00 |
| | Predicted | 6,228 | 8.28 | 60.79 | 0 | 2021.12 |

*Table B6 Summary table of exponential transformed predicted non-green patent applications*

| Sample | R/P | Obs | Mean | Std.Dev. | Min | Max |
|---|---|---|---|---|---|---|
| Full sample | Realistic | 15,714 | 61.95 | 293.54 | 0 | 10045.00 |
| | Predicted | 14,904 | 63.14 | 284.35 | 0 | 7977.02 |
| High-pollution | Realistic | 4,833 | 45.20 | 212.67 | 0 | 4416.33 |
| | Predicted | 4,648 | 46.12 | 216.31 | 0 | 5294.40 |
| Non-pollution | Realistic | 10,881 | 69.39 | 322.76 | 0 | 10045.00 |
| | Predicted | 10,256 | 70.86 | 309.35 | 0 | 7914.95 |
| High-tech | Realistic | 8,361 | 83.58 | 340.96 | 0 | 10045.00 |
| | Predicted | 7,805 | 86.69 | 328.02 | 0 | 7637.60 |
| Low-tech | Realistic | 7,353 | 37.36 | 225.47 | 0 | 4718.00 |
| | Predicted | 7,099 | 37.25 | 222.33 | 0 | 5801.64 |
| Manufacturing | Realistic | 9,207 | 80.43 | 328.57 | 0 | 10045.00 |
| | Predicted | 8,676 | 82.63 | 314.92 | 0 | 7705.33 |
| Non-manufacturing | Realistic | 6,507 | 35.81 | 232.76 | 0 | 4718.00 |
| | Predicted | 6,228 | 35.99 | 232.39 | 0 | 5838.48 |

*Table B7 Mundlak test for CDM and Extended CDM models*

| | CDM | Extended CDM |
|---|---|---|
| $\widehat{\text{lnPATINT}}$ | 0.154* | |
| | (0.066) | |
| $\widehat{\text{lnNECOINT}}$ | | 0.471*** |
| | | (0.050) |
| $\widehat{\text{lnECOINT}}$ | | -0.0729 |
| | | (0.048) |
| lnCAPINT | 0.0604*** | 0.130*** |
| | (0.018) | (0.020) |
| lnEMP | 0.0469 | 0.00573 |
| | (0.053) | (0.024) |
| _cons | 5.336*** | 5.817*** |
| | (0.143) | (0.147) |
| Mean_lnPATINT | -0.173** | |
| | (0.066) | |
| Mean_NECOINT | | -0.529*** |
|  |  | (0.051) |
| Mean_lnECOINT |  | 0.121* |
|  |  | (0.049) |
| Mean_lnCAPINT | 0.177*** | 0.105*** |
|  | (0.029) | (0.031) |
| Mean_lnEMP | -0.0968 | -0.0619* |
|  | (0.055) | (0.030) |
| Chi2 | **44.33** | **211.51** |
| Prob>Chi2 | **(0.000)** | **(0.000)** |

*Table C1 Results of UQR on CDM models (High-pollution sample)*

| High-pollution UQR | Model 75 Q10 | Model 76 Q20 | Model 77 Q30 | Model 78 Q40 | Model 79 Q50 | Model 80 Q60 | Model 81 Q70 | Model 82 Q80 | Model 83 Q90 |
|---|---|---|---|---|---|---|---|---|---|
| $\widehat{\text{lnPATINT}}$ | 0.211*** | 0.209*** | 0.244*** | 0.243*** | 0.320*** | 0.333*** | 0.405*** | 0.423*** | 0.460*** |
|  | (3.850) | (4.726) | (5.654) | (5.773) | (9.305) | (7.517) | (7.389) | (6.692) | (6.958) |
| lnCAPINT | 0.102*** | 0.140*** | 0.120*** | 0.144*** | 0.151*** | 0.115*** | 0.131*** | 0.156*** | 0.097* |
|  | (3.323) | (5.527) | (5.105) | (6.241) | (6.877) | (4.697) | (4.453) | (4.136) | (2.115) |
| lnEMP | 0.104* | 0.036 | 0.061+ | 0.013 | 0.004 | -0.093** | -0.066+ | -0.133** | -0.131* |
|  | (2.325) | (1.048) | (1.813) | (0.412) | (0.125) | (-2.802) | (-1.669) | (-2.834) | (-2.469) |
| _cons | 4.983*** | 5.118*** | 5.395*** | 5.481*** | 5.531*** | 6.078*** | 6.119*** | 6.331*** | 7.063*** |
|  | (21.547) | (26.501) | (29.423) | (30.543) | (35.950) | (34.120) | (27.645) | (22.905) | (21.549) |
| Individual FE | Y | Y | Y | Y | Y | Y | Y | Y | Y |
| Time FE | Y | Y | Y | Y | Y | Y | Y | Y | Y |
| adj. R-sq | 0.612 | 0.612 | 0.612 | 0.612 | 0.612 | 0.612 | 0.612 | 0.612 | 0.612 |
| F-Stat | 56.67 | 56.67 | 56.67 | 56.67 | 56.67 | 56.67 | 56.67 | 56.67 | 56.67 |
| Prob > F | [0.000] | [0.000] | [0.000] | [0.000] | [0.000] | [0.000] | [0.000] | [0.000] | [0.000] |
| Log likelihood | -23266 | -23266 | -23266 | -23266 | -23266 | -23266 | -23266 | -23266 | -23266 |
| N | 4,594 | 4,594 | 4,594 | 4,594 | 4,594 | 4,594 | 4,594 | 4,594 | 4,594 |

Standard errors in parenthesis
\*\*\* p<0.001, \*\* p<0.01, \* p<0.05, + p<0.1

*Table C2 Results of UQR on CDM models (Non-pollution sample)*

| Non-pollution UQR | Model 84 Q10 | Model 85 Q20 | Model 86 Q30 | Model 87 Q40 | Model 88 Q50 | Model 89 Q60 | Model 90 Q70 | Model 91 Q80 | Model 92 Q90 |
|---|---|---|---|---|---|---|---|---|---|
| ln$\widehat{PATINT}$ | 0.334*** | 0.313*** | 0.277*** | 0.285*** | 0.283*** | 0.280*** | 0.319*** | 0.441*** | 0.490*** |
|  | (8.840) | (10.422) | (10.573) | (11.458) | (11.217) | (10.524) | (10.255) | (11.245) | (7.465) |
| lnCAPINT | 0.098*** | 0.091*** | 0.098*** | 0.097*** | 0.085*** | 0.072*** | 0.075*** | 0.103*** | 0.157*** |
|  | (4.944) | (5.730) | (6.924) | (6.913) | (6.043) | (4.910) | (4.389) | (4.173) | (4.096) |
| lnEMP | -0.027 | 0.02 | 0.018 | 0.004 | 0.008 | 0.003 | -0.034 | -0.025 | -0.047 |
|  | (-0.940) | (0.884) | (0.903) | (0.244) | (0.424) | (0.166) | (-1.457) | (-0.833) | (-0.999) |
| _cons | 4.969*** | 5.335*** | 5.595*** | 5.823*** | 6.119*** | 6.410*** | 6.622*** | 6.681*** | 6.944*** |
|  | (39.432) | (51.826) | (60.930) | (64.736) | (67.546) | (66.808) | (58.750) | (43.313) | (29.232) |
| Individual FE | Y | Y | Y | Y | Y | Y | Y | Y | Y |
| Time FE | Y | Y | Y | Y | Y | Y | Y | Y | Y |
| adj. R-sq | 0.612 | 0.612 | 0.612 | 0.612 | 0.612 | 0.612 | 0.612 | 0.612 | 0.612 |
| F-Stat | 56.67 | 56.67 | 56.67 | 56.67 | 56.67 | 56.67 | 56.67 | 56.67 | 56.67 |
| Prob > F | [0.000] | [0.000] | [0.000] | [0.000] | [0.000] | [0.000] | [0.000] | [0.000] | [0.000] |
| Log likelihood | -23266 | -23266 | -23266 | -23266 | -23266 | -23266 | -23266 | -23266 | -23266 |
| N | 10,127 | 10,127 | 10,127 | 10,127 | 10,127 | 10,127 | 10,127 | 10,127 | 10,127 |

Standard errors in parenthesis
*** p<0.001, ** p<0.01, * p<0.05, + p<0.1

*Table C3 Results of UQR on CDM models (High-tech sample)*

| High-tech UQR | Model 111 Q10 | Model 112 Q20 | Model 113 Q30 | Model 114 Q40 | Model 115 Q50 | Model 116 Q60 | Model 117 Q70 | Model 118 Q80 | Model 119 Q90 |
|---|---|---|---|---|---|---|---|---|---|
| ln$\widehat{PATINT}$ | 0.273*** | 0.252*** | 0.226*** | 0.248*** | 0.260*** | 0.283*** | 0.285*** | 0.383*** | 0.545*** |
|  | (6.433) | (7.439) | (7.509) | (8.532) | (8.574) | (8.452) | (7.627) | (7.572) | (8.437) |
| lnCAPINT | 0.093*** | 0.089*** | 0.105*** | 0.107*** | 0.116*** | 0.098*** | 0.068*** | 0.065** | 0.099* |
|  | (3.938) | (4.919) | (6.529) | (6.610) | (7.073) | (5.740) | (3.606) | (2.664) | (2.336) |
| lnEMP | 0.048 | 0.062* | 0.057** | 0.042* | 0.02 | 0.006 | -0.021 | -0.026 | -0.059 |
|  | (1.531) | (2.511) | (2.612) | (2.002) | (0.907) | (0.252) | (-0.745) | (-0.693) | (-1.152) |
| _cons | 4.726*** | 5.073*** | 5.261*** | 5.390*** | 5.512*** | 5.752*** | 6.150*** | 6.208*** | 6.111*** |
|  | (25.364) | (34.363) | (40.047) | (41.379) | (41.068) | (38.738) | (37.536) | (28.516) | (19.786) |
| Individual FE | Y | Y | Y | Y | Y | Y | Y | Y | Y |
| Time FE | Y | Y | Y | Y | Y | Y | Y | Y | Y |
| adj. R-sq | 0.612 | 0.612 | 0.612 | 0.612 | 0.612 | 0.612 | 0.612 | 0.612 | 0.612 |
| F-Stat | 56.67 | 56.67 | 56.67 | 56.67 | 56.67 | 56.67 | 56.67 | 56.67 | 56.67 |
| Prob > F | [0.000] | [0.000] | [0.000] | [0.000] | [0.000] | [0.000] | [0.000] | [0.000] | [0.000] |
| Log likelihood | -23266 | -23266 | -23266 | -23266 | -23266 | -23266 | -23266 | -23266 | -23266 |
| N | 7,714 | 7,714 | 7,714 | 7,714 | 7,714 | 7,714 | 7,714 | 7,714 | 7,714 |

Standard errors in parenthesis
*** p<0.001, ** p<0.01, * p<0.05, + p<0.1

*Table C4 Results of UQR on CDM models (Low-tech sample)*

| Low-tech UQR | Model 120 Q10 | Model 121 Q20 | Model 122 Q30 | Model 123 Q40 | Model 124 Q50 | Model 125 Q60 | Model 126 Q70 | Model 127 Q80 | Model 128 Q90 |
|---|---|---|---|---|---|---|---|---|---|
| ln$\widehat{PATINT}$ | 0.369*** | 0.351*** | 0.319*** | 0.294*** | 0.331*** | 0.339*** | 0.388*** | 0.437*** | 0.591*** |
|  | (7.200) | (9.233) | (9.289) | (9.205) | (11.128) | (10.682) | (10.254) | (9.560) | (8.478) |
| lnCAPINT | 0.103*** | 0.123*** | 0.123*** | 0.117*** | 0.102*** | 0.099*** | 0.127*** | 0.123*** | 0.180*** |
|  | (3.891) | (5.832) | (6.408) | (6.432) | (5.745) | (5.207) | (5.594) | (4.514) | (4.637) |
| lnEMP | -0.049 | -0.005 | -0.005 | -0.007 | -0.017 | -0.051* | -0.051+ | -0.041 | -0.047 |
|  | (-1.188) | (-0.163) | (-0.188) | (-0.293) | (-0.694) | (-2.103) | (-1.795) | (-1.267) | (-0.981) |
| _cons | 5.336*** | 5.595*** | 5.885*** | 6.184*** | 6.542*** | 6.839*** | 6.974*** | 7.393*** | 7.626*** |
|  | (32.428) | (42.104) | (48.377) | (54.538) | (58.863) | (58.108) | (49.313) | (44.312) | (32.307) |
| Individual FE | Y | Y | Y | Y | Y | Y | Y | Y | Y |
| Time FE | Y | Y | Y | Y | Y | Y | Y | Y | Y |
| adj. R-sq | 0.612 | 0.612 | 0.612 | 0.612 | 0.612 | 0.612 | 0.612 | 0.612 | 0.612 |
| F-Stat | 56.67 | 56.67 | 56.67 | 56.67 | 56.67 | 56.67 | 56.67 | 56.67 | 56.67 |
| Prob > F | [0.000] | [0.000] | [0.000] | [0.000] | [0.000] | [0.000] | [0.000] | [0.000] | [0.000] |
| Log likelihood | -23266 | -23266 | -23266 | -23266 | -23266 | -23266 | -23266 | -23266 | -23266 |
| N | 7,007 | 7,007 | 7,007 | 7,007 | 7,007 | 7,007 | 7,007 | 7,007 | 7,007 |

Standard errors in parenthesis
*** $p<0.001$, ** $p<0.01$, * $p<0.05$, + $p<0.1$

*Table C5 Results of UQR on CDM models (Manufacturing sample)*

| Manufacturing UQR | Model 147 Q10 | Model 148 Q20 | Model 149 Q30 | Model 150 Q40 | Model 151 Q50 | Model 152 Q60 | Model 153 Q70 | Model 154 Q80 | Model 155 Q90 |
|---|---|---|---|---|---|---|---|---|---|
| ln$\widehat{PATINT}$ | 0.247*** | 0.232*** | 0.202*** | 0.226*** | 0.256*** | 0.305*** | 0.314*** | 0.414*** | 0.585*** |
|  | (5.482) | (7.047) | (7.152) | (8.458) | (9.313) | (10.874) | (9.522) | (10.368) | (8.929) |
| lnCAPINT | 0.131*** | 0.111*** | 0.119*** | 0.124*** | 0.149*** | 0.150*** | 0.116*** | 0.125*** | 0.156*** |
|  | (4.844) | (5.808) | (7.220) | (7.556) | (9.007) | (8.625) | (6.005) | (5.377) | (4.087) |
| lnEMP | 0.124*** | 0.109*** | 0.103*** | 0.073*** | 0.062** | 0.055** | 0.02 | 0.006 | -0.063 |
|  | (3.500) | (4.345) | (4.892) | (3.651) | (3.055) | (2.609) | (0.827) | (0.209) | (-1.288) |
| _cons | 4.517*** | 4.970*** | 5.202*** | 5.327*** | 5.304*** | 5.381*** | 5.791*** | 5.801*** | 5.738*** |
|  | (21.268) | (32.830) | (40.335) | (42.679) | (42.371) | (41.058) | (38.757) | (32.647) | (19.470) |
| Individual FE | Y | Y | Y | Y | Y | Y | Y | Y | Y |
| Time FE | Y | Y | Y | Y | Y | Y | Y | Y | Y |
| adj. R-sq | 0.621 | 0.621 | 0.621 | 0.621 | 0.621 | 0.621 | 0.621 | 0.621 | 0.621 |
| F-Stat | 28.45 | 28.45 | 28.45 | 28.45 | 28.45 | 28.45 | 28.45 | 28.45 | 28.45 |
| Prob > F | [0.000] | [0.000] | [0.000] | [0.000] | [0.000] | [0.000] | [0.000] | [0.000] | [0.000] |
| Log likelihood | -9193 | -9193 | -9193 | -9193 | -9193 | -9193 | -9193 | -9193 | -9193 |
| N | 8,583 | 8,583 | 8,583 | 8,583 | 8,583 | 8,583 | 8,583 | 8,583 | 8,583 |

Standard errors in parenthesis
*** $p<0.001$, ** $p<0.01$, * $p<0.05$, + $p<0.1$

*Table C6 Results of UQR on CDM models (Non-manufacturing sample)*

| Non-manufacturing UQR | Model 156 Q10 | Model 157 Q20 | Model 158 Q30 | Model 159 Q40 | Model 160 Q50 | Model 161 Q60 | Model 162 Q70 | Model 163 Q80 | Model 164 Q90 |
|---|---|---|---|---|---|---|---|---|---|
| ln$\widehat{PATINT}$ | 0.354*** | 0.395*** | 0.303*** | 0.300*** | 0.304*** | 0.328*** | 0.324*** | 0.366*** | 0.619*** |
|  | (5.679) | (7.747) | (6.957) | (8.474) | (9.023) | (8.530) | (7.706) | (7.631) | (7.583) |
| lnCAPINT | 0.098** | 0.145*** | 0.140*** | 0.104*** | 0.097*** | 0.121*** | 0.115*** | 0.130*** | 0.170*** |
|  | (3.293) | (5.286) | (5.672) | (5.063) | (4.863) | (5.347) | (4.741) | (4.981) | (3.993) |
| lnEMP | -0.096* | -0.014 | -0.014 | -0.027 | -0.043+ | -0.059* | -0.047 | -0.048 | -0.067 |
|  | (-2.011) | (-0.365) | (-0.400) | (-0.968) | (-1.683) | (-2.055) | (-1.624) | (-1.455) | (-1.256) |
| _cons | 5.566*** | 5.727*** | 6.085*** | 6.600*** | 6.909*** | 7.062*** | 7.398*** | 7.698*** | 8.192*** |
|  | (29.791) | (33.116) | (39.488) | (51.315) | (56.053) | (50.693) | (49.987) | (48.473) | (31.778) |
| Individual FE | Y | Y | Y | Y | Y | Y | Y | Y | Y |
| Time FE | Y | Y | Y | Y | Y | Y | Y | Y | Y |
| adj. R-sq | 0.621 | 0.621 | 0.621 | 0.621 | 0.621 | 0.621 | 0.621 | 0.621 | 0.621 |
| F-Stat | 28.45 | 28.45 | 28.45 | 28.45 | 28.45 | 28.45 | 28.45 | 28.45 | 28.45 |
| Prob > F | [0.000] | [0.000] | [0.000] | [0.000] | [0.000] | [0.000] | [0.000] | [0.000] | [0.000] |
| Log likelihood | -9193 | -9193 | -9193 | -9193 | -9193 | -9193 | -9193 | -9193 | -9193 |
| N | 5,214 | 5,214 | 5,214 | 5,214 | 5,214 | 5,214 | 5,214 | 5,214 | 5,214 |

Standard errors in parenthesis
*** p<0.001, ** p<0.01, * p<0.05, + p<0.1

*Table D1 Results of RIF treatment effects without weights (Full sample)*

| Full sample RIF-Treat | Model 183 Q10 | Model 184 Q20 | Model 185 Q30 | Model 186 Q40 | Model 187 Q50 | Model 188 Q60 | Model 189 Q70 | Model 190 Q80 | Model 191 Q90 |
|---|---|---|---|---|---|---|---|---|---|
| ECO | 0.050+ | 0.080*** | 0.062** | 0.080*** | 0.077*** | 0.075*** | 0.042+ | -0.009 | 0.052 |
|  | (1.767) | (3.616) | (2.954) | (3.890) | (3.815) | (3.565) | (1.806) | (-0.307) | (1.331) |
| ln$\widehat{NECOINT}$ | 0.308*** | 0.270*** | 0.288*** | 0.298*** | 0.312*** | 0.330*** | 0.368*** | 0.463*** | 0.547*** |
|  | (9.622) | (10.768) | (12.428) | (13.147) | (13.360) | (13.464) | (12.890) | (13.508) | (10.274) |
| lnCAPINT | 0.093*** | 0.099*** | 0.105*** | 0.115*** | 0.111*** | 0.089*** | 0.089*** | 0.135*** | 0.138*** |
|  | (5.458) | (7.413) | (8.240) | (9.016) | (8.606) | (6.395) | (5.401) | (6.135) | (4.066) |
| lnEMP | 0.022 | 0.039* | 0.026 | 0.006 | -0.007 | -0.022 | -0.048* | -0.050+ | -0.093* |
|  | (0.883) | (1.993) | (1.458) | (0.369) | (-0.392) | (-1.156) | (-2.222) | (-1.915) | (-2.304) |
| _cons | 5.022*** | 5.307*** | 5.523*** | 5.682*** | 5.918*** | 6.259*** | 6.527*** | 6.544*** | 6.986*** |
|  | (43.329) | (58.138) | (64.424) | (66.224) | (67.510) | (66.512) | (59.026) | (45.199) | (32.007) |
| Individual FE | Y | Y | Y | Y | Y | Y | Y | Y | Y |
| Time FE | Y | Y | Y | Y | Y | Y | Y | Y | Y |
| adj. R-sq | 0.609 | 0.609 | 0.609 | 0.609 | 0.609 | 0.609 | 0.609 | 0.609 | 0.609 |
| F-Stat | 41.14 | 41.14 | 41.14 | 41.14 | 41.14 | 41.14 | 41.14 | 41.14 | 41.14 |
| Prob > F | [0.000] | [0.000] | [0.000] | [0.000] | [0.000] | [0.000] | [0.000] | [0.000] | [0.000] |
| Log likelihood | -24481 | -24481 | -24481 | -24481 | -24481 | -24481 | -24481 | -24481 | -24481 |
| N | 14,721 | 14,721 | 14,721 | 14,721 | 14,721 | 14,721 | 14,721 | 14,721 | 14,721 |

Standard errors in parenthesis

*Table D 2 Results of RIF treatment effects without weights (High-pollution sample)*

| High-pollution RIF-Treat | Model 201 Q10 | Model 202 Q20 | Model 203 Q30 | Model 204 Q40 | Model 205 Q50 | Model 206 Q60 | Model 207 Q70 | Model 208 Q80 | Model 209 Q90 |
|---|---|---|---|---|---|---|---|---|---|
| ECO | 0.042 | 0.001 | -0.004 | -0.01 | 0.005 | 0.046 | 0.045 | 0.123** | 0.108* |
|  | (0.985) | (0.031) | (-0.125) | (-0.311) | (0.152) | (1.383) | (1.234) | (2.925) | (2.051) |
| lnNECOINT | 0.200*** | 0.201*** | 0.234*** | 0.292*** | 0.320*** | 0.333*** | 0.393*** | 0.469*** | 0.493*** |
|  | (3.727) | (4.228) | (5.367) | (6.425) | (6.830) | (6.828) | (7.962) | (7.120) | (6.244) |
| lnCAPINT | 0.136*** | 0.148*** | 0.150*** | 0.164*** | 0.155*** | 0.123*** | 0.158*** | 0.161*** | 0.137** |
|  | (3.808) | (5.488) | (6.104) | (6.770) | (6.470) | (5.013) | (5.546) | (4.548) | (2.892) |
| lnEMP | 0.090* | 0.106** | 0.051 | 0.049 | -0.001 | -0.046 | -0.089* | -0.139** | -0.148* |
|  | (2.092) | (2.803) | (1.510) | (1.401) | (-0.031) | (-1.270) | (-2.349) | (-2.846) | (-2.530) |
| _cons | 4.832*** | 5.043*** | 5.266*** | 5.324*** | 5.581*** | 5.988*** | 6.006*** | 6.240*** | 6.775*** |
|  | (17.951) | (24.164) | (27.398) | (27.785) | (28.701) | (30.165) | (29.138) | (23.743) | (19.974) |
| Individual FE | Y | Y | Y | Y | Y | Y | Y | Y | Y |
| Time FE | Y | Y | Y | Y | Y | Y | Y | Y | Y |
| adj. R-sq | 0.546 | 0.546 | 0.546 | 0.546 | 0.546 | 0.546 | 0.546 | 0.546 | 0.546 |
| F-Stat | 24.03 | 24.03 | 24.03 | 24.03 | 24.03 | 24.03 | 24.03 | 24.03 | 24.03 |
| Prob > F | [0.000] | [0.000] | [0.000] | [0.000] | [0.000] | [0.000] | [0.000] | [0.000] | [0.000] |
| Log likelihood | -6274 | -6274 | -6274 | -6274 | -6274 | -6274 | -6274 | -6274 | -6274 |
| N | 4,594 | 4,594 | 4,594 | 4,594 | 4,594 | 4,594 | 4,594 | 4,594 | 4,594 |

Standard errors in parenthesis
*** p<0.001, ** p<0.01, * p<0.05, + p<0.1

*Table D 3 Results of RIF treatment effects without weights (Non-pollution sample)*

| Non-pollution RIF-Treat | Model 219 Q10 | Model 220 Q20 | Model 221 Q30 | Model 222 Q40 | Model 223 Q50 | Model 224 Q60 | Model 225 Q70 | Model 226 Q80 | Model 227 Q90 |
|---|---|---|---|---|---|---|---|---|---|
| ECO | 0.04 | 0.128*** | 0.110*** | 0.086** | 0.109*** | 0.077** | 0.060+ | -0.073* | 0.018 |
|  | (1.110) | (4.359) | (3.966) | (3.170) | (4.144) | (2.777) | (1.945) | (-1.996) | (0.325) |
| lnNECOINT | 0.333*** | 0.299*** | 0.309*** | 0.307*** | 0.298*** | 0.329*** | 0.356*** | 0.431*** | 0.567*** |
|  | (8.794) | (10.007) | (11.032) | (11.222) | (10.814) | (11.406) | (10.909) | (11.090) | (8.324) |
| lnCAPINT | 0.082*** | 0.077*** | 0.096*** | 0.103*** | 0.084*** | 0.078*** | 0.071*** | 0.101*** | 0.155*** |
|  | (4.196) | (4.780) | (5.983) | (6.283) | (5.164) | (4.535) | (3.512) | (3.724) | (3.730) |
| lnEMP | -0.027 | 0.016 | -0.003 | -0.016 | -0.014 | -0.002 | -0.046+ | -0.038 | 0.009 |
|  | (-0.910) | (0.685) | (-0.153) | (-0.743) | (-0.665) | (-0.093) | (-1.822) | (-1.278) | (0.186) |
| _cons | 5.099*** | 5.407*** | 5.573*** | 5.789*** | 6.121*** | 6.364*** | 6.676*** | 6.855*** | 6.941*** |
|  | (40.801) | (52.690) | (55.475) | (56.109) | (59.656) | (58.719) | (51.782) | (40.984) | (27.460) |
| Individual FE | Y | Y | Y | Y | Y | Y | Y | Y | Y |
| Time FE | Y | Y | Y | Y | Y | Y | Y | Y | Y |
| adj. R-sq | 0.622 | 0.622 | 0.622 | 0.622 | 0.622 | 0.622 | 0.622 | 0.622 | 0.622 |
| F-Stat | 23.43 | 23.43 | 23.43 | 23.43 | 23.43 | 23.43 | 23.43 | 23.43 | 23.43 |
| Prob > F | [0.000] | [0.000] | [0.000] | [0.000] | [0.000] | [0.000] | [0.000] | [0.000] | [0.000] |
| Log likelihood | -17716 | -17716 | -17716 | -17716 | -17716 | -17716 | -17716 | -17716 | -17716 |
| N | 10,127 | 10,127 | 10,127 | 10,127 | 10,127 | 10,127 | 10,127 | 10,127 | 10,127 |

*Figure D 1 Non-linear estimations based on RIF treatment effects (High-tech sample)*

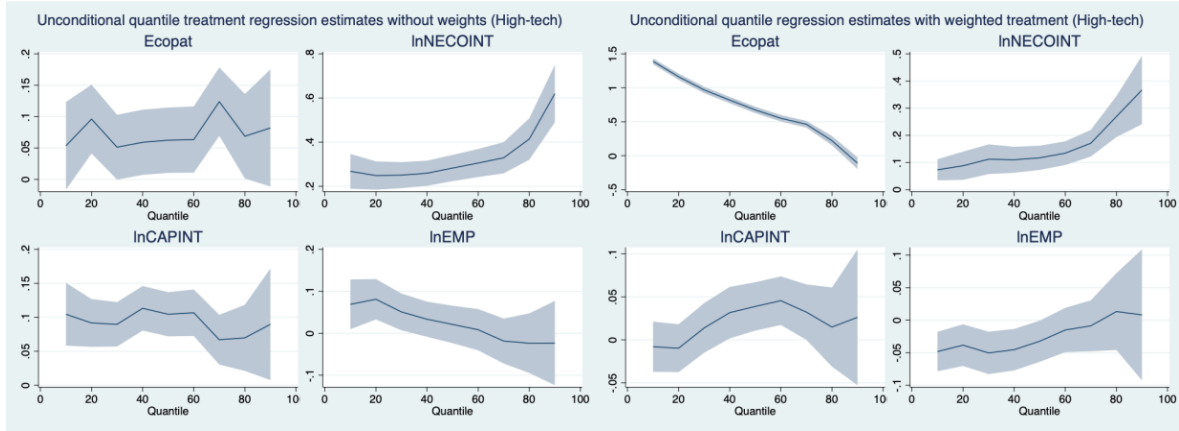

*Table D4 Results of RIF treatment effects without weight (High-tech sample)*

| High-tech RIF-Treat | Model 237 Q10 | Model 238 Q20 | Model 239 Q30 | Model 240 Q40 | Model 241 Q50 | Model 242 Q60 | Model 243 Q70 | Model 244 Q80 | Model 245 Q90 |
|---|---|---|---|---|---|---|---|---|---|
| ECO | 0.054 | 0.096*** | 0.051+ | 0.059* | 0.063* | 0.064* | 0.124*** | 0.069* | 0.082+ |
|  | (1.504) | (3.444) | (1.953) | (2.236) | (2.359) | (2.365) | (4.463) | (2.007) | (1.724) |
| $\widehat{lnNECOINT}$ | 0.268*** | 0.249*** | 0.251*** | 0.259*** | 0.283*** | 0.305*** | 0.329*** | 0.414*** | 0.620*** |
|  | (6.668) | (7.619) | (8.329) | (8.935) | (9.335) | (9.445) | (9.147) | (8.678) | (9.336) |
| lnCAPINT | 0.105*** | 0.092*** | 0.090*** | 0.113*** | 0.104*** | 0.107*** | 0.067*** | 0.070** | 0.090* |
|  | (4.436) | (5.102) | (5.386) | (6.781) | (6.310) | (6.109) | (3.587) | (2.810) | (2.141) |
| lnEMP | 0.069* | 0.081*** | 0.051* | 0.033 | 0.021 | 0.008 | -0.019 | -0.024 | -0.023 |
|  | (2.283) | (3.307) | (2.279) | (1.568) | (0.925) | (0.339) | (-0.679) | (-0.653) | (-0.456) |
| _cons | 4.661*** | 5.008*** | 5.288*** | 5.321*** | 5.513*** | 5.642*** | 5.994*** | 6.103*** | 5.992*** |
|  | (25.465) | (34.877) | (40.315) | (40.963) | (41.613) | (38.886) | (37.720) | (29.269) | (19.519) |
| Individual FE | Y | Y | Y | Y | Y | Y | Y | Y | Y |
| Time FE | Y | Y | Y | Y | Y | Y | Y | Y | Y |
| adj. R-sq | 0.584 | 0.584 | 0.584 | 0.584 | 0.584 | 0.584 | 0.584 | 0.584 | 0.584 |
| F-Stat | 30.29 | 30.29 | 30.29 | 30.29 | 30.29 | 30.29 | 30.29 | 30.29 | 30.29 |
| Prob > F | (0.000) | (0.000) | (0.000) | (0.000) | (0.000) | (0.000) | (0.000) | (0.000) | (0.000) |
| Log likelihood | -11234 | -11234 | -11234 | -11234 | -11234 | -11234 | -11234 | -11234 | -11234 |
| N | 7,714 | 7,714 | 7,714 | 7,714 | 7,714 | 7,714 | 7,714 | 7,714 | 7,714 |

Standard errors in parenthesis

*Table D5 Results of RIF treatment effects with IPW (High-tech sample)*

| High-tech RIF-Weighted Treat | Model 246 Q10 | Model 247 Q20 | Model 248 Q30 | Model 249 Q40 | Model 250 Q50 | Model 251 Q60 | Model 252 Q70 | Model 253 Q80 | Model 254 Q90 |
|---|---|---|---|---|---|---|---|---|---|
| ECO | 1.390*** | 1.163*** | 0.969*** | 0.819*** | 0.674*** | 0.551*** | 0.463*** | 0.221*** | -0.110* |
|  | (62.176) | (50.069) | (40.055) | (34.371) | (28.325) | (22.692) | (18.246) | (6.321) | (-2.502) |
| ln$\widehat{\text{NECOINT}}$ | 0.073*** | 0.088*** | 0.112*** | 0.110*** | 0.117*** | 0.134*** | 0.171*** | 0.269*** | 0.367*** |
|  | (3.708) | (3.325) | (4.027) | (4.508) | (5.163) | (6.063) | (6.839) | (6.994) | (5.713) |
| lnCAPINT | -0.008 | -0.01 | 0.014 | 0.032* | 0.039** | 0.046** | 0.032+ | 0.015 | 0.026 |
|  | (-0.538) | (-0.686) | (0.940) | (2.072) | (2.706) | (3.162) | (1.959) | (0.634) | (0.650) |
| lnEMP | -0.048** | -0.038* | -0.050** | -0.045** | -0.032* | -0.015 | -0.009 | 0.013 | 0.008 |
|  | (-3.118) | (-2.356) | (-3.027) | (-2.794) | (-2.010) | (-0.870) | (-0.435) | (0.440) | (0.158) |
| _cons | 6.189*** | 6.561*** | 6.782*** | 6.958*** | 7.219*** | 7.514*** | 7.829*** | 8.594*** | 9.269*** |
|  | (62.118) | (46.921) | (45.103) | (51.038) | (55.060) | (55.852) | (51.241) | (37.152) | (22.611) |
| Individual FE | Y | Y | Y | Y | Y | Y | Y | Y | Y |
| Time FE | Y | Y | Y | Y | Y | Y | Y | Y | Y |
| adj. R-sq | 0.529 | 0.529 | 0.529 | 0.529 | 0.529 | 0.529 | 0.529 | 0.529 | 0.529 |
| F-Stat | 11.34 | 11.34 | 11.34 | 11.34 | 11.34 | 11.34 | 11.34 | 11.34 | 11.34 |
| Prob > F | (0.000) | (0.000) | (0.000) | (0.000) | (0.000) | (0.000) | (0.000) | (0.000) | (0.000) |
| Log likelihood | 17726 | 17726 | 17726 | 17726 | 17726 | 17726 | 17726 | 17726 | 17726 |
| N | 7,714 | 7,714 | 7,714 | 7,714 | 7,714 | 7,714 | 7,714 | 7,714 | 7,714 |

Standard errors in parenthesis
*** p<0.001, ** p<0.01, * p<0.05, + p<0.1

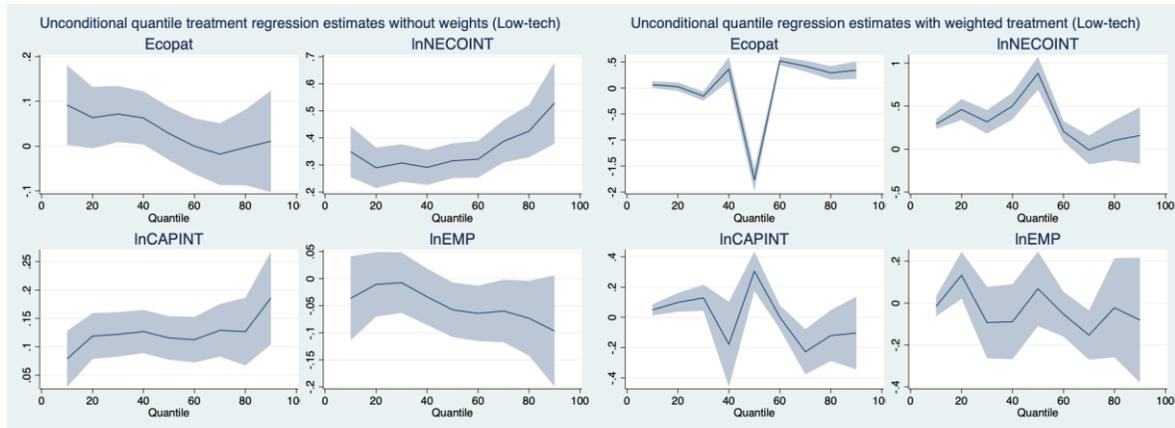

*Figure D2 Non-linear estimations based on RIF treatment effects (low-tech sample)*

*Table D6 Results of RIF treatment effects without weights (Low-tech sample)*

| Low-tech RIF-Treat | Model 255 Q10 | Model 256 Q20 | Model 257 Q30 | Model 258 Q40 | Model 259 Q50 | Model 260 Q60 | Model 261 Q70 | Model 262 Q80 | Model 263 Q90 |
|---|---|---|---|---|---|---|---|---|---|
| ECO | 0.092* | 0.064+ | 0.072* | 0.063* | 0.028 | 0 | -0.018 | -0.003 | 0.011 |
|  | (2.023) | (1.820) | (2.243) | (2.087) | (0.936) | (-0.004) | (-0.504) | (-0.068) | (0.188) |

|  | 0.349*** | 0.289*** | 0.307*** | 0.291*** | 0.315*** | 0.321*** | 0.387*** | 0.425*** | 0.529*** |
| --- | --- | --- | --- | --- | --- | --- | --- | --- | --- |
| ln$\widehat{\text{NECOINT}}$ | (7.254) | (7.584) | (8.724) | (8.826) | (9.614) | (9.332) | (9.753) | (8.609) | (6.885) |
| lnCAPINT | 0.079** | 0.119*** | 0.122*** | 0.127*** | 0.116*** | 0.112*** | 0.129*** | 0.127*** | 0.186*** |
|  | (3.124) | (5.760) | (6.092) | (6.539) | (5.886) | (5.491) | (5.457) | (4.164) | (4.454) |
| lnEMP | -0.036 | -0.011 | -0.008 | -0.034 | -0.058* | -0.064* | -0.060* | -0.073* | -0.097+ |
|  | (-0.921) | (-0.352) | (-0.265) | (-1.269) | (-2.236) | (-2.462) | (-2.029) | (-2.078) | (-1.847) |
| _cons | 5.534*** | 5.643*** | 5.922*** | 6.181*** | 6.551*** | 6.837*** | 7.073*** | 7.504*** | 7.780*** |
|  | (34.418) | (43.485) | (46.822) | (50.667) | (53.384) | (54.136) | (48.194) | (40.316) | (31.040) |
| Individual FE | Y | Y | Y | Y | Y | Y | Y | Y | Y |
| Time FE | Y | Y | Y | Y | Y | Y | Y | Y | Y |
| adj. R-sq | 0.623 | 0.623 | 0.623 | 0.623 | 0.623 | 0.623 | 0.623 | 0.623 | 0.623 |
| F-Stat | 22.1 | 22.1 | 22.1 | 22.1 | 22.1 | 22.1 | 22.1 | 22.1 | 22.1 |
| Prob > F | (0.000) | (0.000) | (0.000) | (0.000) | (0.000) | (0.000) | (0.000) | (0.000) | (0.000) |
| Log likelihood | -12101 | -12101 | -12101 | -12101 | -12101 | -12101 | -12101 | -12101 | -12101 |
| N | 7,007 | 7,007 | 7,007 | 7,007 | 7,007 | 7,007 | 7,007 | 7,007 | 7,007 |

Standard errors in parenthesis
*** p<0.001, ** p<0.01, * p<0.05, + p<0.1

*Table D7 Results of RIF treatment effects with IPW (Low-tech sample)*

| Low-tech RIF-Weighted Treat | Model 264 Q10 | Model 265 Q20 | Model 266 Q30 | Model 267 Q40 | Model 268 Q50 | Model 269 Q60 | Model 270 Q70 | Model 271 Q80 | Model 272 Q90 |
| --- | --- | --- | --- | --- | --- | --- | --- | --- | --- |
| ECO | 0.065* | 0.026 | -0.156*** | 0.365** | -1.764*** | 0.522*** | 0.421*** | 0.291*** | 0.343*** |
|  | (1.988) | (0.631) | (-3.541) | (3.090) | (-15.817) | (12.481) | (7.990) | (4.377) | (4.011) |
| ln$\widehat{\text{NECOINT}}$ | 0.291*** | 0.460*** | 0.317*** | 0.502*** | 0.883*** | 0.206*** | -0.01 | 0.101 | 0.158 |
|  | (9.958) | (7.418) | (4.511) | (6.364) | (8.984) | (3.365) | (-0.116) | (0.852) | (0.946) |
| lnCAPINT | 0.049** | 0.099** | 0.129** | -0.178 | 0.305*** | 0 | -0.227** | -0.121 | -0.103 |
|  | (2.612) | (3.109) | (2.941) | (-1.241) | (4.592) | (0.010) | (-2.981) | (-1.407) | (-0.844) |
| lnEMP | -0.016 | 0.133* | -0.093 | -0.089 | 0.068 | -0.053 | -0.153* | -0.022 | -0.081 |
|  | (-0.644) | (2.321) | (-1.068) | (-0.971) | (0.753) | (-0.970) | (-2.551) | (-0.185) | (-0.533) |
| _cons | 7.128*** | 7.860*** | 7.167*** | 9.597*** | 12.095*** | 8.862*** | 9.110*** | 9.444*** | 10.122*** |
|  | (42.055) | (26.964) | (18.611) | (13.862) | (20.363) | (35.725) | (26.201) | (16.695) | (13.581) |
| Individual FE | Y | Y | Y | Y | Y | Y | Y | Y | Y |
| Time FE | Y | Y | Y | Y | Y | Y | Y | Y | Y |
| adj. R-sq | 0.586 | 0.586 | 0.586 | 0.586 | 0.586 | 0.586 | 0.586 | 0.586 | 0.586 |
| F-Stat | 6.959 | 6.959 | 6.959 | 6.959 | 6.959 | 6.959 | 6.959 | 6.959 | 6.959 |
| Prob > F | (0.000) | (0.000) | (0.000) | (0.000) | (0.000) | (0.000) | (0.000) | (0.000) | (0.000) |
| Log likelihood | -8927 | -8927 | -8927 | -8927 | -8927 | -8927 | -8927 | -8927 | -8927 |
| N | 7,007 | 7,007 | 7,007 | 7,007 | 7,007 | 7,007 | 7,007 | 7,007 | 7,007 |

Standard errors in parenthesis
*** p<0.001, ** p<0.01, * p<0.05, + p<0.1

*Figure D3 Non-linear estimations based on RIF treatment effects (Manufacturing sample)*

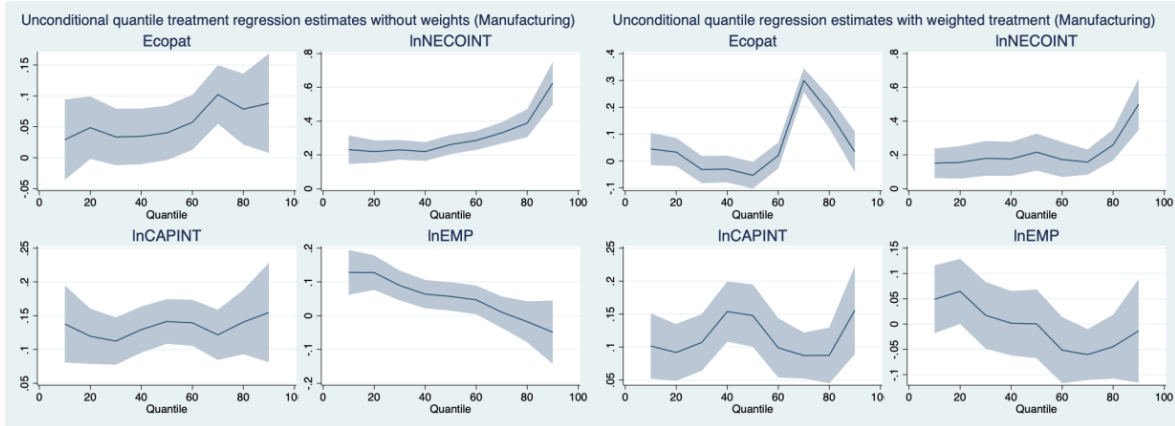

*Table D8 Results of RIF treatment effects without weights (Manufacturing sample)*

| Manufacturing RIF-Treat | Model 273 Q10 | Model 274 Q20 | Model 275 Q30 | Model 276 Q40 | Model 277 Q50 | Model 278 Q60 | Model 279 Q70 | Model 280 Q80 | Model 281 Q90 |
|---|---|---|---|---|---|---|---|---|---|
| ECO | 0.029 | 0.049+ | 0.034 | 0.035 | 0.040+ | 0.057* | 0.102*** | 0.079** | 0.088* |
|  | (0.882) | (1.891) | (1.441) | (1.507) | (1.791) | (2.528) | (4.238) | (2.690) | (2.148) |
| lnNECOINT | 0.231*** | 0.220*** | 0.230*** | 0.220*** | 0.262*** | 0.286*** | 0.331*** | 0.388*** | 0.625*** |
|  | (5.381) | (6.522) | (7.755) | (7.769) | (9.074) | (10.022) | (10.351) | (9.162) | (9.758) |
| lnCAPINT | 0.138*** | 0.120*** | 0.112*** | 0.129*** | 0.141*** | 0.139*** | 0.122*** | 0.141*** | 0.155*** |
|  | (4.735) | (5.735) | (6.302) | (7.429) | (8.397) | (8.019) | (6.420) | (5.783) | (4.121) |
| lnEMP | 0.128*** | 0.128*** | 0.089*** | 0.064** | 0.057** | 0.047* | 0.01 | -0.018 | -0.049 |
|  | (3.772) | (4.885) | (3.979) | (2.985) | (2.658) | (2.163) | (0.438) | (-0.582) | (-1.020) |
| _cons | 4.526*** | 4.920*** | 5.196*** | 5.313*** | 5.342*** | 5.477*** | 5.693*** | 5.773*** | 5.670*** |
|  | (20.095) | (30.295) | (37.849) | (40.119) | (41.073) | (41.069) | (38.764) | (30.585) | (19.640) |
| Individual FE | Y | Y | Y | Y | Y | Y | Y | Y | Y |
| Time FE | Y | Y | Y | Y | Y | Y | Y | Y | Y |
| adj. R-sq | 0.594 | 0.594 | 0.594 | 0.594 | 0.594 | 0.594 | 0.594 | 0.594 | 0.594 |
| F-Stat | 39.23 | 39.23 | 39.23 | 39.23 | 39.23 | 39.23 | 39.23 | 39.23 | 39.23 |
| Prob > F | (0.000) | (0.000) | (0.000) | (0.000) | (0.000) | (0.000) | (0.000) | (0.000) | (0.000) |
| Log likelihood | -12060 | -12060 | -12060 | -12060 | -12060 | -12060 | -12060 | -12060 | -12060 |
| N | 8,583 | 8,583 | 8,583 | 8,583 | 8,583 | 8,583 | 8,583 | 8,583 | 8,583 |

Standard errors in parenthesis
*** $p<0.001$, ** $p<0.01$, * $p<0.05$, + $p<0.1$

*Table D9 Results of RIF treatment effects with IPW (Manufacturing sample)*

| Manufacturing RIF-Weighted Treat | Model 282 Q10 | Model 283 Q20 | Model 284 Q30 | Model 285 Q40 | Model 286 Q50 | Model 287 Q60 | Model 288 Q70 | Model 289 Q80 | Model 290 Q90 |
|---|---|---|---|---|---|---|---|---|---|
| ECO | 0.045 | 0.033 | -0.031 | -0.03 | -0.053* | 0.021 | 0.300*** | 0.182*** | 0.035 |
|  | (1.473) | (1.251) | (-1.218) | (-1.166) | (-2.090) | (0.853) | (13.165) | (6.003) | (0.898) |
| ln$\widehat{NECOINT}$ | 0.151*** | 0.156** | 0.180*** | 0.176*** | 0.216*** | 0.172** | 0.158*** | 0.260*** | 0.500*** |
|  | (3.400) | (3.165) | (3.425) | (3.429) | (3.873) | (3.270) | (4.139) | (5.575) | (6.439) |
| lnCAPINT | 0.102*** | 0.092*** | 0.107*** | 0.154*** | 0.148*** | 0.099*** | 0.087*** | 0.087*** | 0.156*** |
|  | (4.030) | (4.170) | (4.879) | (6.603) | (6.130) | (4.302) | (4.909) | (4.049) | (4.571) |
| lnEMP | 0.049 | 0.065* | 0.017 | 0.002 | 0 | -0.052 | -0.060* | -0.045 | -0.014 |
|  | (1.435) | (1.970) | (0.515) | (0.051) | (0.012) | (-1.546) | (-2.356) | (-1.387) | (-0.260) |
| _cons | 5.057*** | 5.352*** | 5.514*** | 5.467*** | 5.650*** | 6.191*** | 6.369*** | 6.472*** | 5.961*** |
|  | (29.934) | (34.539) | (35.522) | (34.106) | (34.015) | (39.107) | (52.166) | (42.155) | (24.334) |
| Individual FE | Y | Y | Y | Y | Y | Y | Y | Y | Y |
| Time FE | Y | Y | Y | Y | Y | Y | Y | Y | Y |
| adj. R-sq | 0.564 | 0.564 | 0.564 | 0.564 | 0.564 | 0.564 | 0.564 | 0.564 | 0.564 |
| F-Stat | 21.73 | 21.73 | 21.73 | 21.73 | 21.73 | 21.73 | 21.73 | 21.73 | 21.73 |
| Prob > F | (0.000) | (0.000) | (0.000) | (0.000) | (0.000) | (0.000) | (0.000) | (0.000) | (0.000) |
| Log likelihood | -9917 | -9917 | -9917 | -9917 | -9917 | -9917 | -9917 | -9917 | -9917 |
| N | 8,583 | 8,583 | 8,583 | 8,583 | 8,583 | 8,583 | 8,583 | 8,583 | 8,583 |

Standard errors in parenthesis
*** $p<0.001$, ** $p<0.01$, * $p<0.05$, + $p<0.1$

*Figure D4 Non-linear estimations based on RIF treatment effects (Non-manufacturing sample)*

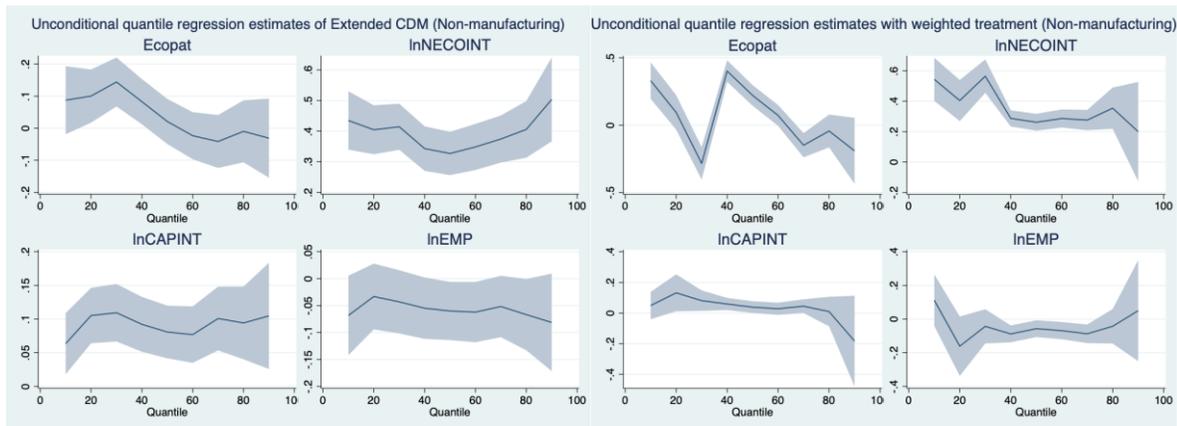

*Table D10 Results of RIF treatment effects without weights (Non-manufacturing sample)*

| Non-manufacturing RIF-Treat | Model 291 Q10 | Model 292 Q20 | Model 293 Q30 | Model 294 Q40 | Model 295 Q50 | Model 296 Q60 | Model 297 Q70 | Model 298 Q80 | Model 299 Q90 |
|---|---|---|---|---|---|---|---|---|---|
| ECO | 0.087 | 0.100* | 0.144*** | 0.083* | 0.021 | -0.023 | -0.041 | -0.01 | -0.031 |

|  | (1.614) | (2.375) | (3.708) | (2.306) | (0.588) | (-0.624) | (-0.982) | (-0.196) | (-0.492) |
|---|---|---|---|---|---|---|---|---|---|
| lnNECOINT̂ | 0.435*** | 0.404*** | 0.414*** | 0.343*** | 0.327*** | 0.348*** | 0.374*** | 0.406*** | 0.504*** |
|  | (8.941) | (9.914) | (10.798) | (9.250) | (9.074) | (9.067) | (9.537) | (8.597) | (7.211) |
| lnCAPINT | 0.063** | 0.105*** | 0.109*** | 0.092*** | 0.081*** | 0.077*** | 0.101*** | 0.094*** | 0.105** |
|  | (2.742) | (5.017) | (5.026) | (4.440) | (4.051) | (3.599) | (4.179) | (3.421) | (2.603) |
| lnEMP | -0.068+ | -0.033 | -0.043 | -0.055+ | -0.060* | -0.062* | -0.052+ | -0.067* | -0.081+ |
|  | (-1.819) | (-1.068) | (-1.443) | (-1.891) | (-2.182) | (-2.177) | (-1.783) | (-1.984) | (-1.763) |
| _cons | 5.725*** | 5.840*** | 6.148*** | 6.541*** | 6.893*** | 7.223*** | 7.417*** | 7.854*** | 8.426*** |
|  | (39.948) | (45.309) | (46.337) | (52.403) | (57.328) | (56.870) | (51.352) | (47.334) | (35.168) |
| Individual FE | Y | Y | Y | Y | Y | Y | Y | Y | Y |
| Time FE | Y | Y | Y | Y | Y | Y | Y | Y | Y |
| adj. R-sq | 0.598 | 0.598 | 0.598 | 0.598 | 0.598 | 0.598 | 0.598 | 0.598 | 0.598 |
| F-Stat | 19.42 | 19.42 | 19.42 | 19.42 | 19.42 | 19.42 | 19.42 | 19.42 | 19.42 |
| Prob > F | (0.000) | (0.000) | (0.000) | (0.000) | (0.000) | (0.000) | (0.000) | (0.000) | (0.000) |
| Log likelihood | -10495 | -10495 | -10495 | -10495 | -10495 | -10495 | -10495 | -10495 | -10495 |
| N | 6,138 | 6,138 | 6,138 | 6,138 | 6,138 | 6,138 | 6,138 | 6,138 | 6,138 |

Standard errors in parenthesis
*** p<0.001, ** p<0.01, * p<0.05, + p<0.1

*Table D11 Results of RIF treatment effects with IPW (Non-manufacturing sample)*

| Non-manufacturing RIF-Weighted Treat | Model 300 Q10 | Model 301 Q20 | Model 302 Q30 | Model 303 Q40 | Model 304 Q50 | Model 305 Q60 | Model 306 Q70 | Model 307 Q80 | Model 308 Q90 |
|---|---|---|---|---|---|---|---|---|---|
| ECO | 0.331*** | 0.095 | -0.284*** | 0.402*** | 0.221*** | 0.072+ | -0.149** | -0.043 | -0.191 |
|  | (4.787) | (1.453) | (-4.636) | (9.906) | (5.660) | (1.827) | (-3.259) | (-0.697) | (-1.524) |
| lnNECOINT̂ | 0.544*** | 0.404*** | 0.565*** | 0.287*** | 0.262*** | 0.286*** | 0.276*** | 0.354*** | 0.2 |
|  | (7.536) | (5.874) | (10.084) | (10.578) | (9.307) | (9.440) | (8.121) | (5.125) | (1.201) |
| lnCAPINT | 0.049 | 0.132* | 0.081* | 0.059** | 0.039* | 0.028 | 0.045* | 0.01 | -0.183 |
|  | (1.081) | (2.149) | (2.383) | (2.918) | (1.981) | (1.384) | (1.964) | (0.195) | (-1.212) |
| lnEMP | 0.112 | -0.161+ | -0.043 | -0.088*** | -0.057* | -0.069** | -0.087** | -0.043 | 0.05 |
|  | (1.430) | (-1.792) | (-0.832) | (-3.448) | (-2.236) | (-2.631) | (-3.107) | (-0.814) | (0.328) |
| _cons | 7.960*** | 6.833*** | 9.293*** | 7.975*** | 8.325*** | 8.734*** | 8.931*** | 9.410*** | 10.588*** |
|  | (19.610) | (14.372) | (33.234) | (54.918) | (55.467) | (54.991) | (50.350) | (31.826) | (12.906) |
| Individual FE | Y | Y | Y | Y | Y | Y | Y | Y | Y |
| Time FE | Y | Y | Y | Y | Y | Y | Y | Y | Y |
| adj. R-sq | 0.743 | 0.743 | 0.743 | 0.743 | 0.743 | 0.743 | 0.743 | 0.743 | 0.743 |
| F-Stat | 3.992 | 3.992 | 3.992 | 3.992 | 3.992 | 3.992 | 3.992 | 3.992 | 3.992 |
| Prob > F | (0.0031) | (0.0031) | (0.0031) | (0.0031) | (0.0031) | (0.0031) | (0.0031) | (0.0031) | (0.0031) |
| Log likelihood | -7894 | -7894 | -7894 | -7894 | -7894 | -7894 | -7894 | -7894 | -7894 |
| N | 6,138 | 6,138 | 6,138 | 6,138 | 6,138 | 6,138 | 6,138 | 6,138 | 6,138 |

Standard errors in parenthesis
*** p<0.001, ** p<0.01, * p<0.05, + p<0.1

*Table D12 Results of CQR with bootstrapping based on CDM models*

| Bootstrap robust CQR | Model 309 Full-sample | Model 310 High-pollution | Model 311 Non-pollution | Model 312 High-tech | Model 313 Low-tech | Model 314 Mnanufacturing | Model 315 Non-manufacturing |
|---|---|---|---|---|---|---|---|
| ln$\widehat{PATINT}$ | 0.412*** | 0.319*** | 0.429*** | 0.421*** | 0.401*** | 0.406*** | 0.396*** |
|  | (15.314) | (4.948) | (11.841) | (9.141) | (10.652) | (10.936) | (10.436) |
| lnCAPINT | 0.110*** | 0.078* | 0.115*** | 0.088*** | 0.120*** | 0.134*** | 0.116*** |
|  | (7.365) | (2.048) | (5.668) | (3.625) | (5.417) | (4.777) | (4.426) |
| lnEMP | -0.106*** | -0.178*** | -0.093*** | -0.054+ | -0.155*** | -0.049+ | -0.153*** |
|  | (-5.236) | (-4.764) | (-4.036) | (-1.926) | (-6.003) | (-1.681) | (-3.833) |
| Individual FE | Y | Y | Y | Y | Y | Y | Y |
| Time FE | Y | Y | Y | Y | Y | Y | Y |
| N | 14,721 | 4,594 | 10,127 | 7,714 | 7,007 | 8,583 | 5,214 |

Standard errors in parenthesis
*** $p<0.001$, ** $p<0.01$, * $p<0.05$, + $p<0.1$

*Table D13 Results of CQR with bootstrapping based on Extended CDM models*

| Bootstrap robust CQR | Model 316 Full-sample | Model 317 High-pollution | Model 318 Non-pollution | Model 319 High-tech | Model 320 Low-tech | Model 321 Mnanufacturing | Model 322 Non-manufacturing |
|---|---|---|---|---|---|---|---|
| ln$\widehat{NECOINT}$ | 0.485*** | 0.192 | 0.580*** | 0.586*** | 0.395*** | 0.323*** | 0.548*** |
|  | (9.067) | (1.199) | (10.939) | (4.385) | (7.702) | (2.856) | (11.01) |
| ln$\widehat{ECOINT}$ | -0.08 | 0.185 | -0.172*** | -0.161 | -0.007 | 0.099 | -0.129*** |
|  | (1.559) | (1.255) | (-3.480) | (-1.274) | (-0.150) | (0.877) | (-3.168) |
| lnCAPINT | 0.111*** | 0.096** | 0.119*** | 0.097*** | 0.123*** | 0.144*** | 0.096*** |
|  | (7.424) | (2.518) | (5.884) | (4.004) | (5.573) | (5.25) | (4.34) |
| lnEMP | -0.101*** | -0.168*** | -0.092*** | -0.042 | -0.154*** | -0.042 | -0.135*** |
|  | (-4.870) | (-4.214) | (-3.950) | (-1.629) | (-6.095) | (-1.498) | (-4.303) |
| Individual FE | Y | Y | Y | Y | Y | Y | Y |
| Time FE | Y | Y | Y | Y | Y | Y | Y |
| N | 14,721 | 4,594 | 10,127 | 7,714 | 7,007 | 8,583 | 6,138 |

Standard errors in parenthesis
*** $p<0.001$, ** $p<0.01$, * $p<0.05$, + $p<0.1$